\documentclass[12pt, a4paper]{article}
 \usepackage[T1]{fontenc}
	 \usepackage{amsmath,amsthm}
	 \usepackage{amsfonts, amssymb}
 \usepackage{float}
 \usepackage{url}
\usepackage{dsfont}
\usepackage{graphicx}
 \usepackage{xcolor}
\usepackage{enumerate}
\usepackage{enumitem}
\usepackage{array}
\usepackage[urlcolor=black,linkcolor=black,citecolor=black,colorlinks=true]{hyperref}
\usepackage[left=2.75cm, right=2.75cm, top=2.25cm, bottom=3.25cm]{geometry}
\usepackage{pict2e}
\usepackage[utf8]{inputenc} 
\usepackage{placeins}
\usepackage{subcaption}

\newtheorem{theoreme}{Th\'{e}or\`{e}me}[section]

\newtheorem{definition}[theoreme]{Definition}

\newtheorem{remarque}[theoreme]{Remark}

\numberwithin{equation}{section}

\title{Constrained deep learning for pricing and hedging european options in incomplete markets}

\author{Nicolas Baradel\footnote{Inria, CMAP, CNRS, \'{E}cole polytechnique, Institut Polytechnique de Paris, 91200 Palaiseau, {nicolas.baradel@polytechnique.edu}.}}

\begin{document}

\maketitle	

\begin{abstract}

In incomplete financial markets, pricing and hedging European options lack a unique no-arbitrage solution due to unhedgeable risks. This paper introduces a constrained deep learning approach to determine option prices and hedging strategies that minimize the Profit and Loss (P\&L) distribution around zero. We employ a single neural network to represent the option price function, with its gradient serving as the hedging strategy, optimized via a loss function enforcing the self-financing portfolio condition.
A key challenge arises from the non-smooth nature of option payoffs (e.g., vanilla calls are non-differentiable at-the-money, while digital options are discontinuous), which conflicts with the inherent smoothness of standard neural networks. To address this, we compare unconstrained networks against constrained architectures that explicitly embed the terminal payoff condition, drawing inspiration from PDE-solving techniques.
Our framework assumes two tradable assets: the underlying and a liquid call option capturing volatility dynamics. Numerical experiments evaluate the method on simple options with varying non-smoothness, the exotic Equinox option, and scenarios with market jumps for robustness. Results demonstrate superior P\&L distributions, highlighting the efficacy of constrained networks in handling realistic payoffs.
This work advances machine learning applications in quantitative finance by integrating boundary constraints, offering a practical tool for pricing and hedging in incomplete markets.

\end{abstract}
\section{Introduction}

In incomplete financial markets, no unique no-arbitrage price exists for derivative securities. This paper proposes a constrained deep-learning framework that simultaneously prices and hedges European options in such environments by determining an initial premium and a dynamic hedging strategy that minimize the dispersion of the terminal profit-and-loss (P\&L) distribution.

\smallbreak

We consider a general incomplete market driven by Brownian motion. The hedging portfolio is required to be self-financing, starts from the initial premium, and aims to replicate the option payoff at maturity as closely as possible. While neural-network approaches to pricing and hedging are not new, see e.g. \cite{hutchinson1994nonparametric}, revisited in \cite{culkin2017machine}. Recent contributions have explored reinforcement learning \cite{halperin2020qlbs}, transaction costs \cite{buehler2019deep}, jumps \cite{agram2024deep}, and other market frictions \cite{fecamp2020deep}.

\smallbreak

Our approach departs from these works in two key respects. First, a \emph{single} neural network outputs both the option price (as its value) and the hedging strategy (as its gradient with respect to the tradable assets), extending the classical complete-market representation in which the delta is the derivative of the price. Second, we explicitly address incomplete markets, whereas many recent deep-learning methods either assume completeness or compute only the price (e.g. BSDE-based solvers in jump-diffusion settings \cite{alasseur2024deep}).

\smallbreak

A well-known practical difficulty when using neural networks for option pricing is enforcing the terminal payoff condition. Vanilla call payoffs are non-differentiable and digital payoffs are discontinuous; standard smooth activations struggle to represent them accurately at maturity without introducing spurious non-smoothness earlier. Even smooth payoffs pose challenges for boundary enforcement in PDE solvers \cite{berg2018unified,liu2019solving}. We therefore adopt and extend constrained architectures that embed the terminal condition by construction.

\smallbreak

Numerical experiments are conducted in a market with two tradable assets (the underlying stock and a liquid vanilla call option written on it). We compare unconstrained networks against several constrained formulations on payoffs of increasing non-smoothness, including an exotic Equinox option. Performance is benchmarked against the classical Black–Scholes delta hedge by comparing terminal P\&L distributions.

\smallbreak

The paper is organized as follows. Section 2 presents the financial setting, the neural-network architecture, the gradient-based hedging strategy, and different strategies for embedding the terminal payoff. Section 3 contains numerical results: we first introduce a stochastic-volatility model with stochastic correlation (making the market incomplete), then compare loss functions and constraint methods on simple and exotic options, and finally assess robustness to jumps and misspecification of the model.

\section{The framework}

This section introduces the mathematical setting, the self-financing condition for the hedging portfolio, the neural network architecture used to simultaneously produce the option price and hedging strategy, and the techniques employed to enforce (possibly non-smooth) terminal payoff conditions.

\subsection{General framework}
Let $\Omega := C([0, T], \mathbb{R}^{d_{\circ}})$ represent the space of continuous functions from $[0, T]$ to $\mathbb{R}^{d_{\circ}}$, where $T > 0$ and functions start at value 0 at time 0. We define the canonical process by $W(\omega) = \omega$, and let $\mathbb{P}$ denote the Wiener measure on the Borel sets of $\Omega$. Consequently, $W = (W^i)_{1 \leq i \leq d_{\circ}}$ consists of $d_{\circ}$ independent Brownian motions.  The filtration $(\mathcal{F}_t)_{0 \leq t \leq T}$ is the (augmented) canonical filtration generated by $W$.

\medbreak

We consider $d < d_{\circ}$ tradable risky assets whose price process $Z = (Z^i)_{1 \leq i \leq d}$ is an $\mathbb{R}^d$-valued, $\mathcal{F}$-adapted semimartingale. The market also contains a risk-free asset with constant interest rate $r \in \mathbb{R}$.

\medbreak

Our goal is to hedge as well as possible a European claim pays $g(Z_T)$ at maturity $T$, where $g: \mathbb{R}^d \to \mathbb{R}$ is a (possibly non-smooth) measurable payoff function.

\begin{definition}[Self-financing portfolio]
A portfolio is specified by its initial value $V_0 \in \mathbb{R}$ and an $\mathbb{R}^d$-valued predictable process $\Delta^Z = (\Delta^{Z,i})_{1 \leq i \leq d}$ representing the number of shares held in each risky asset. The portfolio value process $(V_t)_{0 \leq t \leq T}$ is self-financing if, for all $0 \leq t \leq s \leq T$,
\begin{equation}\label{eqV}
    V_{s} = V_{t} + \int_{t}^{s}r\left(V_{u} - \langle \Delta_u^{Z} \cdot Z_{u}\rangle\right)du + \int_{t}^{s}\langle \Delta_u^{Z} \cdot dZ_{u}\rangle,
\end{equation}
In discrete rebalancing (constant holdings on $[t,s]$), this simplifies to
\begin{equation}
    V_{s} = e^{r(s-t)}\left(V_{t} - \langle \Delta_t^{Z} \cdot Z_{t}\rangle\right) + \langle \Delta_t^{Z} \cdot Z_{s}\rangle.
\end{equation}
\end{definition}

In a complete market with no frictions, there exists a unique no-arbitrage price process $p(t,Z_t)$ and a hedging strategy $\Delta^Z_t = \nabla_z p(t,Z_t)$ such that the self-financing portfolio perfectly replicates the payoff:
\begin{equation}\label{eqp}
    p(t, Z_t) + \int_{t}^{T}r\left(p(u, Z_u) - \langle \nabla^{z} p(u, Z_u) \cdot Z_{u}\rangle\right)du + \int_{t}^{T}\langle \nabla^{z} p(u, Z_u) \cdot dZ_{u}\rangle = g(Z_T).    
\end{equation}

In an incomplete market, no such perfect replication is possible in general. The price is not unique and perfect hedging cannot be achieved. Our goal is therefore to approximate a pricing function $p(t,z)$ whose spatial gradient $\nabla_z p(t,z)$ serves as a hedging strategy, chosen so that the self-financing portfolio yields a terminal value as close as possible to $g(Z_T)$ in a suitable sense. The next subsections introduce a deep neural network to achieve this objective while properly enforcing the (possibly non-smooth) terminal condition $p(T,z) = g(z)$.

\subsection{Neural network}

We represent the candidate pricing function using a fully connected (feedforward) deep neural network. We define the input dimension as $d_{in} \geq 1$, representing the number of variables in the input $x \in \mathbb{R}^{d_{in}}$. For simplicity, we set the output dimension to 1, meaning the network produces a single output value. The network consists of $L \geq 2$ layers, with the number of neurons in each layer denoted by $(D_{\ell})_{1 \leq \ell \leq L}$. Specifically:
\begin{itemize}
    \item The first layer (input layer) has $D_1 = d_{in}$ neurons.
    \item The last layer (output layer) has $D_{L} = 1$ neuron.
    \item The $L-2$ layers in between are hidden layers, each with $D_\ell = D$ neurons for simplicity, where $2 \leq \ell \leq L-1$.
\end{itemize}

A feedforward neural network $N$ is a function that maps an input $x \in \mathbb{R}^{d_{in}}$ to an output in $\mathbb{R}$. It is defined as a composition of transformations:
\begin{equation}\label{defNN}
        x \mapsto A_L \circ \phi \circ A_{L-1} \circ  \ldots \circ \phi \circ A_{1}(x). 
\end{equation}
where:
\begin{itemize}
    \item $A_{\ell}$ for $1 \leq \ell \leq L$ are affine transformations, defined as $A_{\ell}(x) = \mathcal{W}_{\ell} x + \beta_{\ell}$. Here, $\mathcal{W}_{\ell}$ is a weight matrix, and $\beta_{\ell}$ is a bias vector.
    \item The dimensions of these transformations are: $A_1$ maps from $\mathbb{R}^{d_{in}}$ to $\mathbb{R}^{D}$;\\$A_2, \ldots, A_{L-1}$ map from $\mathbb{R}^{D}$ to $\mathbb{R}^D$; and $A_{L}$ maps from $\mathbb{R}^{D}$ to $\mathbb{R}$.
    \item $\phi : \mathbb{R} \mapsto I \subset \mathbb{R}$ is a nonlinear activation function (where $I$ is either $\mathbb{R}$ or a subinterval), applied element-wise to the output of each affine transformation. For example, for a vector $(x_{1}, \ldots, x_D)$ we have $\phi(x_{1}, \ldots, x_D) = (\phi(x_1), \ldots, \phi(x_D))$.
\end{itemize}

The parameters of the neural network are the weight matrices $(W_\ell)_{1 \leq \ell \leq  L}$ and bias vectors $(\beta_\ell)_{1 \leq \ell \leq L}$ collectively denoted as $\theta$. The total number of parameters, $m_{_{L, D}}$, is calculated as
\[
   m_{_{L, D}} = \sum_{\ell=1}^{L}D_{\ell}(1+D_\ell)= (1+d_{in})D + (L-2)D(1+D) + (1+D)  
\]
accounting for the weights and biases across all layers, thus $\theta \in \mathbb{R}^{m_{_{L, D}}}$. We denote the neural network of \eqref{defNN} with parameter $\theta$ as $N_{\theta}$, and the set of all such networks with $L$ layers and $D$ neurons per hidden layer as $(N_{\theta})_{\theta \in \mathbb{R}^{m_{_{L, D}}}}$. 

\smallbreak

We also define the set of all neural networks with $L$ layers and varying hidden layer sizes:
\[
    \mathcal{N}_{L} := \bigcup_{D \geq 0} (N_{\theta})_{\theta \in \mathbb{R}^{m_{_{L, D}}}}.
\]

\paragraph{Universal approximation theorem \mdseries{(\cite[Theorem 3.1]{pinkus1999approximation})}.} Let $\phi$ be a continuous activation function. If $\phi$ is non-polynomial, then for any $L \geq 3$, on any compact set $K \subset \mathbb{R}^{d_{in}}$, the set $\mathcal{N}_{L}$ is dense in $C(K)$, the set of continuous functions defined on $K$ equipped the uniform norm.

\smallbreak

The smoothness of $N_\theta$ is inherited from $\phi$: if $\phi \in C^\infty(\mathbb{R})$, then $N_\theta \in C^\infty(\mathbb{R}^{d_{\text{in}}})$. Standard smooth activations (tanh, sigmoid, softplus, etc.) therefore yield infinitely differentiable pricing functions: desirable away from maturity but problematic at $t=T$, where many option payoffs are only continuous (vanilla call/put) or even discontinuous (digital, barrier, and many exotic options).

\smallbreak

Using non-smooth activations such as ReLU resolves the terminal issue to some extent, but introduces non-differentiability everywhere, including far before maturity, where the true price function is typically smooth.

\smallbreak

These conflicting requirements (smoothness before maturity and possible non-smoothness exactly at maturity) motivate the constrained architectures introduced in the next subsection.

\subsection{Deep learning}

We consider a European claim with (possibly parameterised) payoff $g(Z_T, P)$, where $P \in \mathbb{R}^{k_1}$ collects contractual parameters (strike, barrier, etc.). To increase expressive power and allow the network to reference liquid hedging instruments with their own parameters, we introduce an auxiliary input vector $K \in \mathbb{R}^{k_2}$.

The candidate price at time $t$ is represented by a neural network:
    \begin{equation}\label{NN}
        (t, z, K, P) \mapsto N_{\theta}(T-t, z, K, P),
    \end{equation}

where $T-t$ is time-to-maturity, $z \in \mathbb{R}^d$ are the current levels of the tradable risky assets, and $\theta$ denotes the trainable parameters.

\medbreak

As in the complete-market case, the hedging strategy in the risky assets is obtained directly from the network via automatic differentiation:

\begin{itemize}
    \item $N_{\theta}$ represents the option price,
    \item $\nabla^{z}N_{\theta}$ provides the hedging strategy.
\end{itemize}

In a complete, frictionless market, there exists $\theta^*$ such that the self-financing portfolio initialised at $V_0 = N_{\theta^*}(T, Z_0, K, P)$ and rebalanced according to \eqref{eqp} perfectly replicates the payoff:
\begin{equation}
    N_{\theta^*}(0, Z_T, K, P) = g(Z_T, P) \quad \mathbb{P}\text{-a.s.}
\end{equation}

In an incomplete market with discrete hedging, $N_{\theta}$ satisfies the following approximation for $\Delta t > 0$:
	\begin{equation}\label{V_N_approx}
		\begin{aligned}
		N_{\theta}(T-(t+\Delta t), Z_{t+\Delta t}, K, P) &\approx e^{r\Delta t}\left(N_{\theta}(T-t, Z_{t}, K, P) - \langle \Delta_t^{Z} \cdot Z_{t}\rangle\right) \\
        &\quad + \langle \Delta_t^{Z} \cdot Z_{t+\Delta t}\rangle,\\
		N_{\theta}(0, Z_{T}, K, P) &\approx g(Z_{T}, P).
		\end{aligned}
	\end{equation}
We therefore train the network by minimizing the expected deviation from the discrete-time self-financing condition. In \eqref{V_N_approx} forces the network to discover a price $N_{\theta}$ and corresponding hedge $\nabla^z N_{\theta}$, enabling the neural network to simultaneously learn the option price and hedging strategy by leveraging their interdependence.

\subsection{Training objective and loss functions}

We generate $n$ independent Monte Carlo paths of the tradable assets $Z$ on a time grid $0 = t_0 < t_1 < \cdots < t_m = T$, together with fixed or randomly drawn contract parameters $P^i$ and auxiliary hedging-instrument parameters $K^i$. This yields the simulated dataset:
\begin{equation}\label{simulations}
    \left(t_j, Z_{t_{j}}^{i}, K^{i}, P^{i}\right)_{0 \leq j \leq m}^{1 \leq i \leq n}.
	\end{equation}
The network is trained by minimizing a composite loss that simultaneously enforces the discrete-time self-financing condition along each path and the terminal payoff condition. The latter loss is:
\[
\ell_{T}(\theta) = \sum_{i=1}^{n}\left(N_{\theta}(0, Z_T^i, K^i, P^i) - g(Z_T^i, P^i)\right)^2.
\]
To measure hedging performance, we introduce the following two loss functions.

\begin{definition}[Self-financing approach]\label{def_af}
Based on \eqref{V_N_approx}, we define the associated loss over $n$ paths:
\[
    \begin{aligned}
    \ell_{\texttt{path}}^{SF}(\theta) &:= \sum_{i=1}^{n}\sum_{j=0}^{m-1}\left(\left(N_{\theta}(T-t_{j}, Z_{t_{j}}^i, K^i, P^i) - \langle\Delta_{t_{j}}^{Z} \cdot Z_{t_{j}}^i\rangle\right)e^{r(t_{j+1}-t_{j})} \right. \\
    &\qquad \left. + \langle\Delta_{t_{j}}^{Z} \cdot Z_{t_{j+1}}^i\rangle - N_{\theta}(T-t_{j+1}, Z_{t_{j+1}}^i, K^i, P^i)\right)^2.
    \end{aligned}
\]
where $\Delta_{t_j}^Z := \nabla^z N_{\theta}(T - t_j, Z_{t_j}^i,K^i, P^i)$.    
\end{definition}

\begin{definition}[Profit and Loss approach]\label{def_pl}
    Based on \eqref{eqp}, we define the associated loss over $n$ paths:
\[
    \begin{aligned}
    \ell_{\texttt{path}}^{PL}(\theta) &:= \sum_{i=1}^{n}\left(\sum_{j=0}^{m-1}N_{\theta}(T, Z_{0}^i, K^i, P^i)e^{rT} - \langle\Delta_{t_{j}}^{Z} \cdot Z_{t_{j}}^i\rangle e^{r(T-t_{j})}  \right. \\
    &\qquad \left. + \langle\Delta_{t_{j}}^{Z} \cdot Z_{t_{j+1}}^i\rangle e^{r(T-t_{j+1})} - g(Z_T^i, P^i)\right)^2.
    \end{aligned}
\]
where $\Delta_{t_j}^Z := \nabla^z N_{\theta}(T - t_j, Z_{t_j}^i,K^i, P^i)$.
\end{definition}

The full training objective is:
\begin{equation}\label{loss_discrete_af}
    \ell(\theta) := \ell_{\texttt{path}}^{D}(\theta) + \lambda_{T}\ell_{T}(\theta),
\end{equation}
where $\lambda_T > 0$ is a hyperparameter controlling the strength of terminal payoff enforcement and $D \in \{SF, PL\}$ according to the context.

\smallbreak

Parameters $\theta$ are estimated by stochastic gradient descent on mini-batches of paths and time steps. At iteration $k$, a random subset $\Lambda(k) \subset \{1,\dots,n\} \times \{0,\dots,m-1\}$ of size $Card(\Lambda(k)) \ll n \times m$ is sampled uniformly, and gradients are computed via automatic differentiation and backpropagation through time.

\subsection{Constrained architectures for non-smooth payoffs}

The function $g$ for the terminal condition is not always smooth, i.e. not always differentiable or even not continuous. For the simple vanilla call option, the payoff is continuous but not differentiable in the strike price. For the digital option, it is not even continuous. In general, the price is smooth before the maturity but not at the maturity.

Standard feedforward networks with smooth activations produce infinitely differentiable functions everywhere in their domain. This poses a fundamental difficulty when pricing derivatives whose payoff is not differentiable or even discontinuous. The true price is typically smooth on time-to-maturity $T-t \in (0,T]$ but inherits the payoff’s irregularity at $T-t=0$.

\medbreak

Using non-smooth activations such as ReLU removes global differentiability and therefore destroys the desirable pre-maturity smoothness that diffusion-driven prices exhibit.

\medbreak

For a payoff $g$, we fix a reference function
\[
f(t,z,P),\quad (t, z, P) \in[0,T] \times \mathbb{R}^{d} \times \mathbb{R}^{k_1},
\]
that is continuously differentiable on $(0,T]\times\mathbb{R}^d\times\mathbb{R}^{k_1}$ the terminal condition: 
\[
f(0,z,P)=g(z,P),\quad (z, P) \in  \mathbb{R}^{d} \times \mathbb{R}^{k_1}.
\]
Typical choices for $f$ are closed-form prices from tractable complete-market proxies as Black-Scholes or more sophisticated models. The candidate price is then represented as the convex combination
\begin{equation}\label{N_poids}
    \overline{N}_{\theta}(T-s, z, K, P) := w(s, T)f(T-s, z, P) + w'(s, T)N_{\theta}(T-s, z, K, P),
\end{equation}
where $w$ and $w'$ user-chosen smooth weighting functions. The hedging strategy is obtained by automatic differentiation of $\overline{N}_\theta$:
\[
\Delta^Z_t = \nabla_z \overline{N}_\theta(T-t,z,K,P).
\]
We investigate the four architectures summarized below:
\begin{definition}\label{defw}
We introduce four specific cases of \eqref{N_poids}.
\begin{itemize}
    \item \textbf{Unconstrained}: $w := 0$ and $w' := 1$ corresponds to the plain neural network; no terminal condition enforced by construction.
    \item \textbf{Zero-target}: $w(s, T) := \frac{s}{T}$ and $w' := 1$ corresponds to the case where the network $N_\theta$ is only required to output zero at $T-t=0$, considerably easing the fitting of non-smooth payoffs while leaving pre-maturity behavior nearly unconstrained.
    \item \textbf{Control-variate}: $w := 1$ and $w' := 1$ corresponds to the classical control-variate Monte Carlo: $f$ persistently contributes and $N_\theta$ learns the residual correction.
    \item \textbf{Constrained}: $w(s, T) := \frac{s}{T}$ and $w' := 1-w$ corresponds to the case where the terminal condition $\overline{N}_\theta(0,z,K,P)=g(z,P)$ is satisfied exactly for any parameters $\theta$, eliminating terminal payoff error entirely.
\end{itemize}
\end{definition}

We analyze all cases outlined in Definition \ref{defw}. The \textbf{Unconstrained} approach imposes no constraints, representing a standard neural network. The \textbf{Zero Target} method mitigates the non-smoothness of the terminal condition without significantly affecting the solution away from maturity. The \textbf{Control Variate} approach is similar to Zero Target but assigns a constant weight of 1 to the function $f$, which serves as a control-variate. Finally, the \textbf{Constrained} approach enforces a strict condition at maturity, ensuring the payoff is always satisfied for any neural network.

\section{Numerical results}

We now assess the practical performance of the proposed framework in an incomplete market featuring stochastic volatility and stochastic correlation between the underlying and its volatility process. The market is driven by $d_{\circ}=3$ independent Brownian motions but only $d=2$ assets are tradable: the underlying stock and one liquid vanilla option used as a hedging instrument, making perfect replication impossible, even in continuous-time.

\smallbreak

We first demonstrate that the P\&L loss (Definition \ref{def_pl}) systematically dominates the self-financing loss (Definition \ref{def_af}) in terms of out-of-sample P\&L distribution sharpness. We then systematically compare the four architectures of Definition \ref{defw} (Unconstrained, Zero-target, Control-variate, and Constrained) on a range of simple payoffs of different irregularity, and an exotic one: the Equinox option.

\subsection{The market model}

We define a market model through the following triplet of stochastic processes.

\begin{definition}\label{model_croissant}
Let $\mu \in \mathbb{R}, a > 0, \sigma_{\circ} > 0, \xi > 0, \gamma \in [0.5, 1], b > 0, p_{\circ} \in \mathbb{R}, \chi > 0$. For initial conditions $(x, \sigma, p) \in (\mathbb{R}_+)^2 \times \mathbb{R}$ at time $t \in [0, T]$, and for $s \in [t, T]$, the processes are given by:
\begin{equation}\label{eq_model_croissant}
    \begin{aligned}
        X_{s}^{t, x} &= x + \int_{t}^{s}\mu X_{u}^{t, x}du + \int_{t}^{s}\Sigma_{u}^{t, \sigma}X_{u}^{t, x}dW_u^1,\\
        \Sigma_{s}^{t, \sigma} &= \sigma + \int_{t}^{s}-a(\Sigma_{u}^{t, \sigma} - \sigma_{\circ})du + \int_{t}^{s}\xi (\Sigma_u^{t, \sigma})^\gamma d(\rho_u^{t, p}W_u^1 + \sqrt{1-(\rho_u^{t, p})^2}W_u^2),  \\
        P_{s}^{t, p} &= p + \int_{t}^{s}-b(P_{u}^{t, p} - p_{\circ})du + \int_{t}^{s}\chi dW_u^3, 
    \end{aligned}
\end{equation}
    where the correlation process is defined as $\rho^{t, p} := \tanh(P^{t, p})$.
\end{definition}

In this model, $X^{t, x}$ represents the underlying price with stochastic volatility $\Sigma^{t, \sigma}$, which reverts to a long-term mean $\sigma_{\circ}$ at rate $a$. The volatility $\Sigma^{t, \sigma}$ is correlated with $X^{t, x}$ via a stochastic correlation $\rho^{t, p}$, driven by the process $P^{t, p}$, which itself exhibits mean reversion to $p_{\circ}$. The term $\rho^{t, p}_u W_u^1 + \sqrt{1 - (\rho^{t, p}_u)^2} W_u^2$ ensures a unit-variance Brownian motion with correlation $\rho^{t, p}$ to $W^1$.

\begin{remarque}
The process $(X^{t, x}, \Sigma^{t, \sigma}, P^{t, p})$ in Definition \ref{model_croissant} admits a unique strong solution under the specified conditions.
\end{remarque}

\noindent For simplicity, we henceforth denote $(X, \Sigma, P)$ as $(X^{t, x}, \Sigma^{t, \sigma}, P^{t, p})$.

\subsubsection{The tradable assets}

Of the processes introduced in Definition \ref{model_croissant}, not all are tradable assets. We now define those that are.

\begin{definition}[Tradable assets]
The market includes two tradable assets:
\begin{itemize}
\item The underlying $X$.
\item A European call option $C(K)$ with strike $K > 0$.
\end{itemize}
The price of the call option $C(K)$ at time $s \in [t, T]$ is modeled as the Black-Scholes price with instantaneous volatility $\Sigma_s$:
    \[
        C_{s}(K) := BS(T-s, X_{s}, \Sigma_{s}, r, K),
    \]
where $BS(u, x, \sigma, r, K)$  denotes the Black-Scholes price of a European call option with time to maturity $u \in [0, T]$, underlying price $x > 0$, volatility $\sigma > 0$, interest rate $r \in \mathbb{R}$, and strike $K > 0$.
\end{definition}

This framework establishes an incomplete market, as there are two tradable assets but three independent sources of randomness (the Brownian motions $W^1, W^2, W^3$). Consequently, perfect hedging of a derivative is unattainable, though our goal is to minimize the hedging error.

\subsubsection{The neural network}

Recall that we seek to determine the price and hedging strategy for a European option with payoff $g(X_{T}, P)$ where $P \in \mathbb{R}^{k_1}$ for $k_1 \geq 1$ represents the option's parameters (e.g., strike price or barrier level). The neural network of \eqref{NN} is in our context:
    \[
    (t, x, c, K, P) \mapsto N_{\theta}(T-t, x, c, K, P),
    \]
where $\theta$ denotes the trainable parameters, $T-t$ is the time to maturity, $x$ is the price of the underlying asset $X_t$, $c$ is the price of the tradable call option $C_t$, and $K > 0$ is the strike price of the tradable call option $C(K)$.

\smallbreak

Finally,

\begin{itemize}
    \item $N_{\theta}$ represents the option price,
    \item $\partial_{x}N_{\theta}$ and $ \partial_{c}N_{\theta}$ provide the hedging strategies for the underlying asset and the tradable call option, respectively.
\end{itemize}

\subsubsection{Evaluation criterion}

To assess the performance of the trained neural network in pricing and hedging, we evaluate its out-of-sample hedging error on paths simulated from the true data-generating process \eqref{eq_model_croissant}.

Consider an option with time to maturity $T$, initial underlying price $X_0 = x$, initial call price $C_0 = c$, strike $K$, and additional option parameter $P$. The neural network approximates the price function as $N_\theta(T, x, c, K, P)$.

\smallbreak

Given a discretized time grid $0 = t_0 < t_1 < \cdots < t_m = T$, the network defines the hedging strategy at each rebalancing date $t_j$ as the gradients of the approximated value function:
\[
\Delta_{t_j}^x := \partial_{x}N_{\theta}(T-t_j, X_{t_j}, C_{t_j}, K, P), \quad  \Delta_{t_j}^c := \partial_{c}N_{\theta}(T-t_j, X_{t_j}, C_{t_j}, K, P).
\]
At maturity $T$, the option payoff $g(X_T, P)$ is delivered.
\smallbreak

The discounted Profit-and-Loss of the hedged position for a single simulated path $i$ is defined as:
\[
    \begin{aligned}
    P\&L_i := N_{\theta}(T, x, c, K, P) &+ \sum_{j=0}^{m-1}\left(\Delta_{t_j}^x(X_{t_{j+1}} - X_{t_j}) + \Delta_{t_j}^c(C_{t_{j+1}} - C_{t_j})\right)e^{-rt_{j+1}} \\
    &- g(X_T, P)e^{-rT}.
    \end{aligned}
\]
Alternatively, the P\&L could be computed at maturity by multiplying by $e^{r T}$.

\smallbreak

For $n$ independent Monte Carlo trajectories $(X_{t_j}^i, C_{t_j}^i)_{0 \leq j \leq m, 1 \leq i \leq n}$ from the model in \eqref{eq_model_croissant}, we obtain $(P\&L_i)_{1 \leq i \leq n}$. The empirical distribution of these P\&L values is analyzed to assess hedging effectiveness, using metrics such as mean, standard-deviation, and quantiles.

\subsection{Shortcomings of the self-financing approach compared to profit and loss}\label{call}

Consider a European call option with maturity $T$ and terminal payoff $g(X_T, P) := (X_T-P)^+$, where $P > 0$ is the strike price. We simulate trajectories using the stochastic volatility model from Definition \ref{model_croissant}, with the parameters reported in Table \ref{model_param} for the dynamics of the underlying asset, volatility and the stochastic correlation.

\begin{table}[H]
\[
\begin{array}{|c|c|}
\hline
\text{Parameters} & \text{Value} \\
\hline
\mu & 0 \\
a & 5 \\
\sigma_{\circ} & 0.2 \\
\xi & 0.5 \\
\gamma & 0.7 \\
b & 5 \\
p_{\circ} & -0.3 \\
\chi & 0.5 \\
\hline
\end{array}
\]
\caption{Parameters for the model in Definition \ref{model_croissant}.}
\label{model_param}
\end{table}

\subsubsection{Comparison of the different loss functions}

The neural network is a fully-connected multilayer perceptron \cite{rosenblatt1958perceptron} with three layers, each containing 32 neurons, and $\tanh$ activation functions. The network parameters $\theta$ are trained using the Adam optimiser \cite{Kingma2014AdamAM} implemented in PyTorch \cite{paszke2017automatic}.

\medbreak

For each market configuration considered in Definition~\ref{defw}, we train the network separately using two different objective functions:

For each case of Definition \ref{defw}, we train the network separately using two different objective functions:
\begin{itemize}
  \item the \emph{self-financing loss} introduced in Definition~\ref{def_af},
  \item the \emph{P\&L loss} introduced in Definition~\ref{def_pl}.
\end{itemize}
The purpose of this section is to demonstrate that, \emph{across all specifications in Definition~\ref{defw}}, the P\&L loss systematically delivers superior out-of-sample hedging performance compared with the classical self-financing loss.

\medbreak

In the figures reporting the neural network prices, we also display the Black–Scholes price (computed with the initial volatility $\sigma_{\circ}$) purely as a familiar benchmark. This Black–Scholes value is neither the true theoretical price (which does not admit a closed form in our incomplete market with discrete hedging) nor the target of the training procedure. 

\smallbreak

Trained parameters are denoted $\widehat{\theta}$ and are obtained after $10^5$ optimisation epochs.

\paragraph{The unconstrained case}
~
\smallbreak
\noindent In this first setting, we use the neural network directly as the pricing and hedging function:  
$\overline{N}_\theta = N_\theta$.

\smallbreak

\begin{figure}[H]
\centering
\begin{subfigure}[b]{0.48\textwidth} 
    \hspace*{-0.5cm}\includegraphics[scale=0.36]{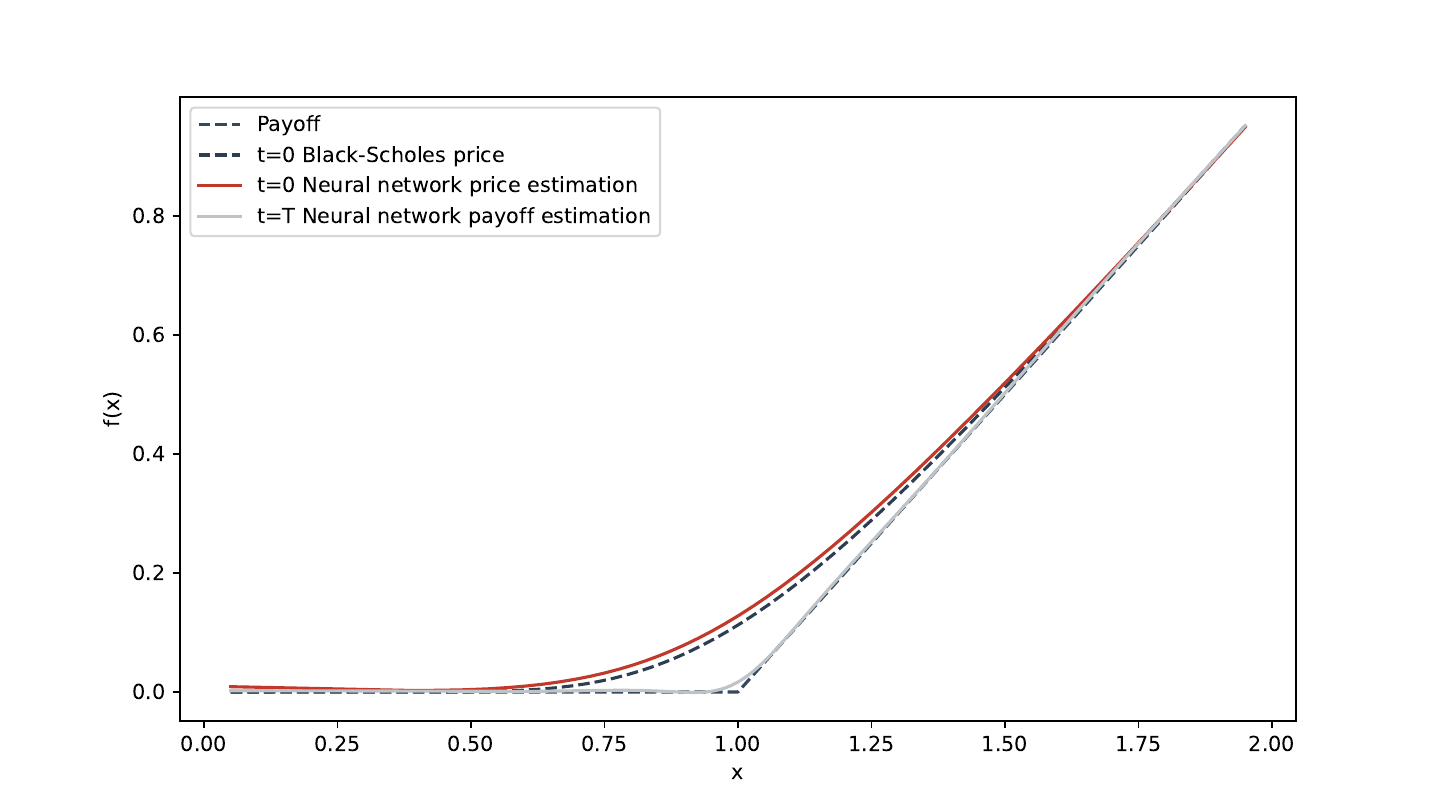}
    \caption{Self-financing loss.}
\end{subfigure}
\hfill
\begin{subfigure}[b]{0.48\textwidth}
    \hspace*{-0.6cm}\includegraphics[scale=0.36]{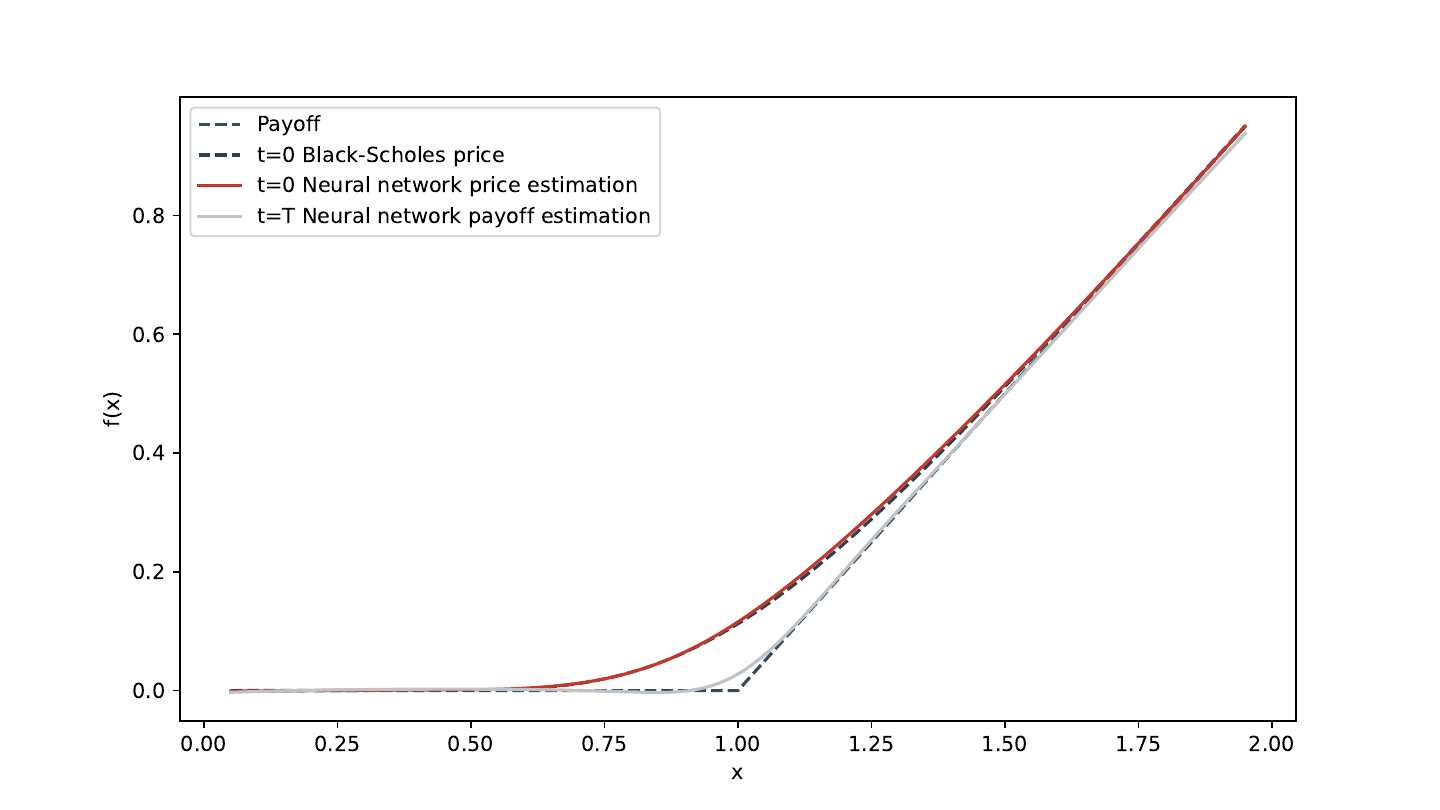}
    \caption{P\&L loss.}
\end{subfigure}
\caption{Pricing functions learned in the unconstrained case with $T=2$, $K=1.2$, and $P=1$.}
\label{fig:price_call_unconstrained}
\end{figure}

Figure~\ref{fig:price_call_unconstrained} displays the learned pricing functions together with the true terminal payoff for a fixed initial volatility $\sigma_\circ$. The left panel corresponds to the network trained with the self-financing (replication) loss, and the right panel to the network trained with the direct P\&L loss.

Both networks struggle to fit the kink of the payoff exactly at-the-money because $N_\theta$ is continuously differentiable while the call payoff is not. This well-known limitation of smooth approximations is visible near $X_T = K$.

\begin{figure}[H]
\centering
\begin{subfigure}[b]{0.48\textwidth} 
    \includegraphics[scale=0.39]{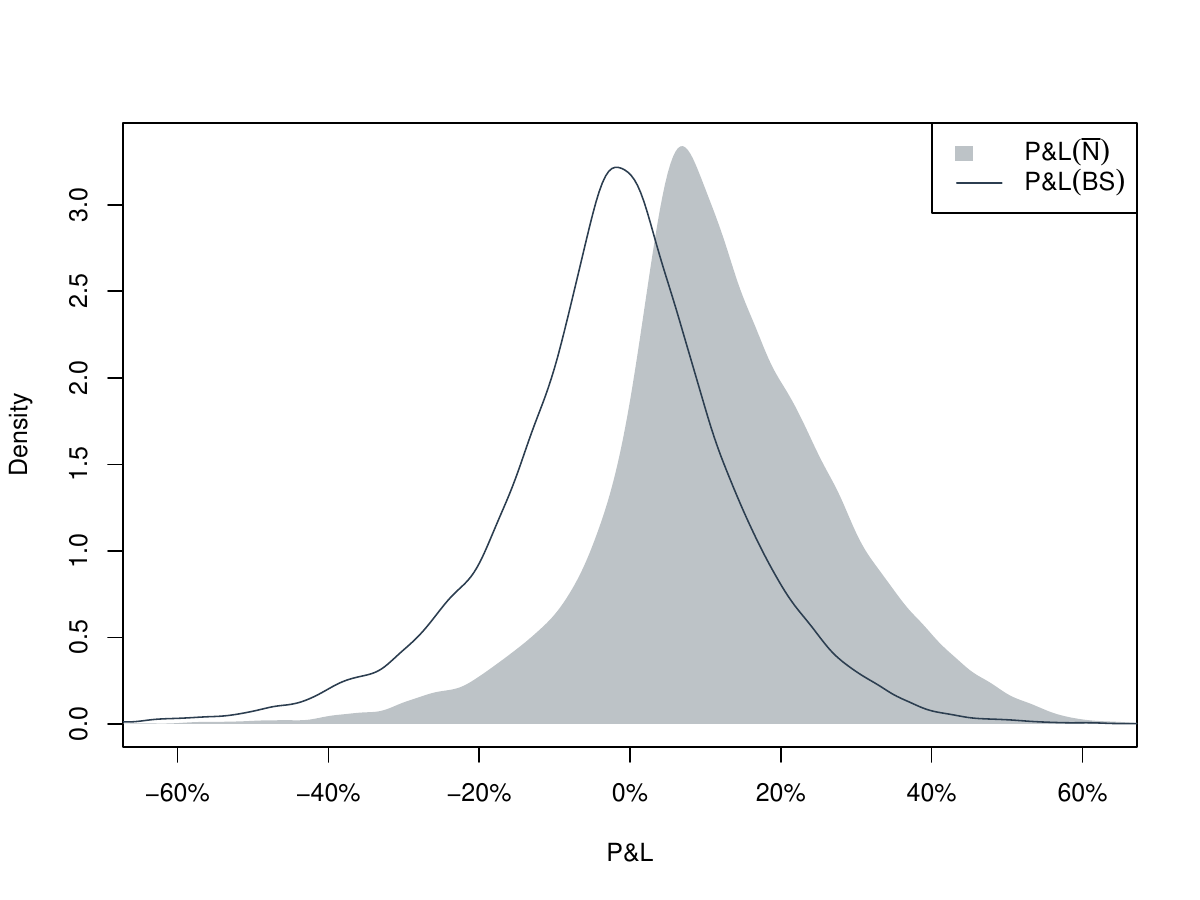}
    \caption{Self-financing loss}
    \label{fig:pl_call_af_unconstrained}
\end{subfigure}
\hfill
\begin{subfigure}[b]{0.48\textwidth}
    \hspace*{-0.25cm}\includegraphics[scale=0.39]{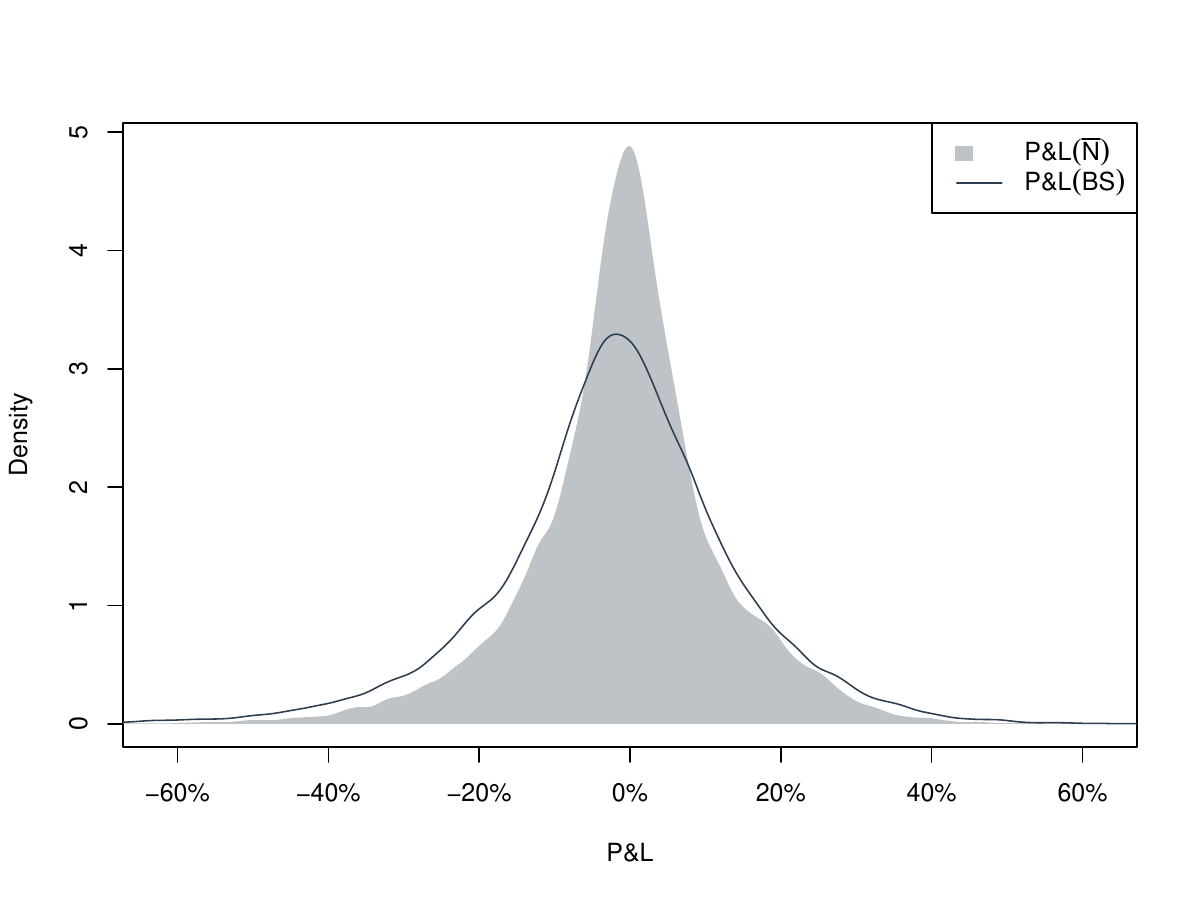}
    \caption{P\&L loss}
    \label{fig:pl_call_pl_unconstrained}
\end{subfigure}
\caption{Empirical P\&L distributions for the neural network hedge and the Black–Scholes delta hedge in the unconstrained case. $T=2, K=1.2, P=1$, and $X_{0} = 1$.}
\label{fig:pl_call_unconstrained}
\end{figure}

The out-of-sample hedging performance is shown in Figure~\ref{fig:pl_call_unconstrained} and quantified in Table~\ref{stats_call_unconstrained}. All P\&L figures are expressed as percentages of the Black-Scholes initial price.

\begin{table}[H]
\centering
\resizebox{0.9\textwidth}{!}{%
\begin{tabular}{|c|c|c|c|}
\hline
\text{Statistic} & \text{Black-Scholes hedging} & \text{Neural Network hedging (SF)} & \text{Neural Network hedging (PL)}\\
\hline
\text{Mean} & -1.839 \% & 11.739 \% & -0.572 \%\\
\text{S.D.} & 16.11 \% & 15.80 \% & 12.84 \%\\
\hline
\text{Quantile $1\%$} & -47.93 \% & -30.64 \% & -37.14 \%\\
\text{Quantile $10\%$} & -21.26 \% & -6.37 \% & -15.51 \%\\
\text{Quantile $90\%$} & 17.11 \% & 31.97 \% & 14.82 \%\\
\text{Quantile $99\%$} & 35.61 \% & 49.67 \% & 31.46 \%\\
\hline
\end{tabular}
}
\caption{Summary statistics of the P\&L distributions shown in Figure~\ref{fig:pl_call_unconstrained}.}
\label{stats_call_unconstrained}
\end{table}

The network trained with the self-financing loss tends to overestimate the option price and delivers hedging errors comparable to a Black–Scholes delta hedge. In contrast, the network trained with the direct P\&L loss produces a nearly unbiased price and significantly reduces the dispersion of the terminal P\&L.

\medbreak

Despite these gains, the difficulty of fitting the non-differentiable payoff at maturity remains in both cases (Figure~\ref{fig:price_call_unconstrained}). For more exotic options, the payoff can even be discontinuous, making the problem considerably harder.

\smallbreak

The next three approaches are specifically designed to better handle non-smooth terminal conditions.

\paragraph{The zero-target case}
~
\smallbreak
\noindent This framework corresponds to:
    \[
    \overline{N}_{\theta}(T-t, x, c, K, P) = \frac{s}{T}f(T-t, x, c, K, P) + N_{\theta}(T-t, x, c, K, P),
    \]
    where $f(0, x, c, K, P) = g(x, P)$. We use the Black-Scholes formula with constant volatility $\sigma_{\circ}$ from Table \ref{model_param}, i.e. in our context:
    \begin{equation}\label{fbs}
        \begin{aligned}
        f(T-t, x, c, K, P) &:= BS(T-t, x, P) \\
        &= x \Phi[d_1(T-t, x, P)] - Pe^{-r(T-t)}\Phi[d_1(T-t, x, P) - \sigma_{\circ}\sqrt{T-t}],
        \end{aligned}
    \end{equation}
where 
    \[
    d_1(T-t, x, P) := \frac{1}{\sigma_{\circ}\sqrt{T-t}}\left[\log\left(\frac{x}{P}\right) + \left(r + \frac{\sigma_{\circ}^2}{2}\right)(T-t)\right].
    \]

\begin{figure}[H]
\centering
\begin{subfigure}[b]{0.48\textwidth} 
    \hspace*{-0.5cm}\includegraphics[scale=0.36]{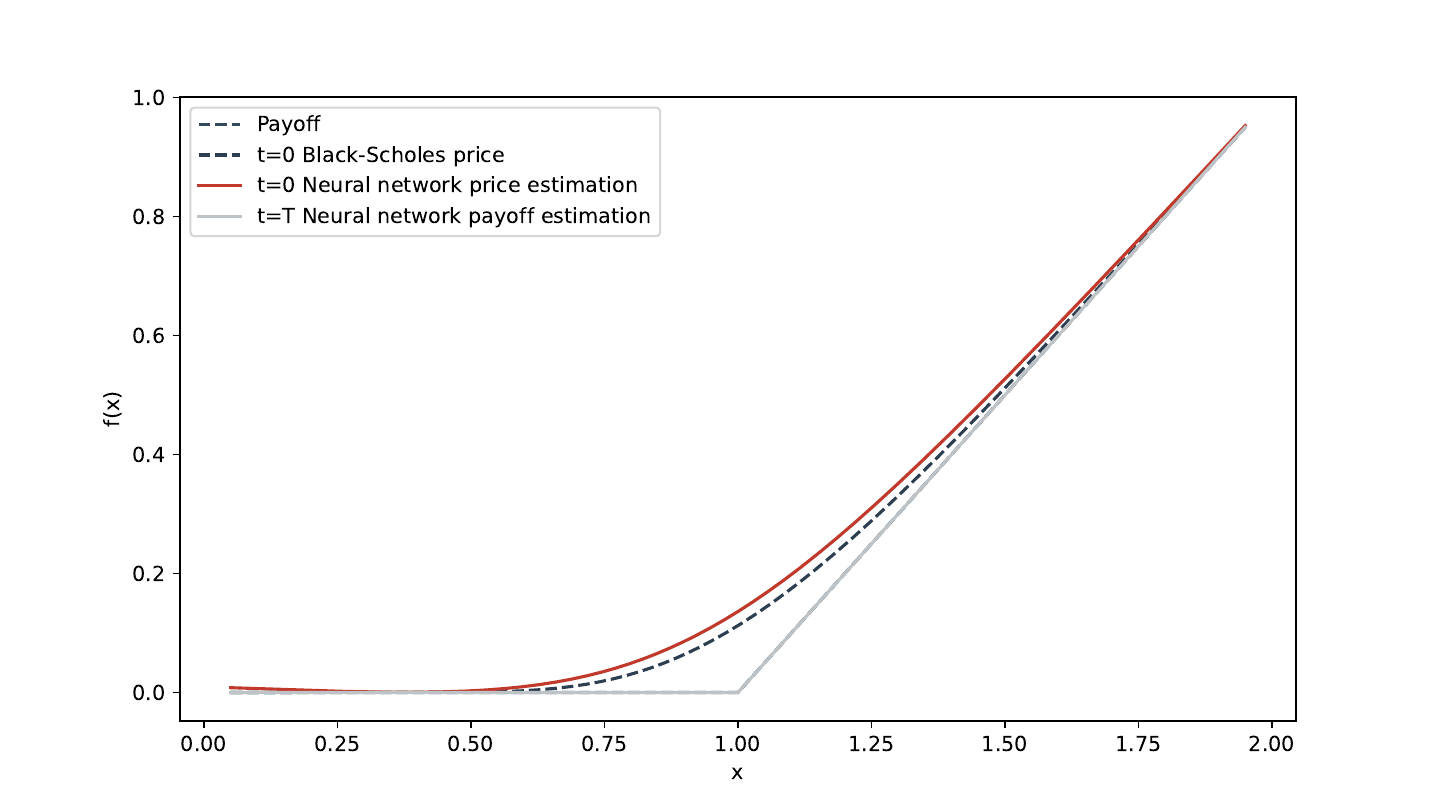}
    \caption{Self-financing loss}
\end{subfigure}
\hfill
\begin{subfigure}[b]{0.48\textwidth}
    \hspace*{-0.6cm}\includegraphics[scale=0.36]{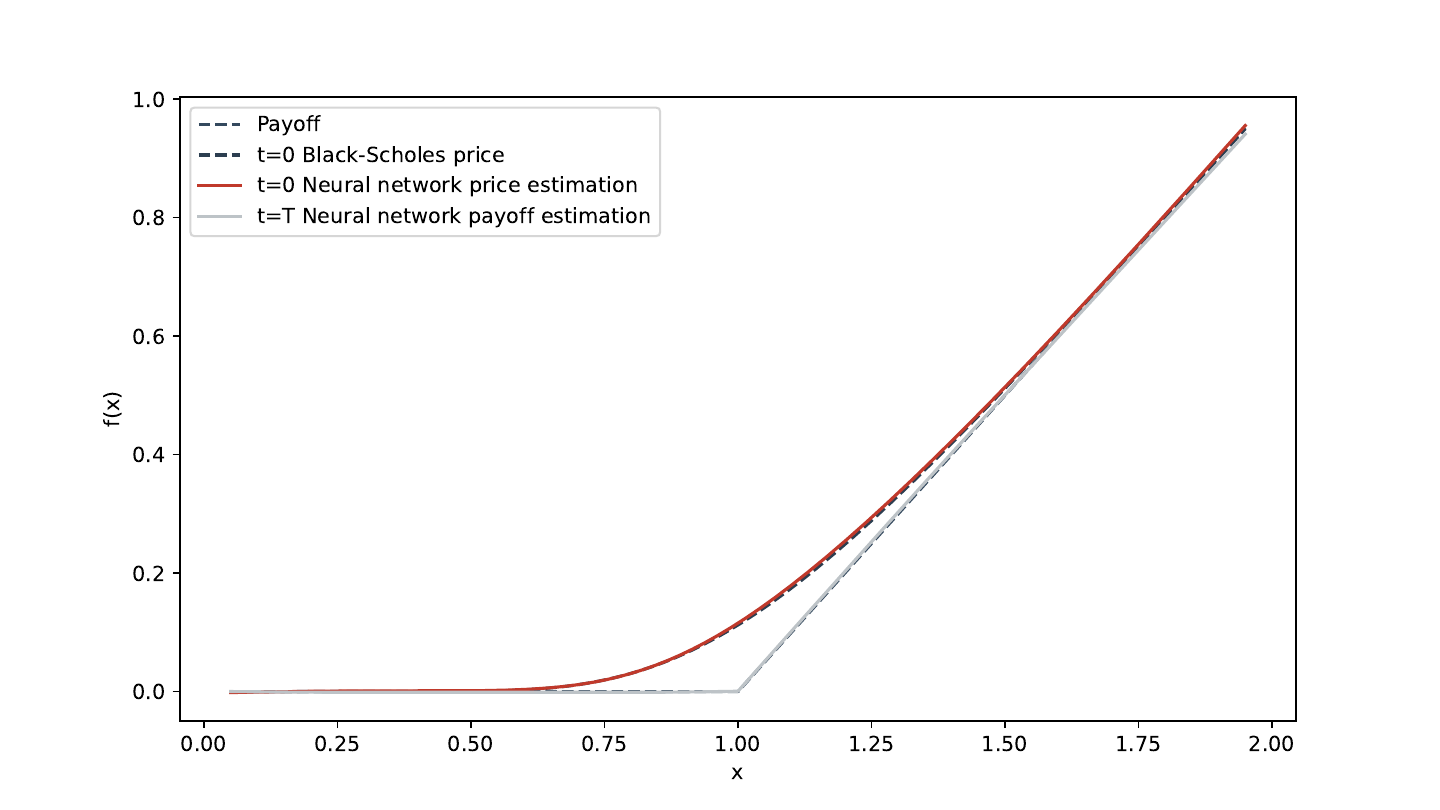}
    \caption{P\&L loss}
\end{subfigure}
\caption{Pricing functions in the zero-target case, $T=2, K=1.2, P=1$.}
\label{fig:price_call_zerotarget}
\end{figure}

Figure~\ref{fig:price_call_zerotarget} shows the resulting pricing functions. Thanks to the embedding, the terminal condition is now almost perfectly matched in both training objectives.

\begin{figure}[H]
\centering
\begin{subfigure}[b]{0.48\textwidth} 
    \includegraphics[scale=0.39]{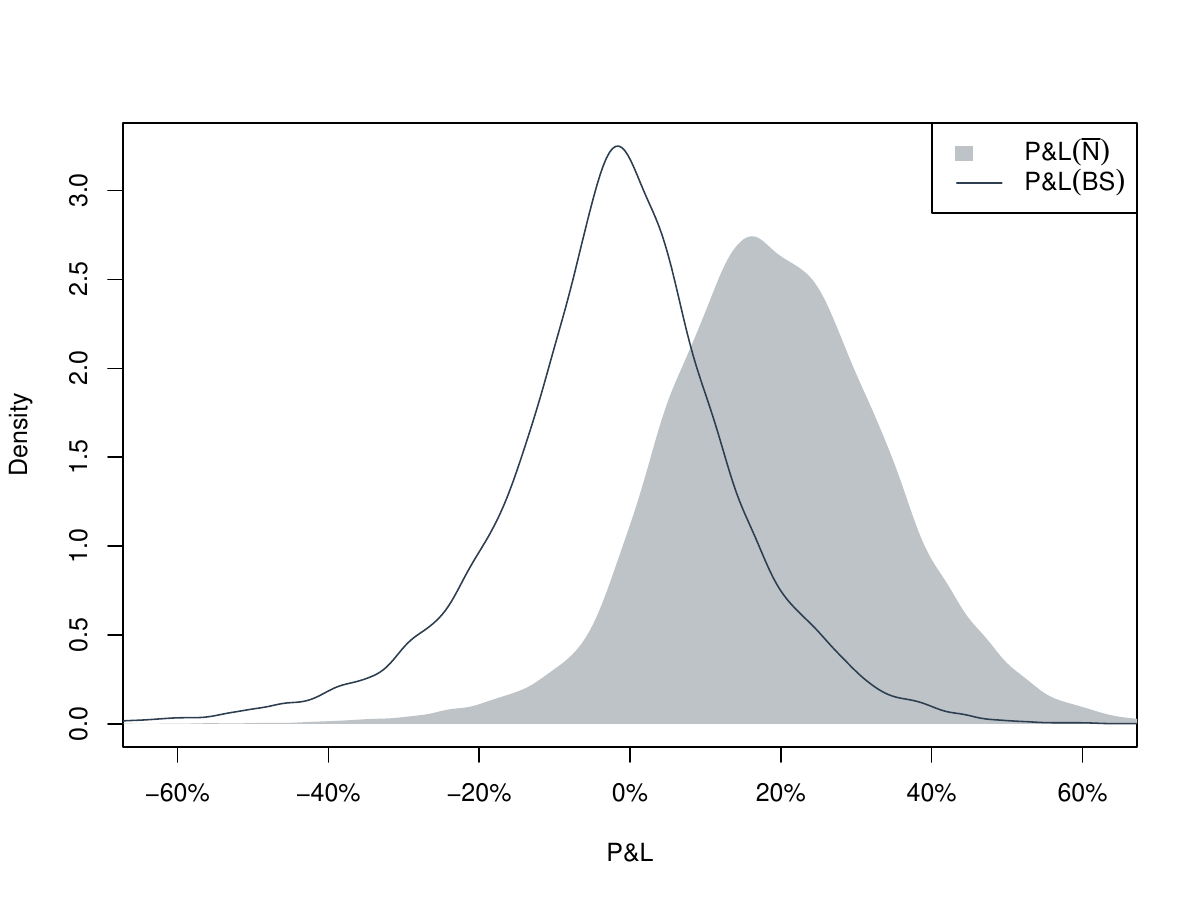}
    \caption{Self-financing loss}
\end{subfigure}
\hfill
\begin{subfigure}[b]{0.48\textwidth}
    \hspace*{-0.25cm}\includegraphics[scale=0.39]{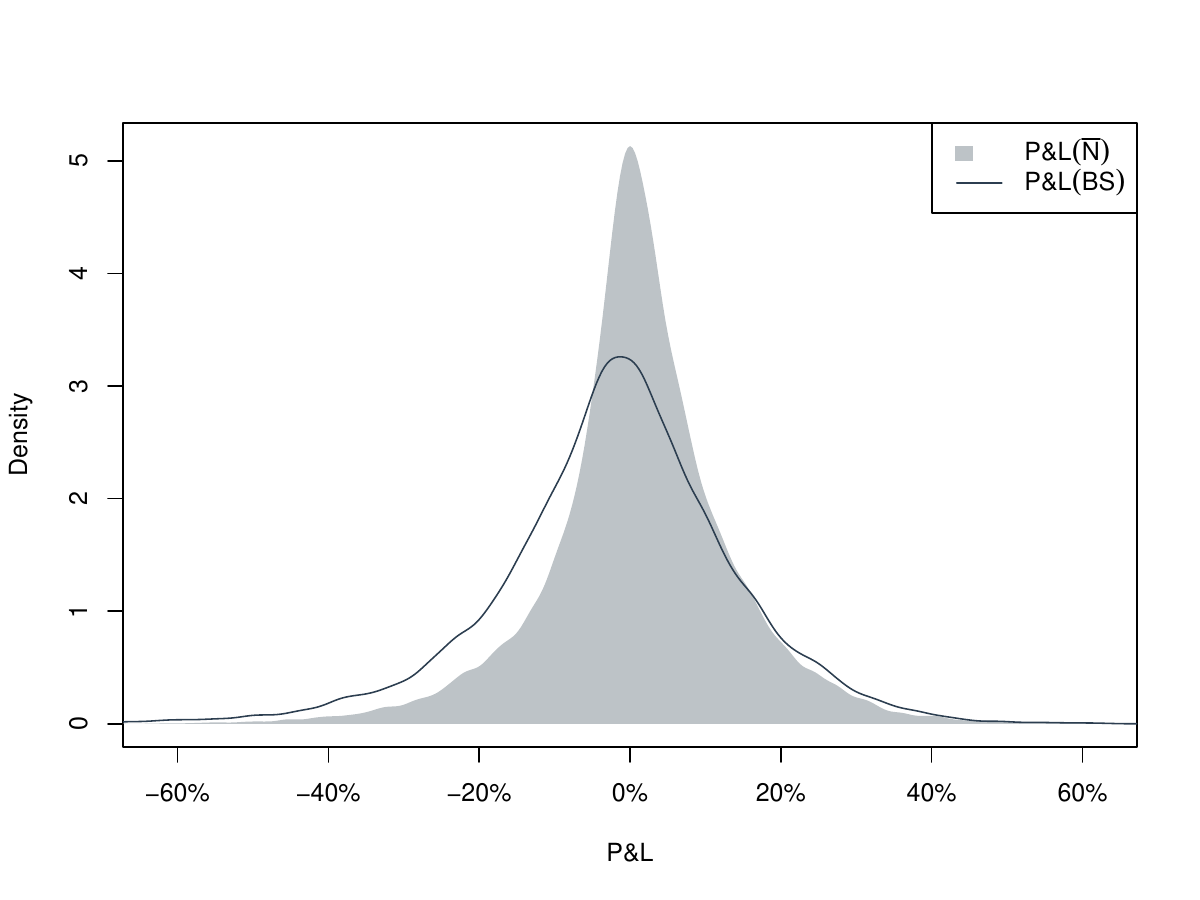}
    \caption{P\&L loss}
\end{subfigure}
\caption{Empirical P\&L distributions for the neural network hedge and the Black–Scholes delta hedge in the zero-target case. $T=2, K=1.2, P=1$ and $X_{0} = 1$.}
\label{fig:pl_call_zerotarget}
\end{figure}

The out-of-sample hedging performance is presented in Figure~\ref{fig:pl_call_zerotarget} and Table~\ref{stats_call_zerotarget} (expressed as percentages of the Black-Scholes price model).

To evaluate the hedging strategy's effectiveness, we compare its Profit and Loss (P\&L) distribution to that of the Black-Scholes model in Figure \ref{fig:pl_call_zerotarget} (with the Self-Financing loss on the left and the P\&L loss on the right) and we provide some statistics about the distributions in Table \ref{stats_call_zerotarget}.

\begin{table}[H]
\centering
\resizebox{0.9\textwidth}{!}{%
\begin{tabular}{|c|c|c|c|}
\hline
\text{Statistic} & \text{Black-Scholes hedging} & \text{Neural Network hedging (SF)} & \text{Neural Network hedging (PL)}\\
\hline
\text{Mean} & -1.839 \% & 19.11 \% & 1.209 \%\\
\text{S.D.} & 16.11 \% & 15.40 \% & 12.57 \%\\
\hline
\text{Quantile $1\%$} & -47.93 \% & -20.71 \% & -34.42 \%\\
\text{Quantile $10\%$} & -21.26 \% & 0.87 \% & -12.98 \%\\
\text{Quantile $90\%$} & 17.11 \% & 38.38 \% & 15.87 \%\\
\text{Quantile $99\%$} & 35.61 \% & 56.01 \% & 34.42 \%\\
\hline
\end{tabular}
}
\caption{Summary statistics of the P\&L distributions shown in Figure~\ref{fig:pl_call_zerotarget}.}
\label{stats_call_zerotarget}
\end{table}

Again, the neural network's hedging strategy and price using the Self-Financing loss yields modest results. The price is overestimated and the quality of the hedge is similar to the one of Black-Scholes. The P\&L loss approach based on Definition \ref{def_pl} outperforms the self-financing approach.

\paragraph{The control-variate case}
~\smallbreak
\noindent We set :
\[
\overline{N}_{\theta}(T-t, x, c, K, P) = f(T-t, x, c, K, P) + N_{\theta}(T-t, x, c, K, P)
\]
where $f$ is chosen as the Black-Scholes formula in \eqref{fbs}.

\smallbreak

\begin{figure}[H]
\centering
\begin{subfigure}[b]{0.48\textwidth} 
    \hspace*{-0.5cm}\includegraphics[scale=0.36]{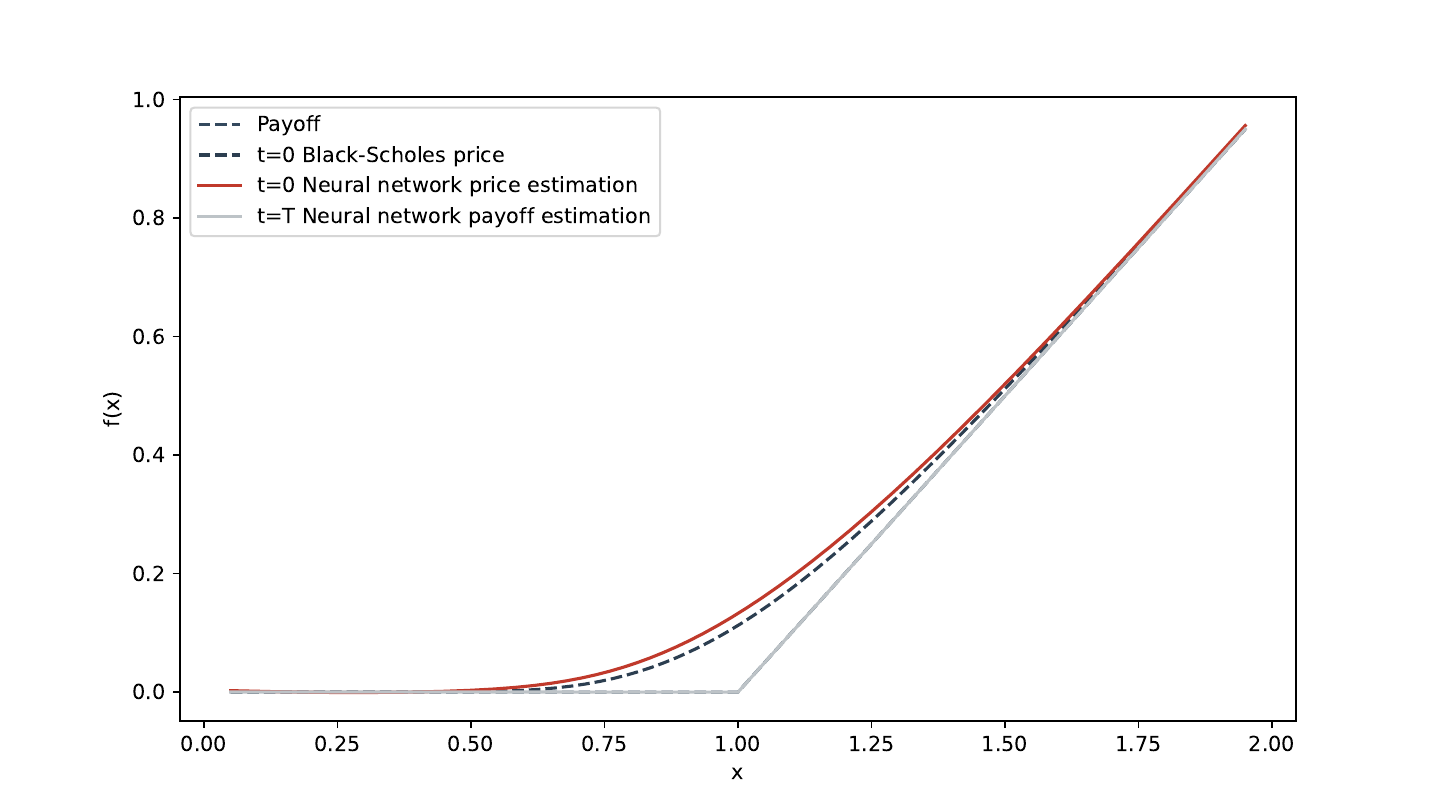}
    \caption{Self-financing loss.}
    \label{fig:price_call_af_controlvariate}
\end{subfigure}
\hfill
\begin{subfigure}[b]{0.48\textwidth}
    \hspace*{-0.6cm}\includegraphics[scale=0.36]{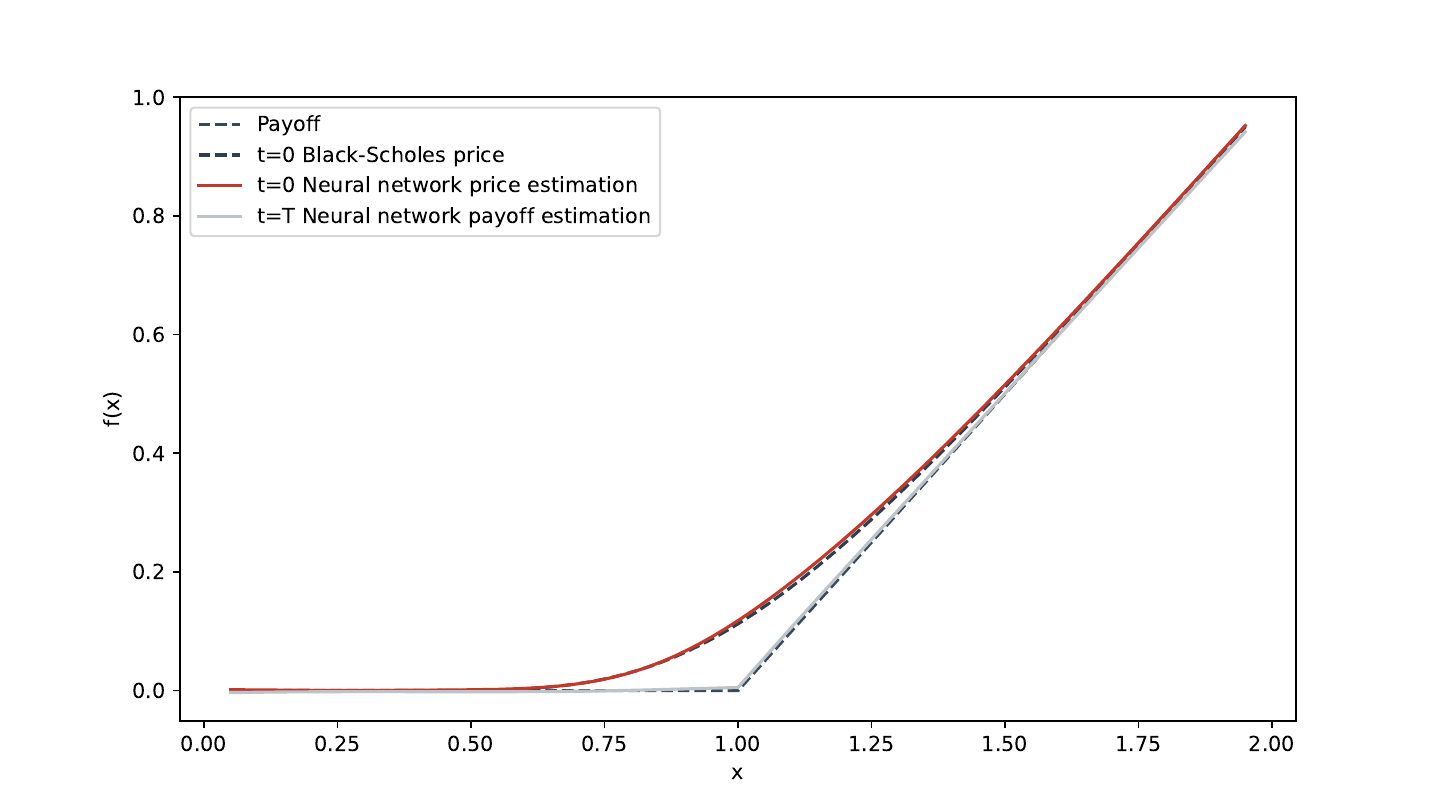}
    \caption{P\&L loss.}
    \label{fig:price_call_pl_controlvariate}
\end{subfigure}
\caption{Pricing functions in the control-variate case, $T=2, K=1.2, P=1$.}
\label{fig:price_call_controlvariate}
\end{figure}

Figure \ref{fig:price_call_controlvariate} shows the prices obtained, for a fixed volatility of $\sigma_{\circ}$, with the Self Financing loss (Figure \ref{fig:price_call_af_controlvariate} on the left) and with the P\&L loss (Figure \ref{fig:price_call_pl_controlvariate} on the right).

The terminal condition is well fitted. We observe again that the Self-Financing approach gives higher prices.

\smallbreak

\begin{figure}[H]
\centering
\begin{subfigure}[b]{0.48\textwidth} 
    \includegraphics[scale=0.39]{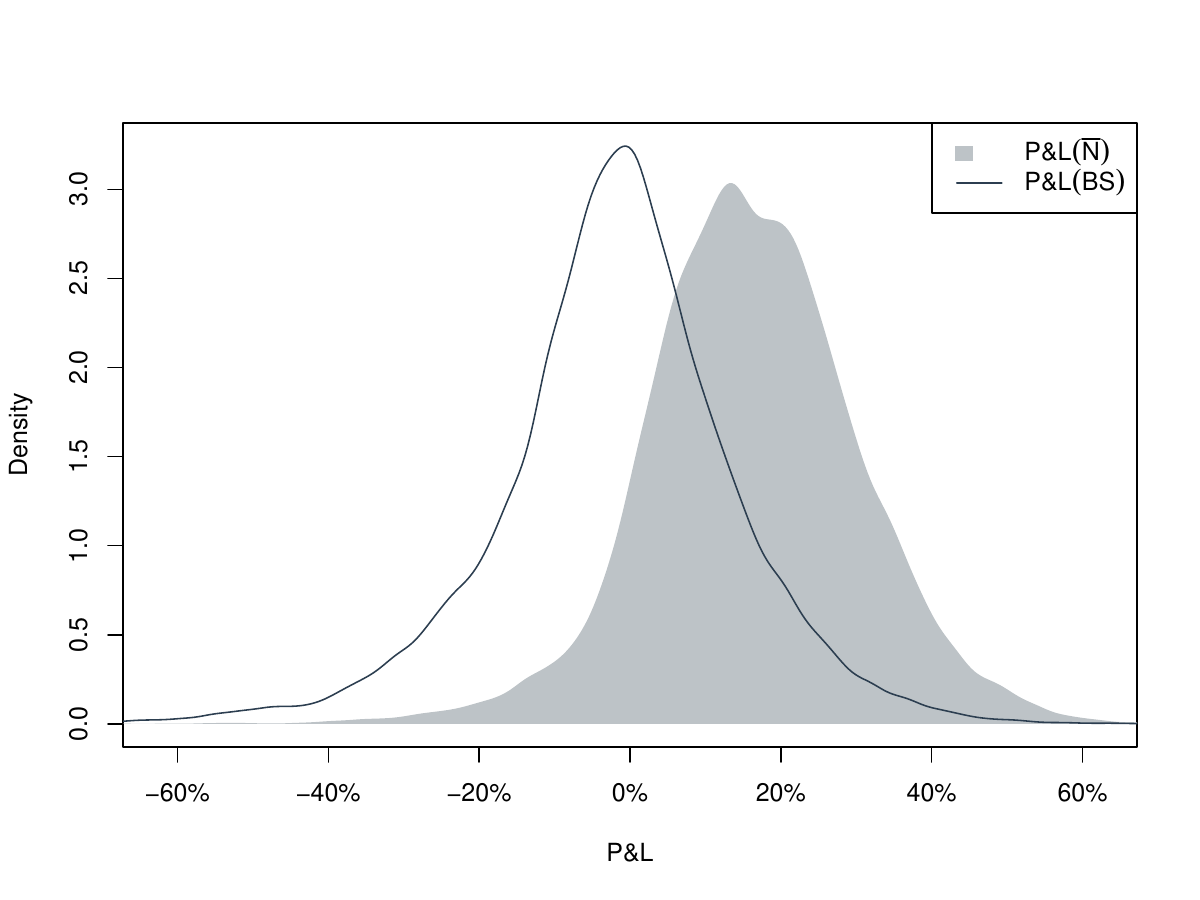}
    \caption{Self-financing loss.}
\end{subfigure}
\hfill
\begin{subfigure}[b]{0.48\textwidth}
    \hspace*{-0.25cm}\includegraphics[scale=0.39]{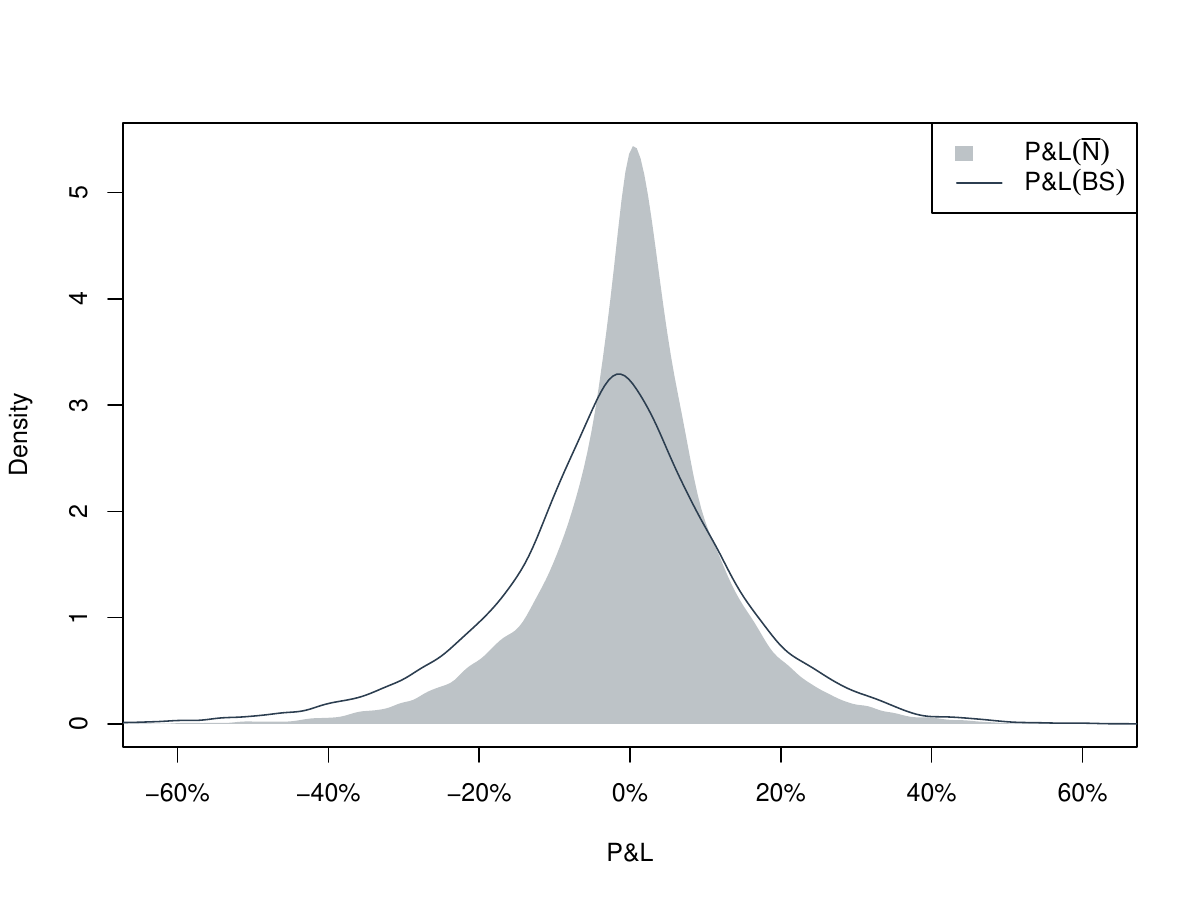}
    \caption{P\&L loss.}
\end{subfigure}
\caption{Empirical P\&L distributions for the neural network hedge and the Black–Scholes delta hedge in the control-variate case. $T=2, K=1.2, P=1$ and $X_{0} = 1$.}
\label{fig:pl_call_controlvariate}
\end{figure}

To evaluate the hedging strategy's effectiveness, we compare its Profit and Loss (P\&L) distribution to that of the Black-Scholes model in Figure \ref{fig:pl_call_controlvariate} (with the Self-Financing loss on the left and the P\&L loss on the right) and we provide some statistics about the distributions in Table \ref{stats_call_controlvariate}.

\begin{table}[H]
\centering
\resizebox{0.9\textwidth}{!}{%
\begin{tabular}{|c|c|c|c|}
\hline
\text{Statistic} & \text{Black-Scholes hedging} & \text{Neural Network hedging (SF)} & \text{Neural Network hedging (PL)}\\
\hline
\text{Mean} & -1.839 \% & 16.181 \% & 0.407 \%\\
\text{S.D.} & 16.11 \% & 15.80 \% & 12.21 \%\\
\hline
\text{Quantile $1\%$} & -47.93 \% & -21.19 \% & -34.45 \%\\
\text{Quantile $10\%$} & -21.26 \% & -0.17 \% & -14.04 \%\\
\text{Quantile $90\%$} & 17.11 \% & 34.01 \% & 14.22 \%\\
\text{Quantile $99\%$} & 35.61 \% & 50.61 \% & 32.40 \%\\
\hline
\end{tabular}
}
\caption{Summary statistics of the P\&L distributions shown in Figure~\ref{fig:pl_call_controlvariate}.}
\label{stats_call_controlvariate}
\end{table}

Again, the neural network's hedging strategy and price using the Self-Financing loss yields modest results. The price is overestimated and the quality of the hedge is similar to the one of Black-Scholes. The P\&L loss approach based on Definition \ref{def_pl} outperforms the self-financing approach.

\paragraph{The constrained case}
~\smallbreak
\noindent In this approach,
\[
\overline{N}_{\theta}(T-t, x, c, K, P) = \frac{s}{T}f(T-t, x, c, K, P) + \left(1-\frac{s}{T}\right)N_{\theta}(T-t, x, c, K, P)
\]
where $f$ is the Black-Scholes formula in \eqref{fbs}.

\smallbreak

\begin{figure}[H]
\centering
\begin{subfigure}[b]{0.48\textwidth} 
    \hspace*{-0.5cm}\includegraphics[scale=0.36]{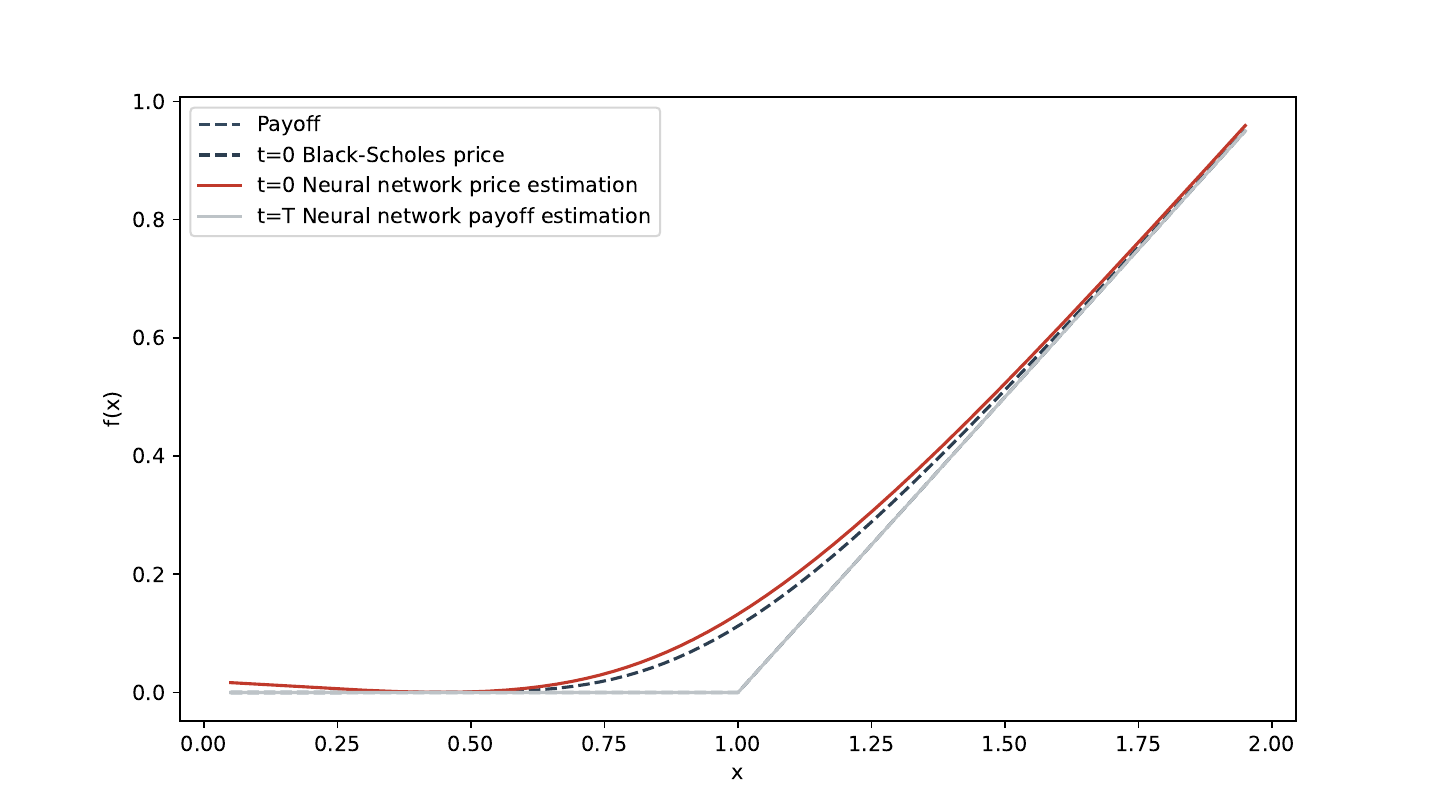}
    \caption{Self-financing loss.}
\end{subfigure}
\hfill
\begin{subfigure}[b]{0.48\textwidth}
    \hspace*{-0.6cm}\includegraphics[scale=0.36]{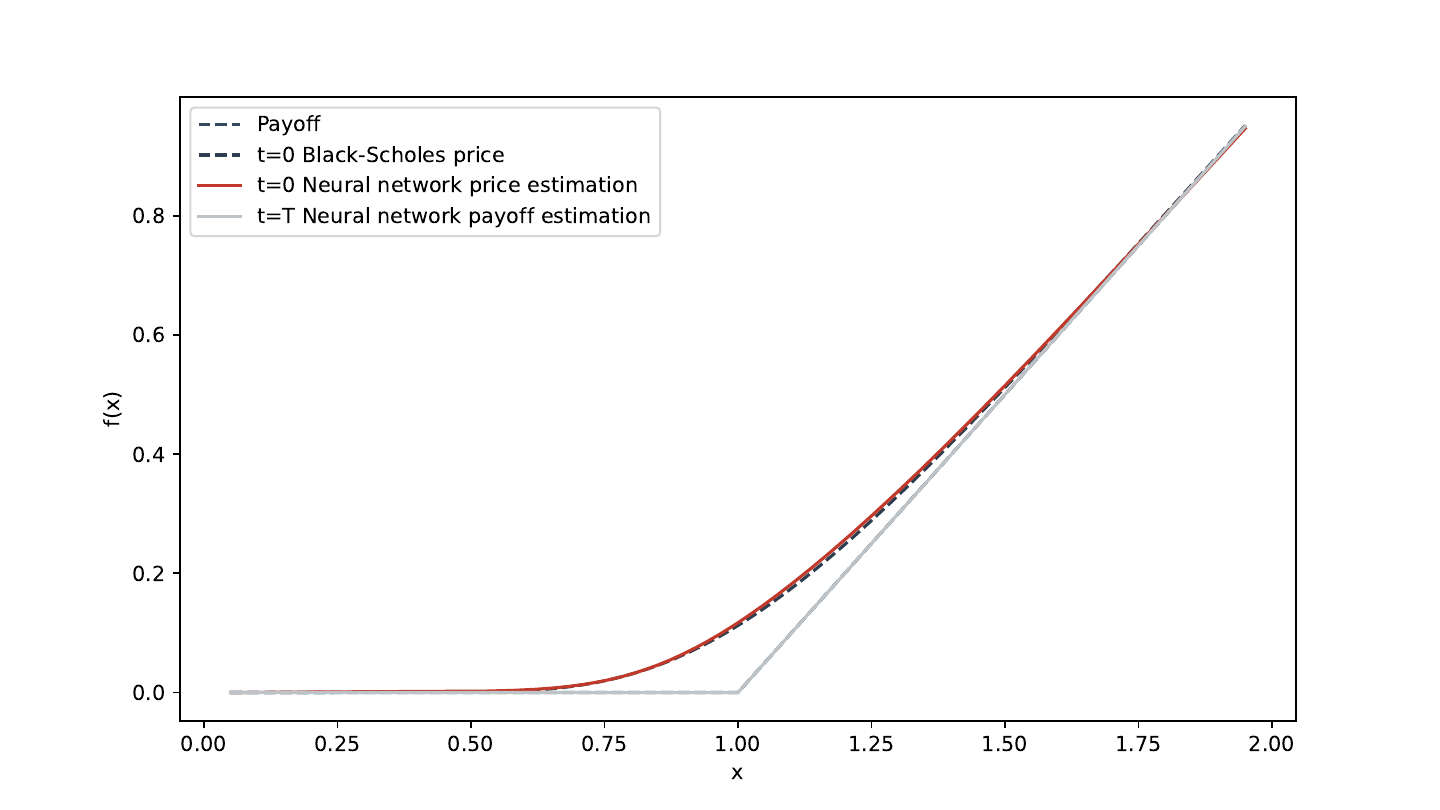}
    \caption{P\&L loss.}
\end{subfigure}
\caption{Pricing functions in the control-variate case, $T=2, K=1.2, P=1$.}
\label{fig:price_call_constrained}
\end{figure}

Figure \ref{fig:price_call_constrained} shows the prices obtained, for a fixed volatility of $\sigma_{\circ}$, with the Self Financing loss on the left and the P\&L loss on the right.

\begin{figure}[H]
\centering
\begin{subfigure}[b]{0.48\textwidth} 
    \includegraphics[scale=0.39]{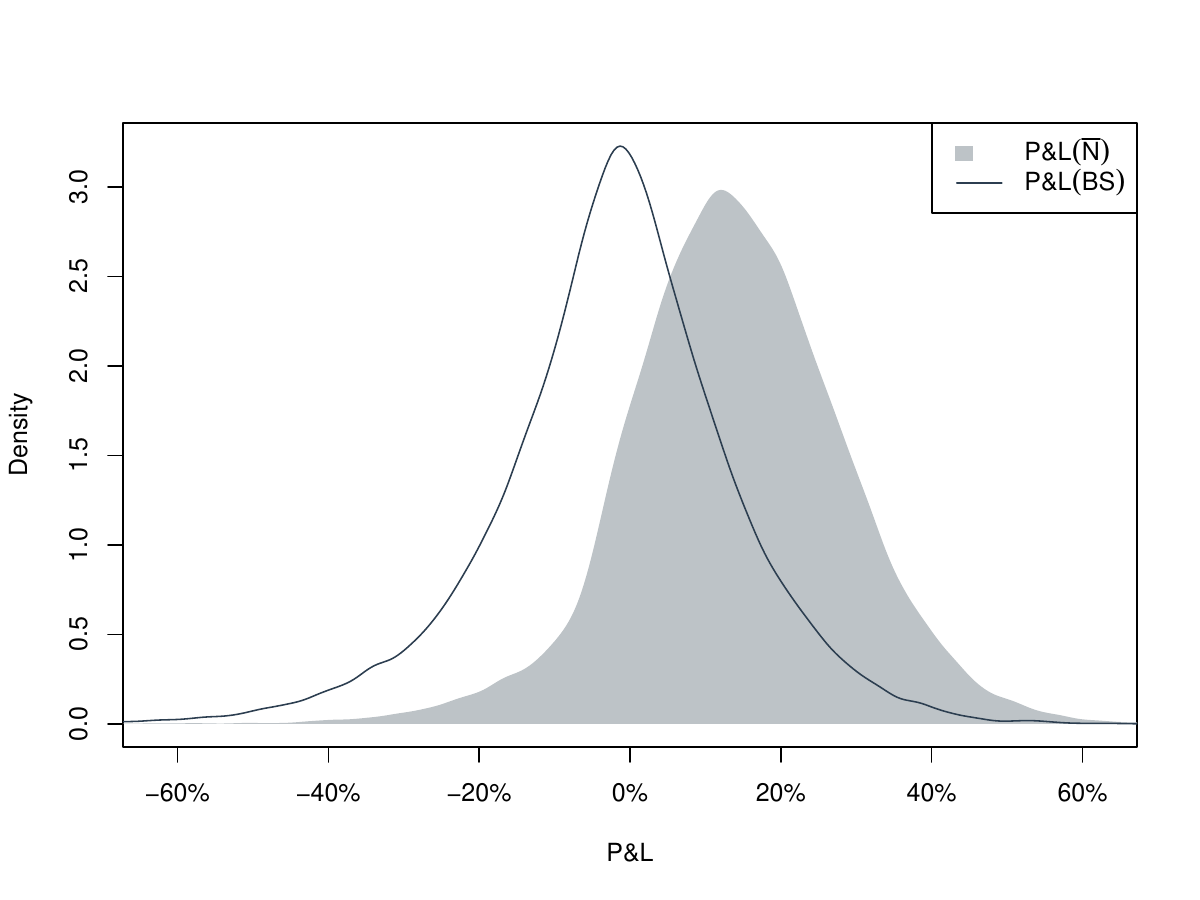}
    \caption{Self-financing loss.}
\end{subfigure}
\hfill
\begin{subfigure}[b]{0.48\textwidth}
    \hspace*{-0.25cm}\includegraphics[scale=0.39]{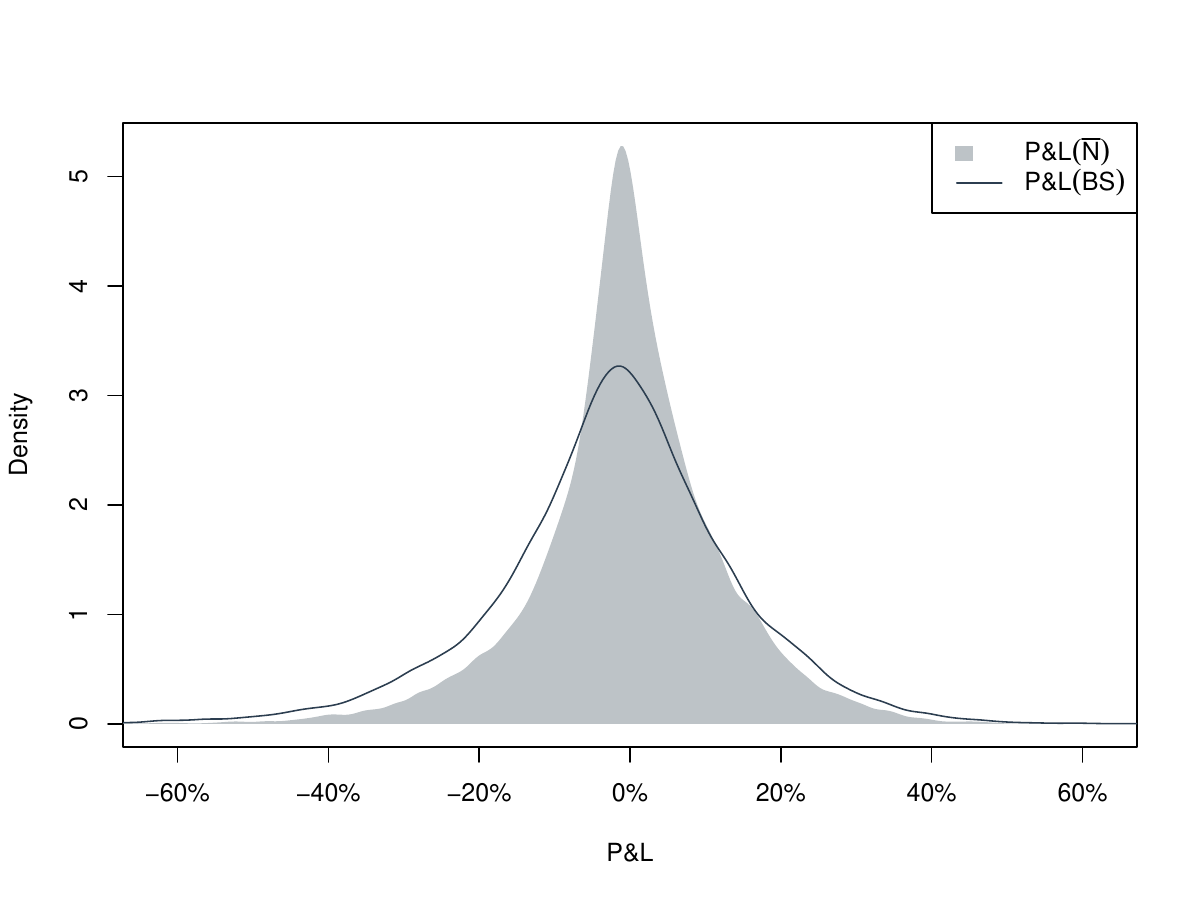}
    \caption{P\&L loss.}
\end{subfigure}
\caption{Empirical P\&L distributions for the neural network hedge and the Black–Scholes delta hedge in the control-variate case. $T=2, K=1.2, P=1$ and $X_{0} = 1$.}
\label{fig:pl_call_constrained}
\end{figure}

The terminal condition is perfectly reproduced in both training settings. We observe again that the Self-Financing approach gives higher prices. Out-of-sample hedging performance is shown in Figure~\ref{fig:pl_call_constrained} and summarized in Table~\ref{stats_call_constrained} (P\&L expressed as percentages of the initial model price).

\begin{table}[H]
\centering
\resizebox{0.9\textwidth}{!}{%
\begin{tabular}{|c|c|c|c|}
\hline
\text{Statistic} & \text{Black-Scholes hedging} & \text{Neural Network hedging (SF)} & \text{Neural Network hedging (PL)}\\
\hline
\text{Mean} & -1.839 \% & 13.960 \% & -0.183 \%\\
\text{S.D.} & 16.11 \% & 14.54 \% & 12.48 \%\\
\hline
\text{Quantile $1\%$} & -47.93 \% & -24.27 \% & -35.54 \%\\
\text{Quantile $10\%$} & -21.26 \% & -2.90 \% & -14.58 \%\\
\text{Quantile $90\%$} & 17.11 \% & 32.06 \% & 14.68 \%\\
\text{Quantile $99\%$} & 35.61 \% & 49.25 \% & 31.67 \%\\
\hline
\end{tabular}
}
\caption{Summary statistics of the P\&L distributions shown in Figure~\ref{fig:pl_call_constrained}.}
\label{stats_call_constrained}
\end{table}

The pattern observed in the previous cases repeats: the self-financing loss produces a significantly upward-biased price and hedging errors comparable to those of a plain Black–Scholes strategy. By contrast, the P\&L loss yields an almost unbiased initial price and markedly lower P\&L dispersion, confirming its clear superiority also when a control-variate structure is used.

\subsubsection{Conclusion}

Across all architectures considered in Definition~\ref{defw}, the direct P\&L loss (Definition~\ref{def_pl}) consistently delivers superior out-of-sample hedging performance compared with the classical self-financing loss. We therefore discard the self-financing loss as a standalone objective and retain only the P\&L loss for the remainder of the study.

\smallbreak

Nevertheless, the pure P\&L loss has one important limitation: the loss function only directly constrains the price at maturity. Formally, if $\overline{N}_{\widehat{\theta}}$ is the trained network and $b(t)$ is any deterministic function such that $b(T) = 0$, then the modified function  
\[
\overline{N}_{\widehat{\theta}}(t, x, c, K, P) + b(t)
\] 
yields the same P\&L loss as $\overline{N}_{\widehat{\theta}}$. This behavior is observed empirically. To obtain coherent prices at all times while preserving strong hedging performance, we combine the two losses with suitable weights. In our experiments, assigning a weight of 5 to the self-financing loss and 1 to the P\&L loss proves effective.

\smallbreak

An alternative simple regularization consists of including paths that start at random intermediate dates. In that case, dates close to maturity naturally receive higher weight in the Monte Carlo average.

\smallbreak

Table~\ref{stats_call} summarises the out-of-sample P\&L performance of the P\&L-trained networks under the four terminal-condition treatments.

\begin{table}[H]
\centering
\resizebox{0.9\textwidth}{!}{%
\begin{tabular}{|c|c|c|c|c|c|}
\hline
\text{Statistic} & \text{Black-Scholes} & \text{Unconstrained} & \text{Zero-target} & \text{Control-variate} & \text{Constrained} \\
\hline
\text{Mean} & -1.839 \% & -0.572 \% & 1.209 \% & 0.407 \% & -0.183 \%\\
\text{S.D.} & 16.11 \% & 12.84 \% & 12.57 \% & 12.21 \% & 12.48 \%\\
\hline
\text{Quantile $1\%$} & -47.93 \% & -37.14 \% & -34.42 \% &-34.45 \% & -35.54 \%\\
\text{Quantile $10\%$} & -21.26 \% & -15.51 \% & -12.98 \% & -14.04 \% & -14.58 \%\\
\text{Quantile $90\%$} & 17.11 \% & 14.82 \% & 15.87 \% & 14.22 \% & 14.68 \%\\
\text{Quantile $99\%$} & 35.61 \% & 31.46 \% & 34.42 \% & 32.40 \%  & 31.67 \%\\
\hline
\end{tabular}
}
\caption{Summary statistics of the P\&L distributions.}
\label{stats_call}
\end{table}

We also compare the four methods for handling the terminal condition. The pricing charts show that the Unconstrained approach struggles to fit the payoff kink at-the-money at maturity. Nevertheless, Table~\ref{stats_call} reveals very similar overall hedging performance across all variants. 
The only noticeable difference is that the Unconstrained case has a slightly higher P\&L standard deviation (roughly 2\% to 5\%) than the three methods that explicitly enforce the terminal payoff.

\subsection{Results on other simple options}

Having established in the previous section that the direct P\&L loss (Definition~\ref{def_pl}) systematically dominates the self-financing loss, we now retain only the P\&L loss and compare the four terminal condition treatments of Definition~\ref{defw} on two additional elementary payoffs :

\begin{itemize}
    \item the square option: $g(X_T,K) = (X_T-K)^2$ (smooth payoff),
    \item the digital (binary) option: $g(X_T,K) = \mathbf{1}_{\{X_T > K\}}$ (discontinuous payoff).
\end{itemize}
These two contracts provide useful contrasts with the vanilla call studied earlier: the square payoff is infinitely differentiable, while the digital payoff is discontinuous at the strike.  
When a baseline function $f$ is required (zero-target, control-variate, and constrained cases), we use the exact Black–Scholes price of the corresponding payoff assuming constant volatility $\sigma_\circ$ (Table~\ref{model_param}).

\subsubsection{The square option}

The square option has payoff $g(X_T,P) = (X_T-P)^2$. Because the terminal condition is smooth, we expect the embedding of $f$ to be less critical than for non-smooth payoffs.

\medbreak

\begin{figure}[H]
\centering
\begin{subfigure}[b]{0.48\textwidth} 
    \hspace*{-0.5cm}\includegraphics[scale=0.36]{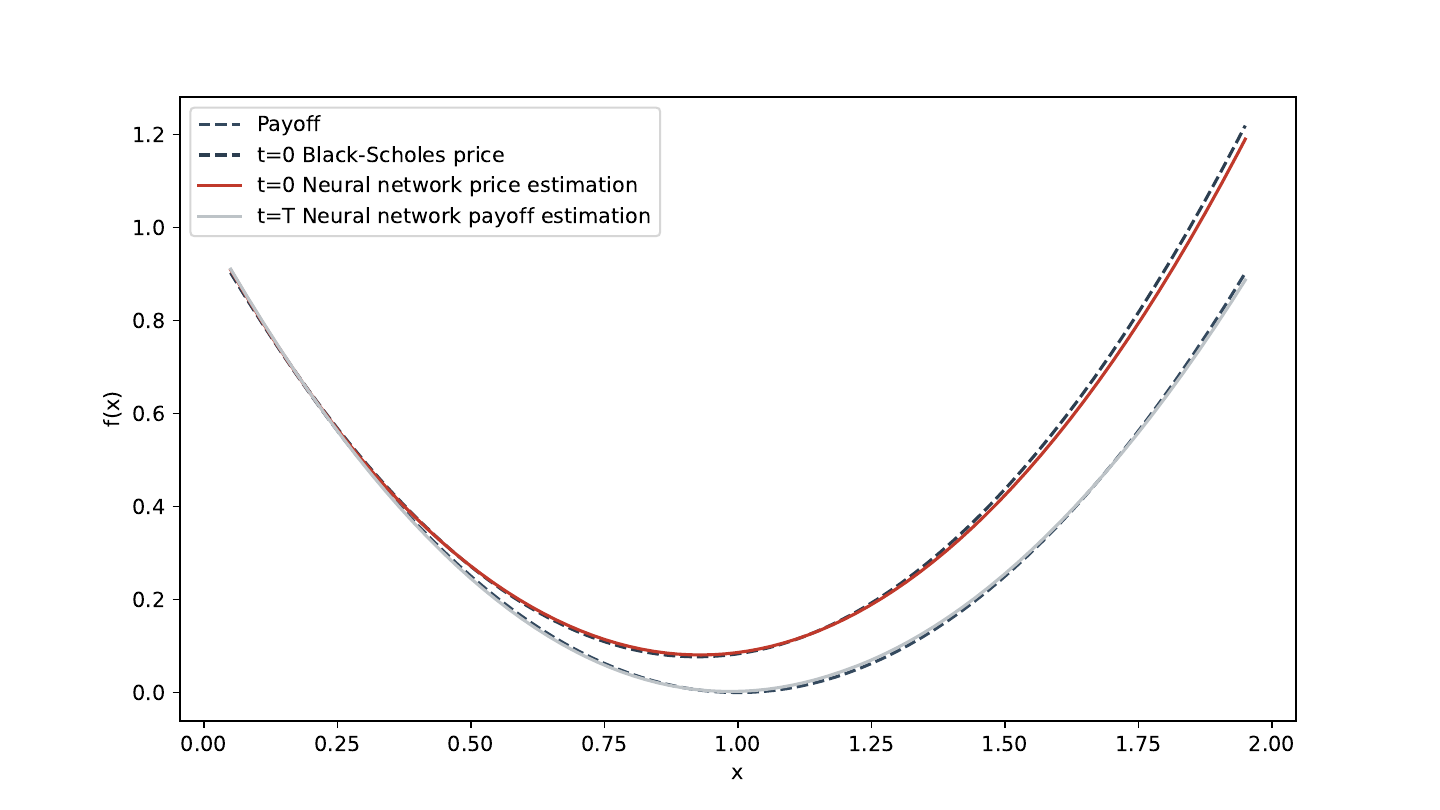}
    \caption{Unconstrained.}
\end{subfigure}
\hfill
\begin{subfigure}[b]{0.48\textwidth}
    \hspace*{-0.6cm}\includegraphics[scale=0.36]{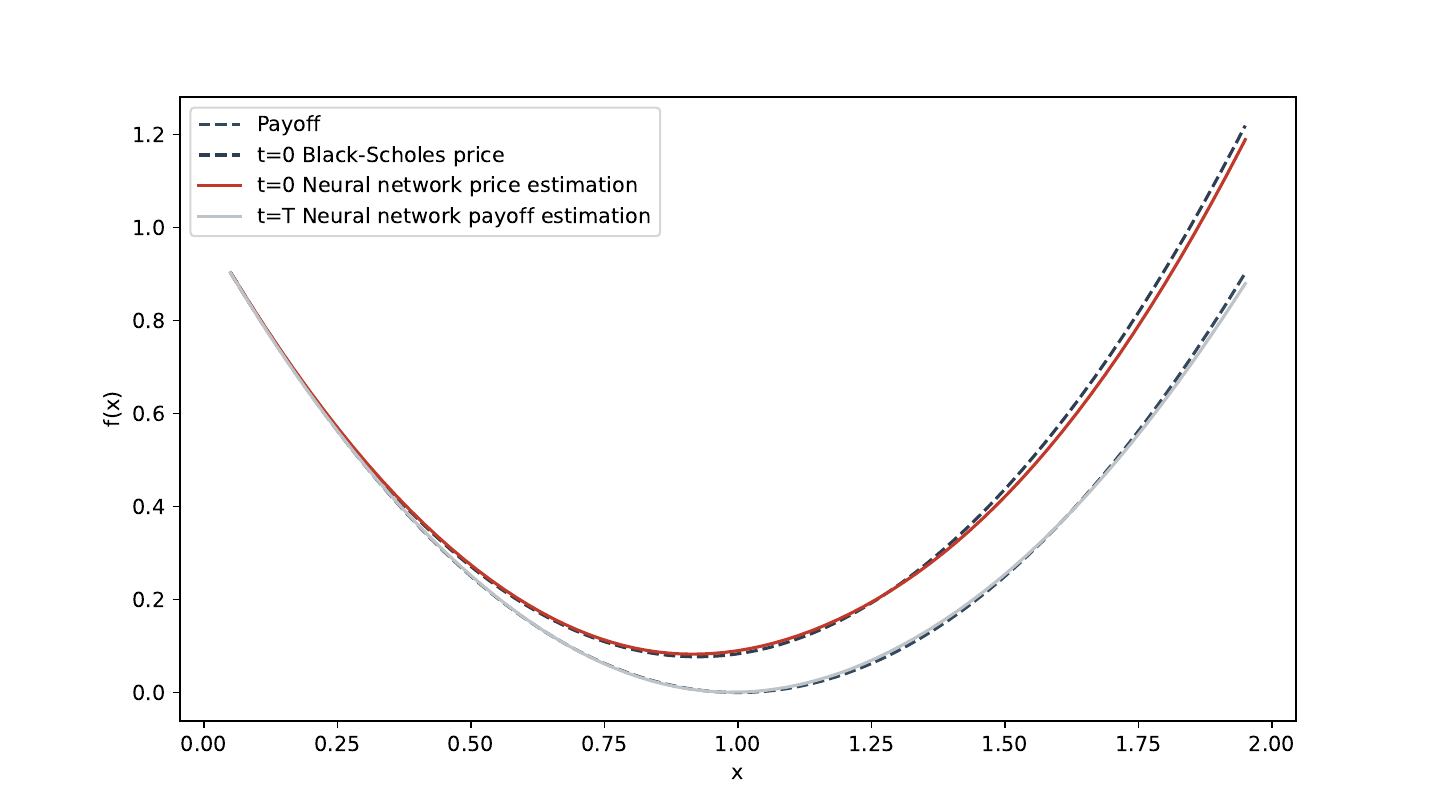}
    \caption{Zero-target.}
\end{subfigure}
\begin{subfigure}[b]{0.48\textwidth} 
    \hspace*{-0.5cm}\includegraphics[scale=0.36]{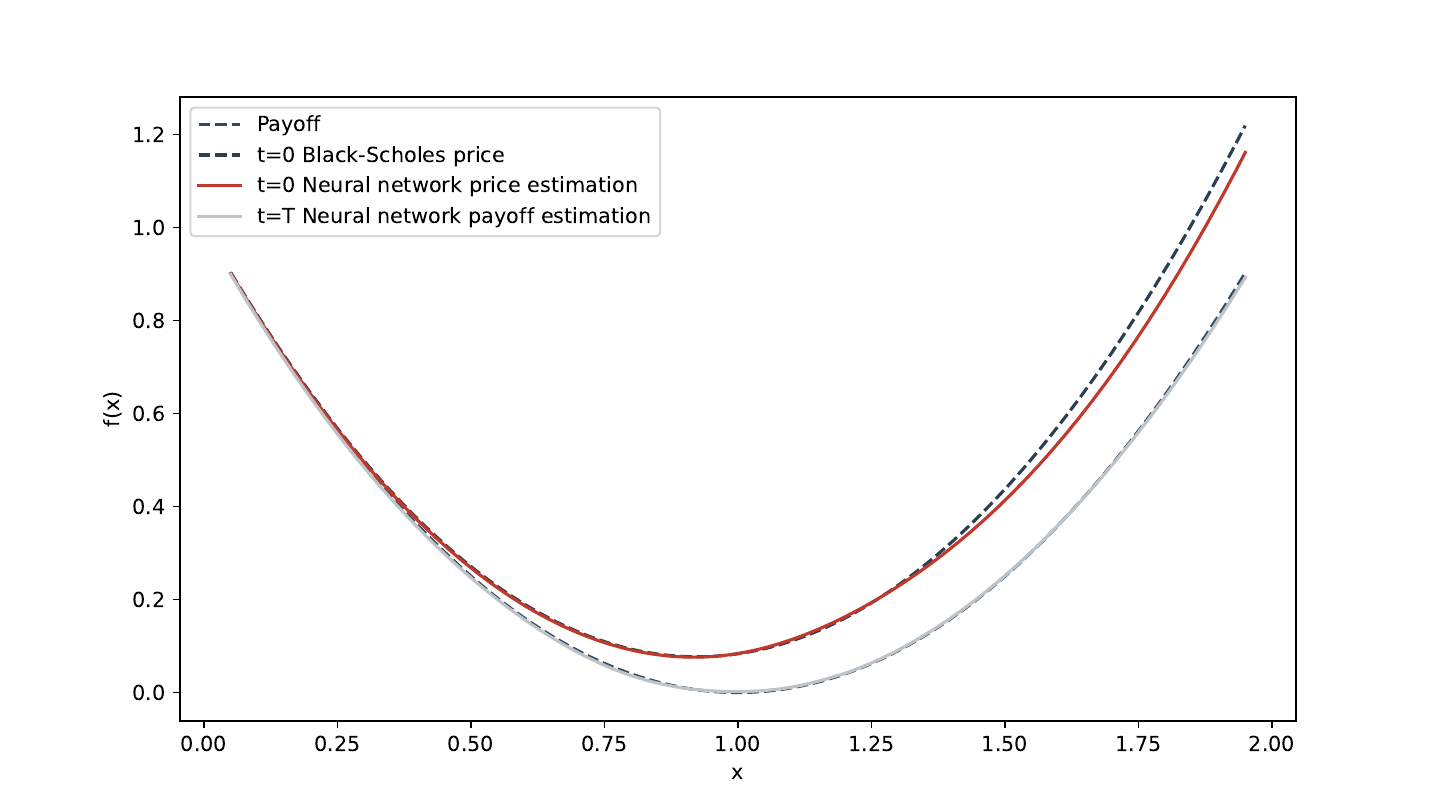}
    \caption{Control-variate.}
\end{subfigure}
\hfill
\begin{subfigure}[b]{0.48\textwidth}
    \hspace*{-0.6cm}\includegraphics[scale=0.36]{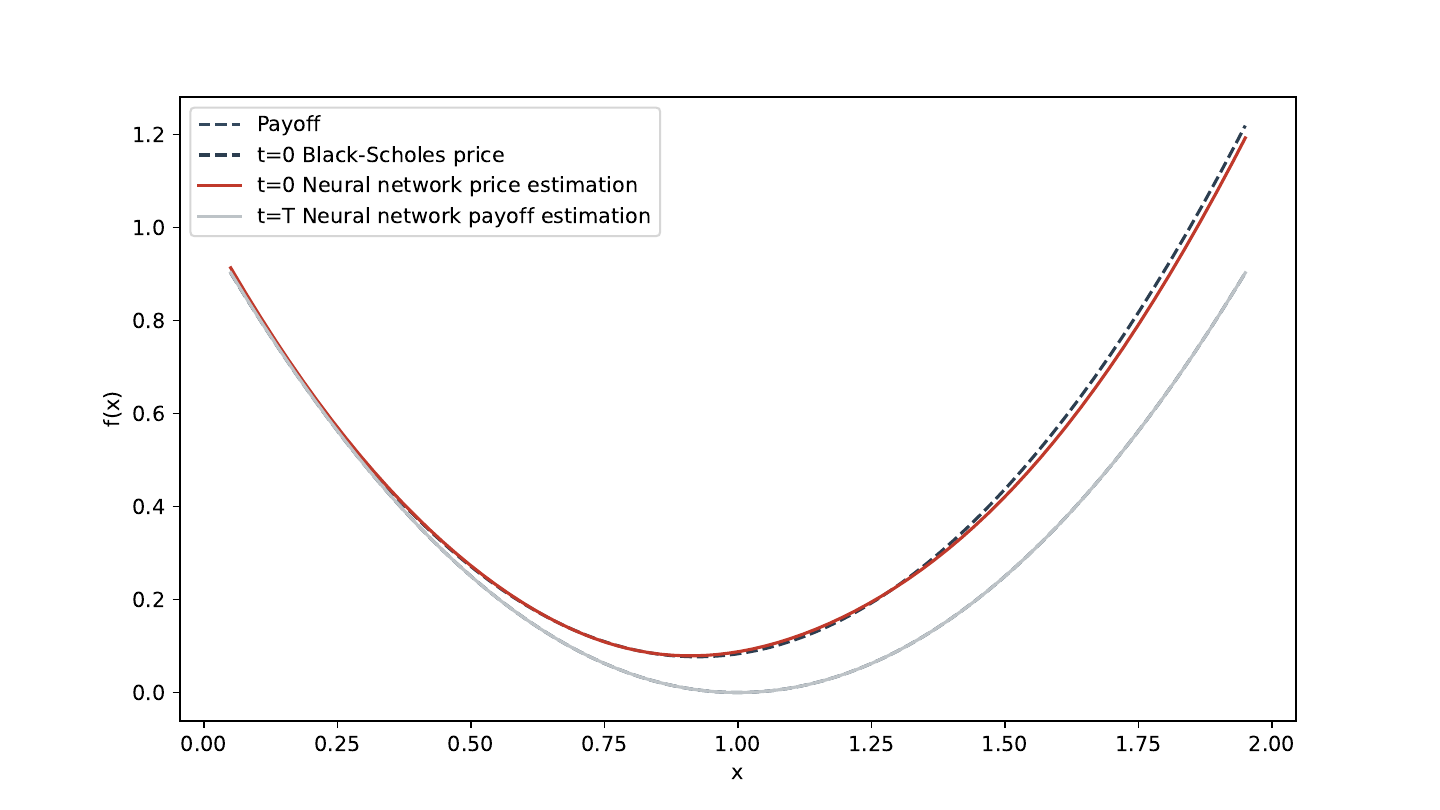}
    \caption{Constrained.}
\end{subfigure}
\caption{Pricing functions for the square option with $T=2, K=1.2, P=1$.}
\label{fig:price_x2}
\end{figure}

\begin{figure}[H]
\centering
\begin{subfigure}[b]{0.48\textwidth} 
    \includegraphics[scale=0.39]{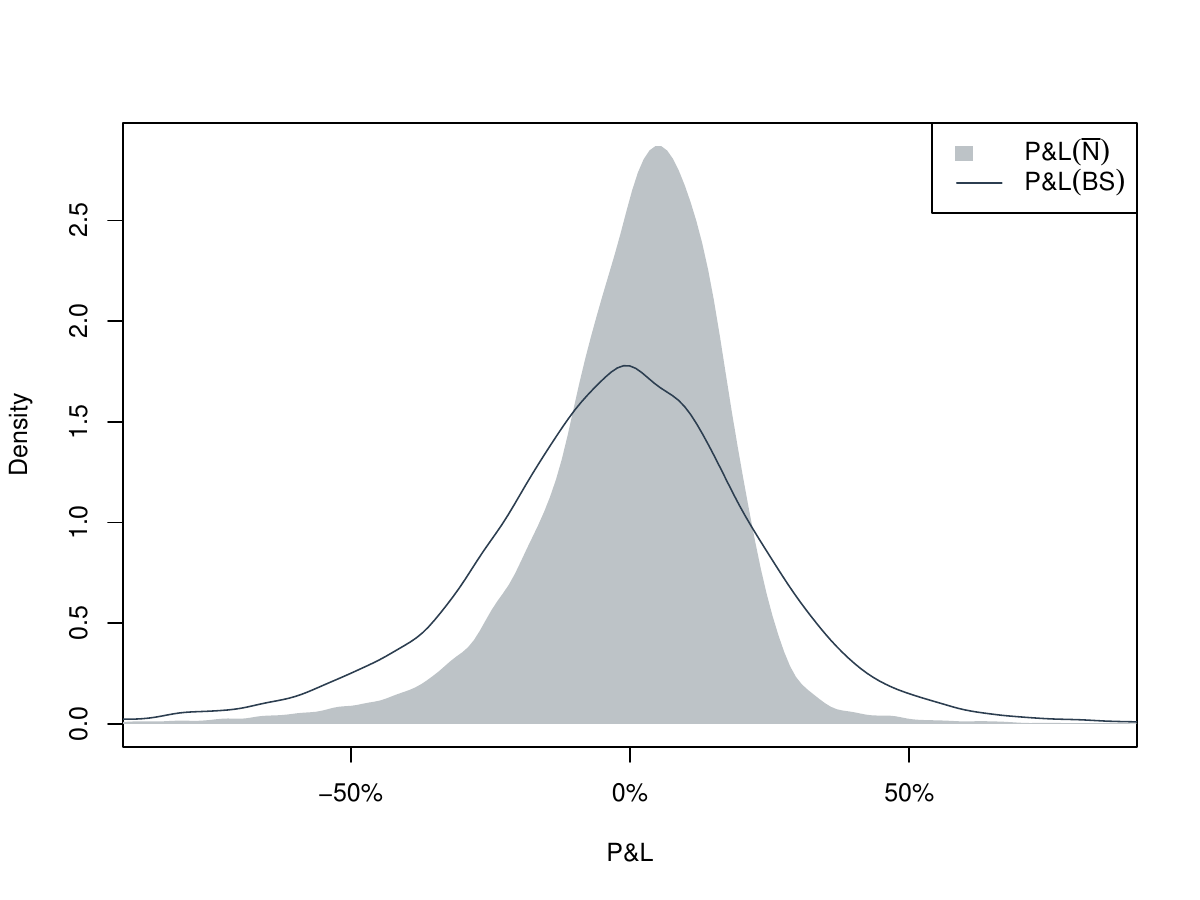}
    \caption{Unconstrained.}
\end{subfigure}
\hfill
\begin{subfigure}[b]{0.48\textwidth}
    \hspace*{-0.25cm}\includegraphics[scale=0.39]{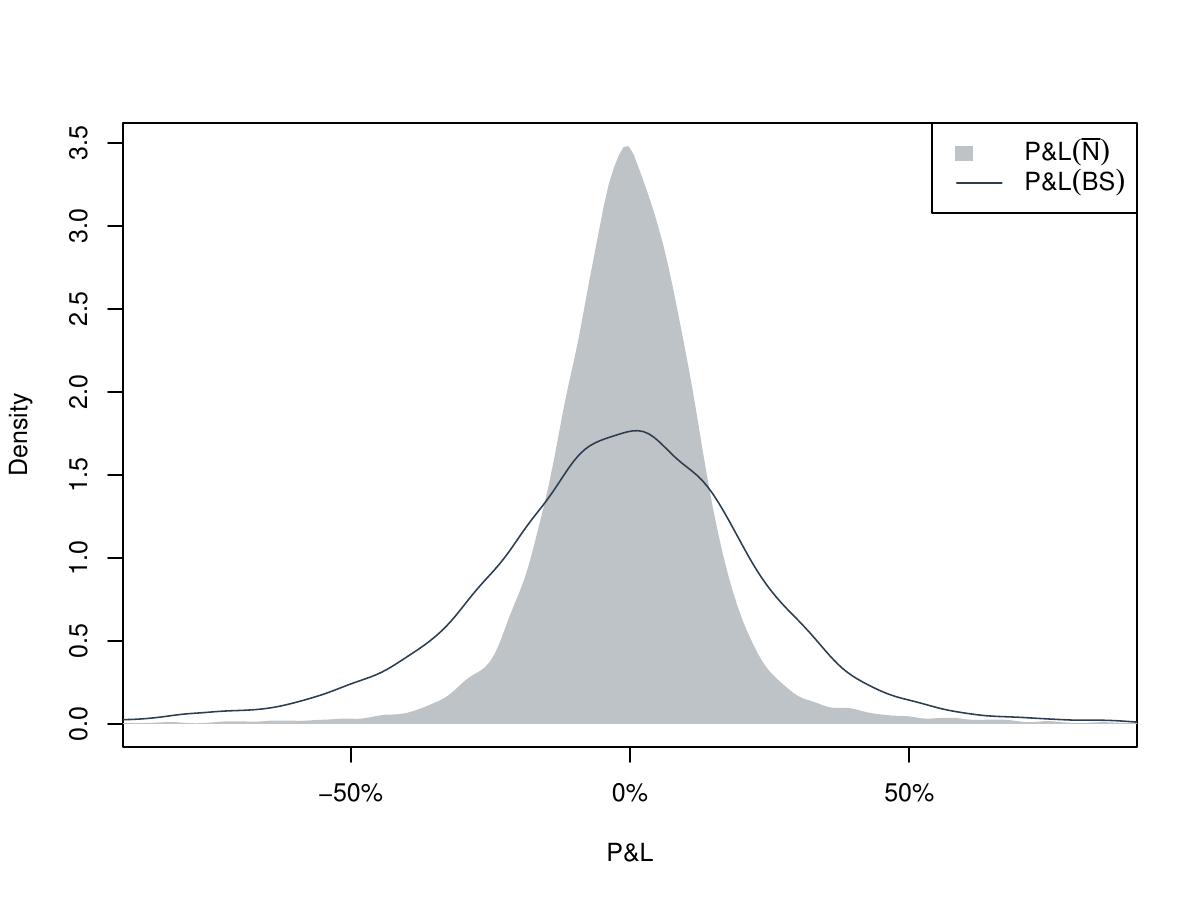}
    \caption{Zero-target.}
\end{subfigure}
\begin{subfigure}[b]{0.48\textwidth} 
    \includegraphics[scale=0.39]{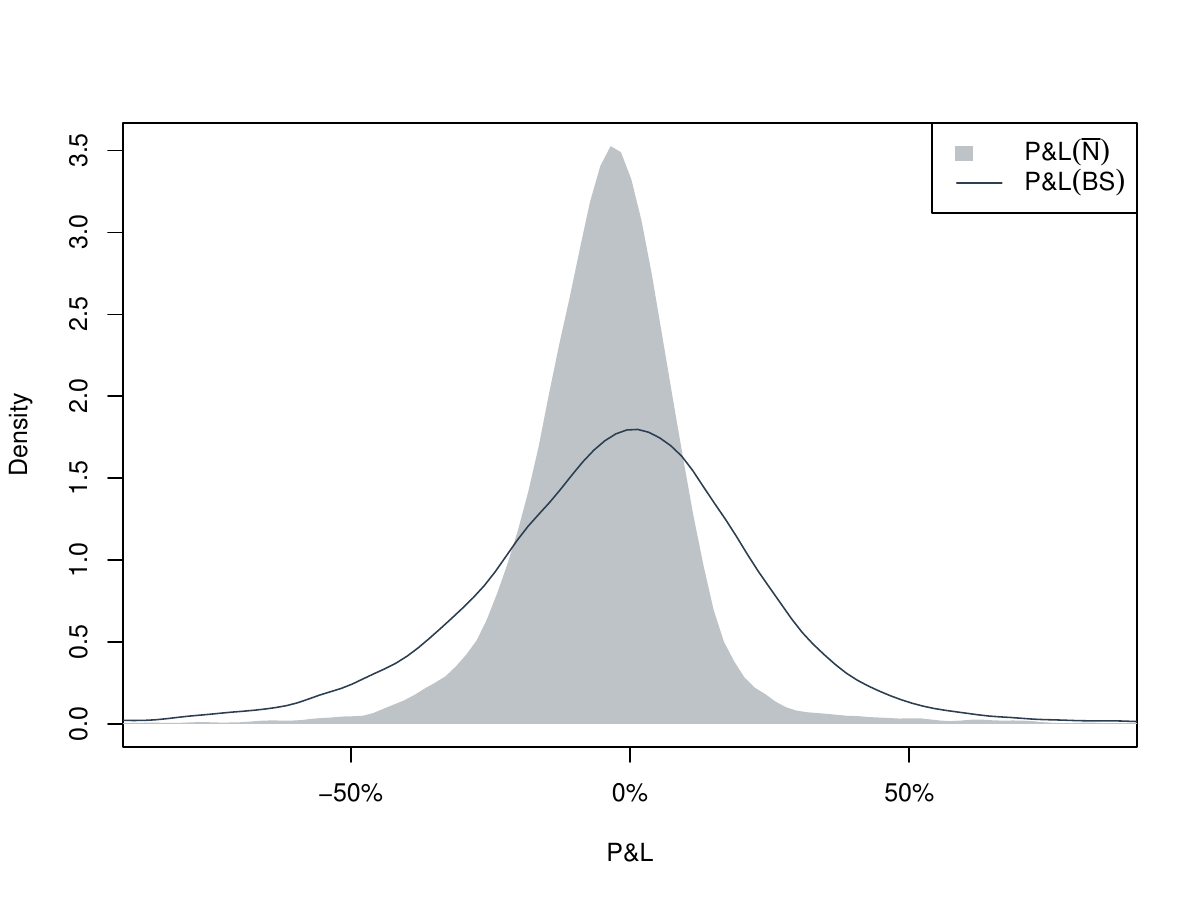}
    \caption{Control-variate.}
\end{subfigure}
\hfill
\begin{subfigure}[b]{0.48\textwidth}
    \hspace*{-0.25cm}\includegraphics[scale=0.39]{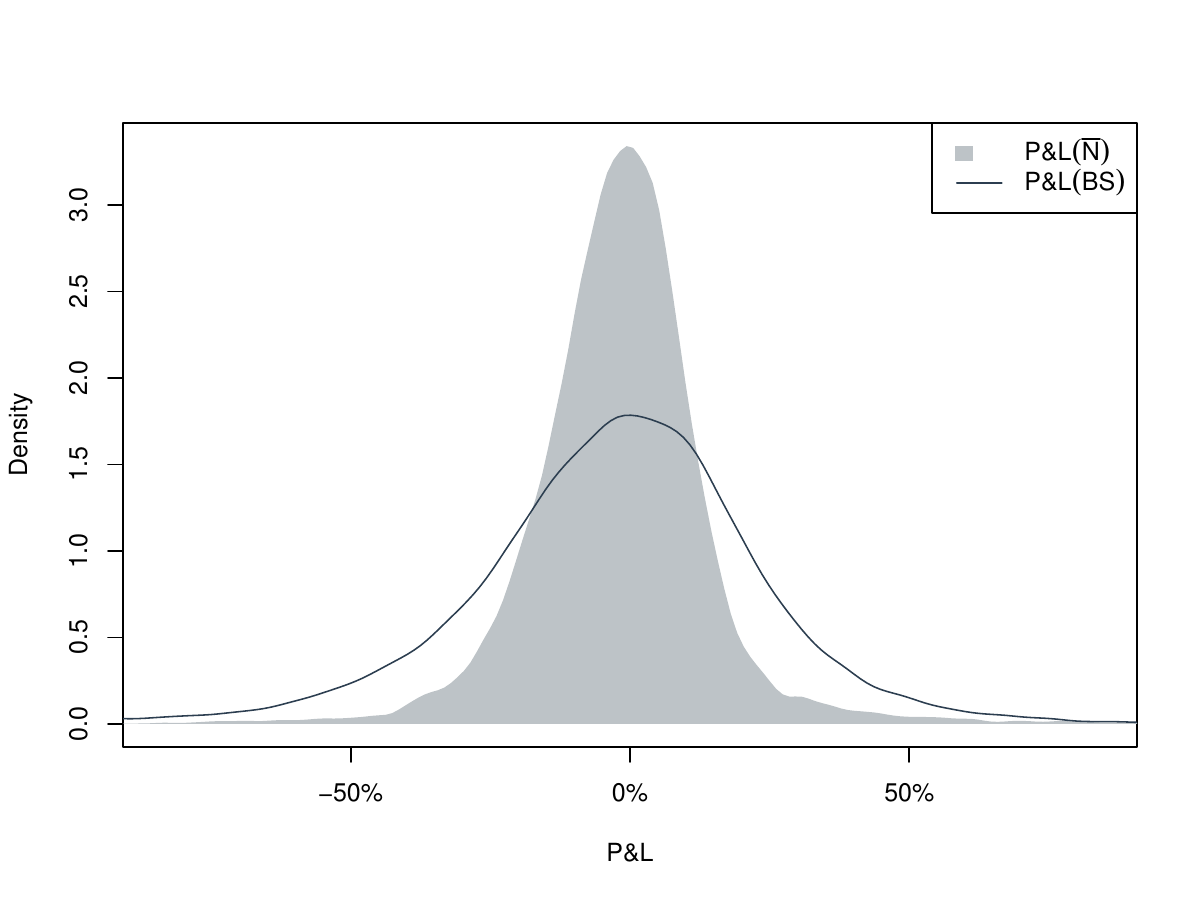}
    \caption{Constrained.}
\end{subfigure}
\caption{Empirical P\&L distributions for the neural network hedge and the Black–Scholes delta hedge for the square option with $T=2, K=1.2, P=1$ and $X_{0} = 1$.}
\label{fig:pl_x2}
\end{figure}

Figure~\ref{fig:price_x2} shows the learned pricing function, while Figure~\ref{fig:pl_x2} displays the corresponding out-of-sample P\&L distributions. Numerical summary statistics are reported in Table~\ref{stats_x2} (P\&L expressed as percentages of the Black–Scholes benchmark price of the square option).

\begin{table}[H]
\centering
\resizebox{0.9\textwidth}{!}{%
\begin{tabular}{|c|c|c|c|c|c|}
\hline
\text{Statistic} & \text{Black-Scholes} & \text{Unconstrained} & \text{Zero-target} & \text{Control-variate} & \text{Constrained} \\
\hline
\text{Mean} & -2.994 \% & 0.046 \% & -0.315 \% & -4.547 \% & -1.895 \%\\
\text{S.D.} & 28.09 \% & 19.77 \% & 16.46 \% & 17.21 \% & 17.19 \%\\
\hline
\text{Quantile $1\%$} & -82.82 \% & -62.23 \% & -43.71 \% & -47.40 \% & -45.84 \%\\
\text{Quantile $10\%$} & -34.47 \% & -21.26 \% & -16.62 \% & -21.16 \% & -19.03 \%\\
\text{Quantile $90\%$} & 27.19 \% & 18.65 \% & 15.47 \% & 10.63 \% & 13.86 \%\\
\text{Quantile $99\%$} & 62.29 \% & 38.14 \% & 47.79 \% & 43.69 \%  & 47.98 \%\\
\hline
\end{tabular}
}
\caption{Summary P\&L statistics for the square option.}
\label{stats_x2}
\end{table}

Even though the payoff is perfectly smooth, the Unconstrained approach, while producing the most accurate initial price, still delivers the poorest hedging performance, exactly as observed for the vanilla call. Its P\&L standard deviation is approximately 15–20\% higher than that of the three methods that explicitly embed the terminal condition, and its 1\% quantile is substantially worse. Enforcing the exact payoff at maturity is critical for robust out-of-sample hedging, regardless of the smoothness of the terminal condition.

\subsubsection{The digital option}\label{digital}

We finally consider a digital call with payoff $g(X_T, P) = \mathbf{1}_{\{X_T > P\}}$. This contract is the most challenging of the three because the terminal payoff is discontinuous at the strike. As before, whenever a baseline function $f$ is required (zero-target, control-variate, and constrained cases), we use the exact Black–Scholes digital price computed with constant volatility $\sigma_\circ$.

\medbreak

\begin{figure}[H]
\centering
\begin{subfigure}[b]{0.48\textwidth} 
    \hspace*{-0.5cm}\includegraphics[scale=0.36]{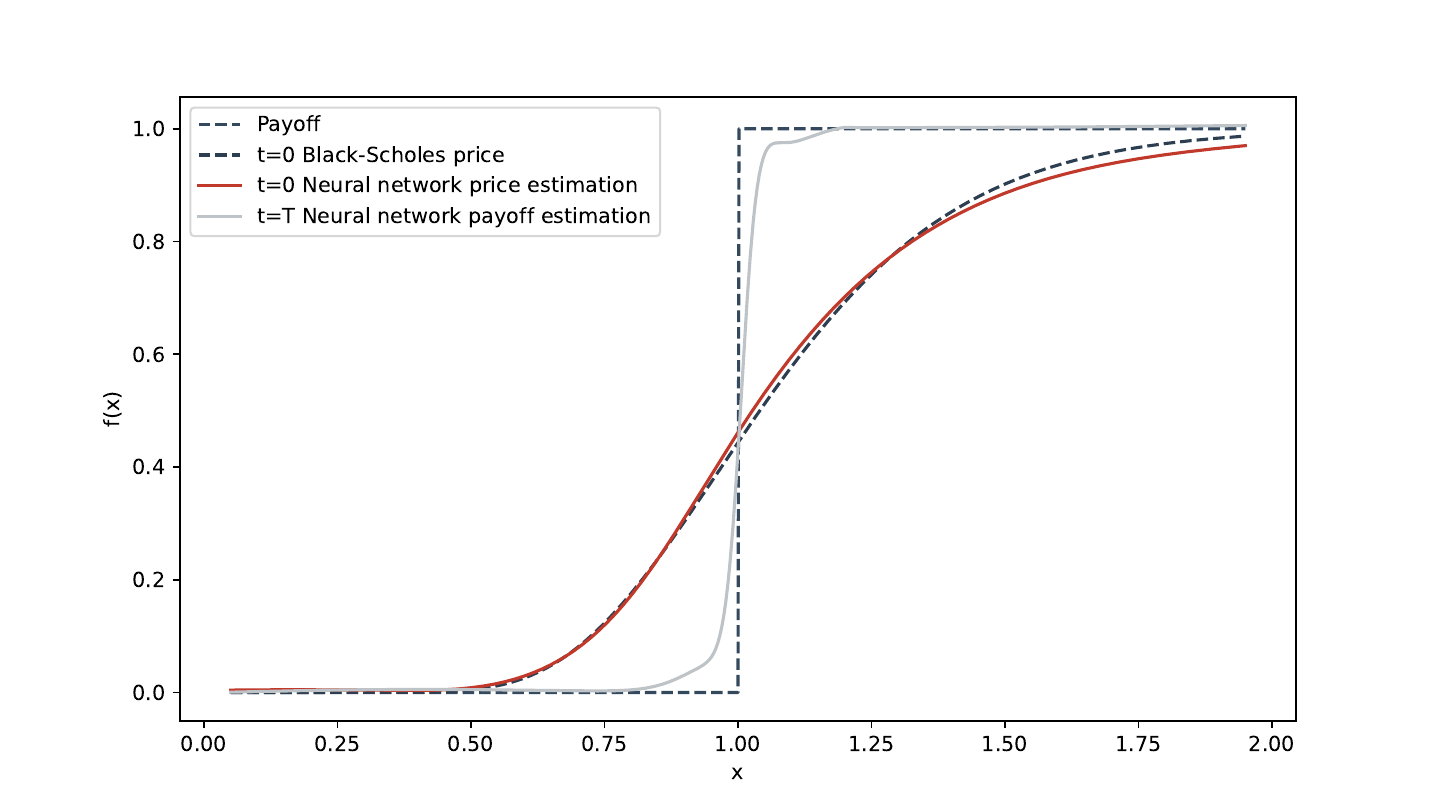}
    \caption{Unconstrained.}
    \label{fig:price_bin_unconstrained}
\end{subfigure}
\hfill
\begin{subfigure}[b]{0.48\textwidth}
    \hspace*{-0.6cm}\includegraphics[scale=0.36]{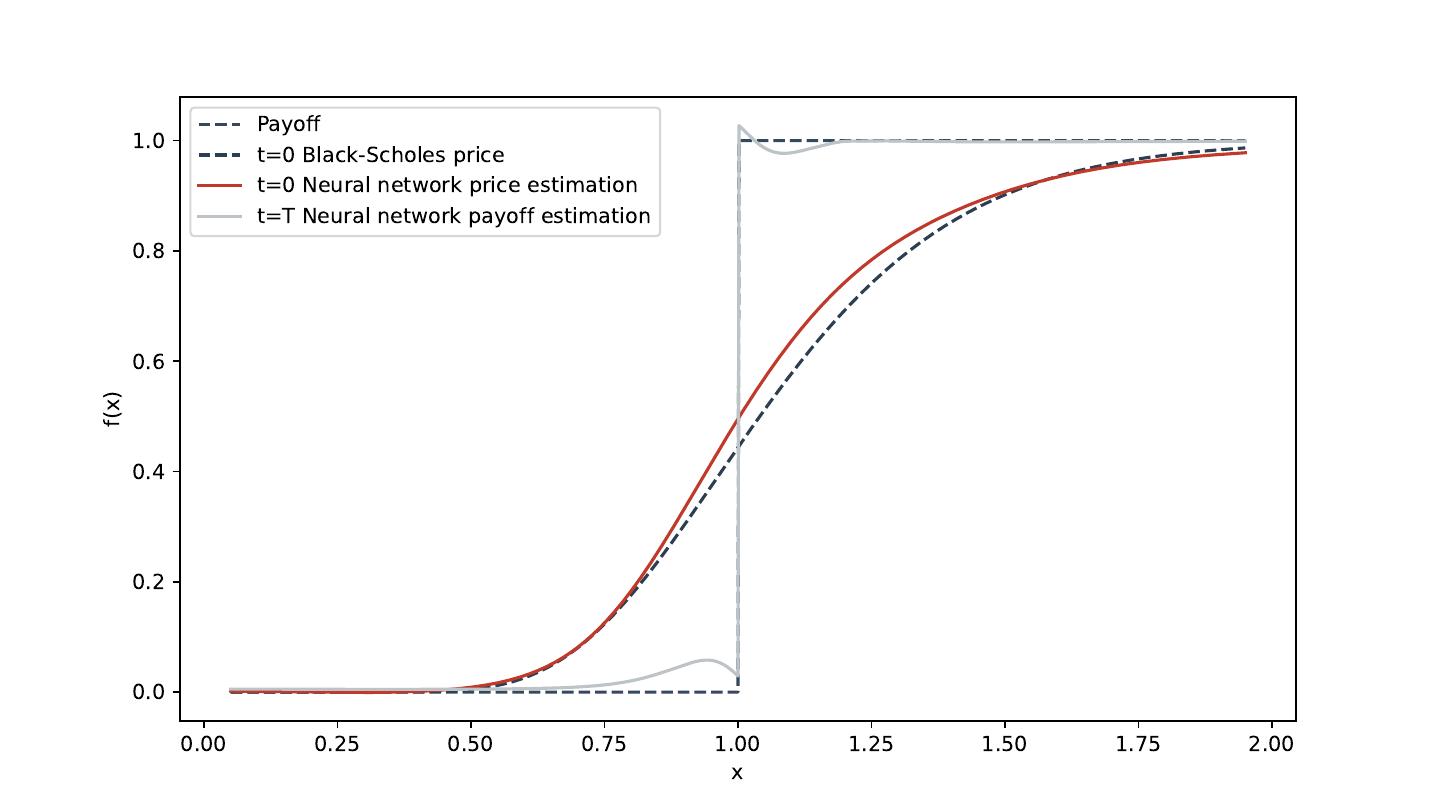}
    \caption{Zero-target.}
    \label{fig:price_bin_zerotarget}
\end{subfigure}
\begin{subfigure}[b]{0.48\textwidth} 
    \hspace*{-0.5cm}\includegraphics[scale=0.36]{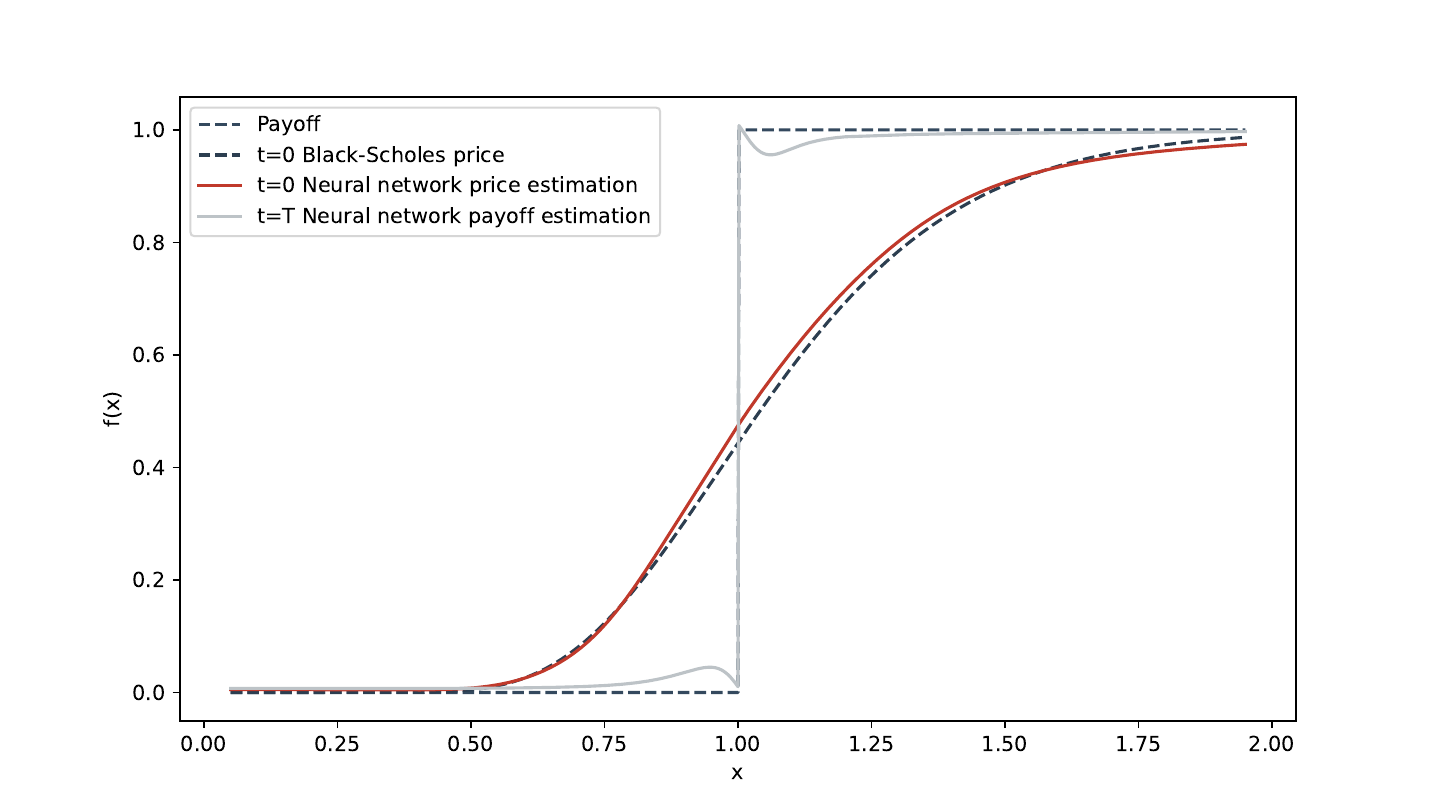}
    \caption{Control-variate.}
    \label{fig:price_bin_controlvariate}
\end{subfigure}
\hfill
\begin{subfigure}[b]{0.48\textwidth}
    \hspace*{-0.6cm}\includegraphics[scale=0.36]{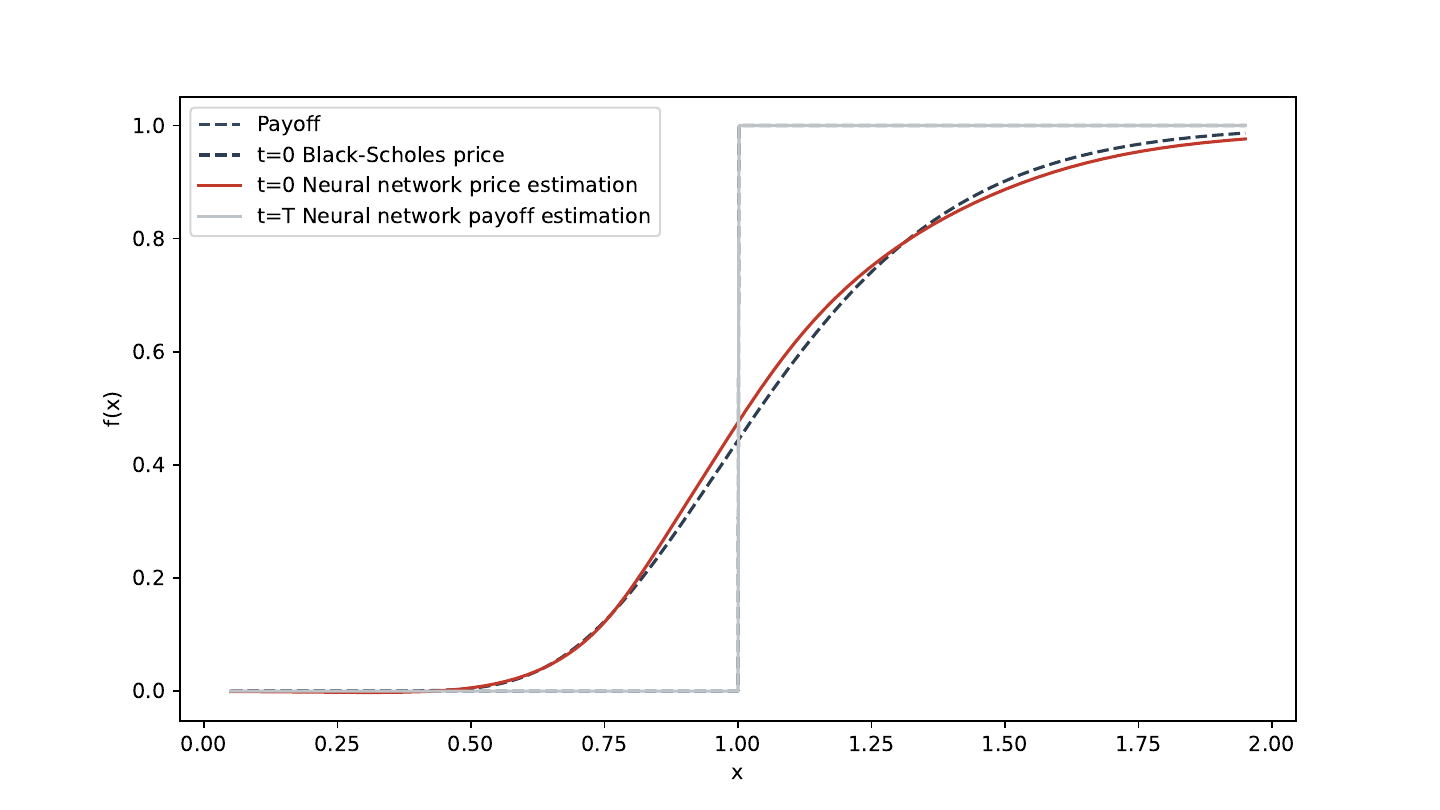}
    \caption{Constrained.}
    \label{fig:price_bin_constrained}
\end{subfigure}
\caption{Pricing functions for the digital option with $T=2, K=1.2, P=1$.}
\label{fig:price_bin}
\end{figure}

\begin{figure}[H]
\centering
\begin{subfigure}[b]{0.48\textwidth} 
    \includegraphics[scale=0.39]{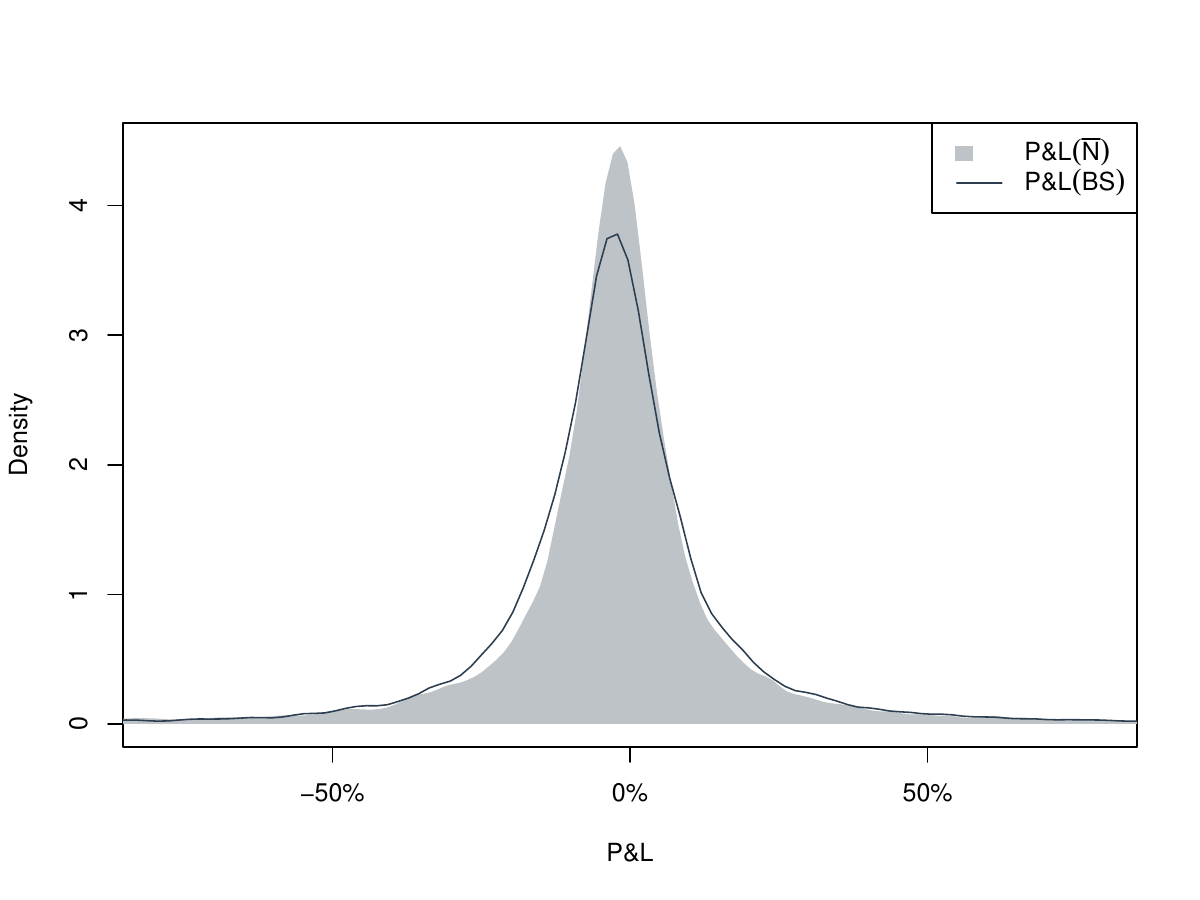}
    \caption{Unconstrained.}
    \label{fig:pl_bin_unconstrained}
\end{subfigure}
\hfill
\begin{subfigure}[b]{0.48\textwidth}
    \hspace*{-0.25cm}\includegraphics[scale=0.39]{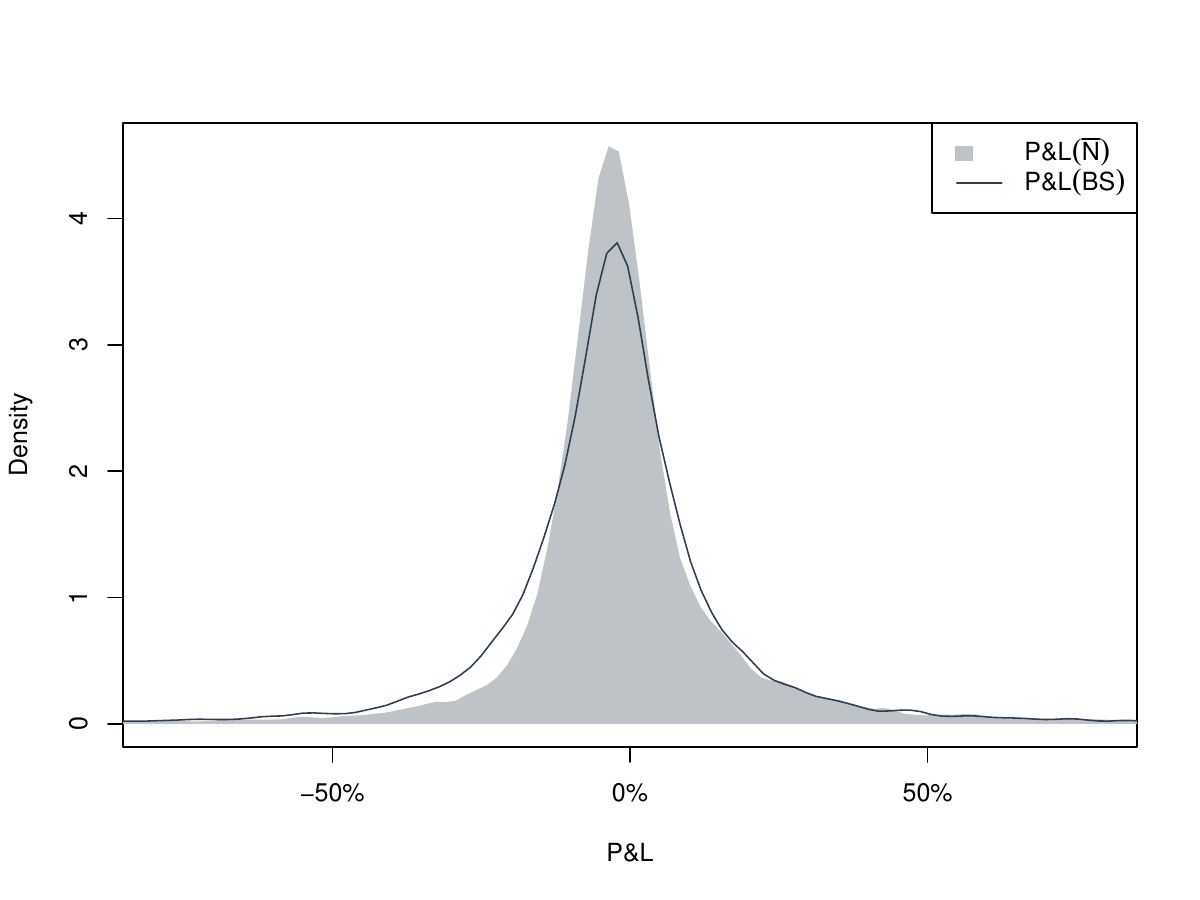}
    \caption{Zero-target.}
    \label{fig:pl_bin_zerotarget}
\end{subfigure}
\begin{subfigure}[b]{0.48\textwidth} 
    \includegraphics[scale=0.39]{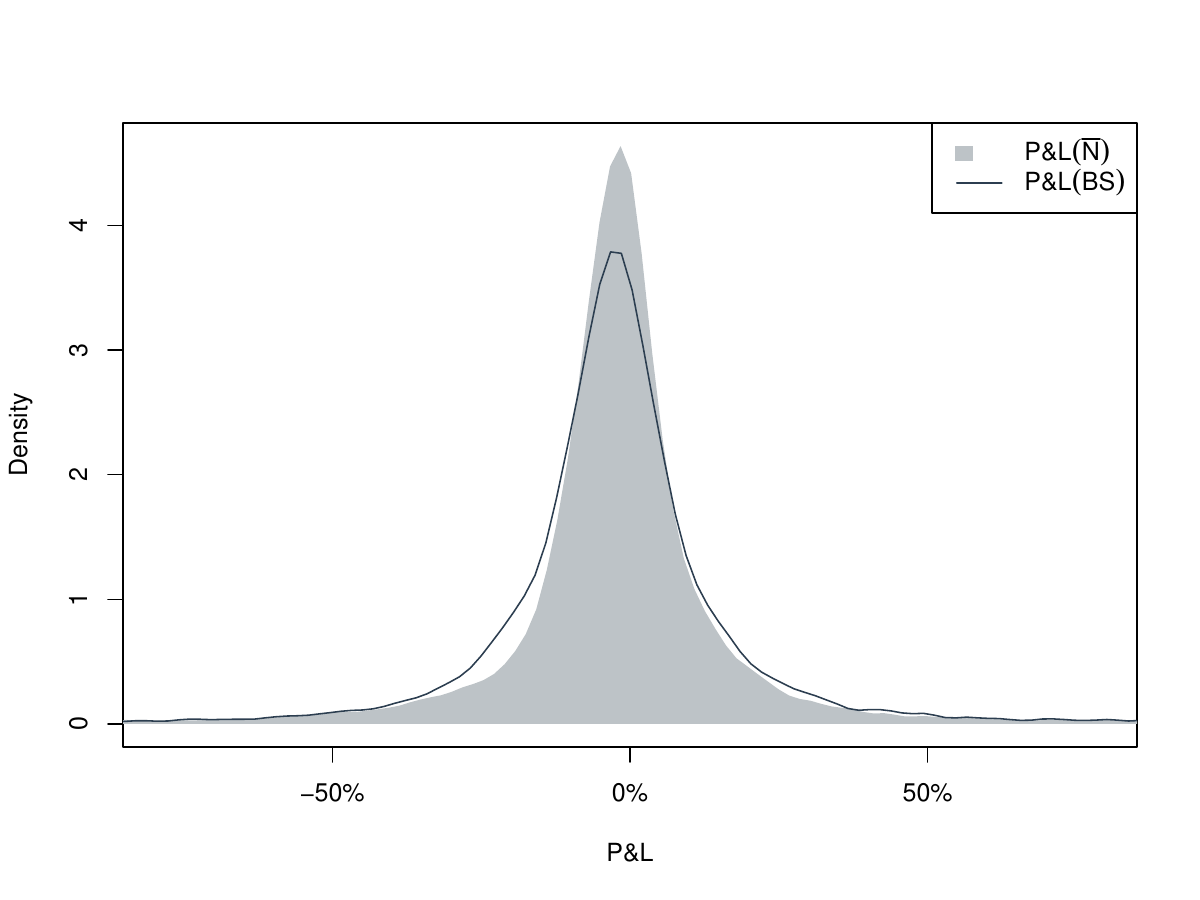}
    \caption{Control-variate.}
    \label{fig:pl_bin_controlvariate}
\end{subfigure}
\hfill
\begin{subfigure}[b]{0.48\textwidth}
    \hspace*{-0.25cm}\includegraphics[scale=0.39]{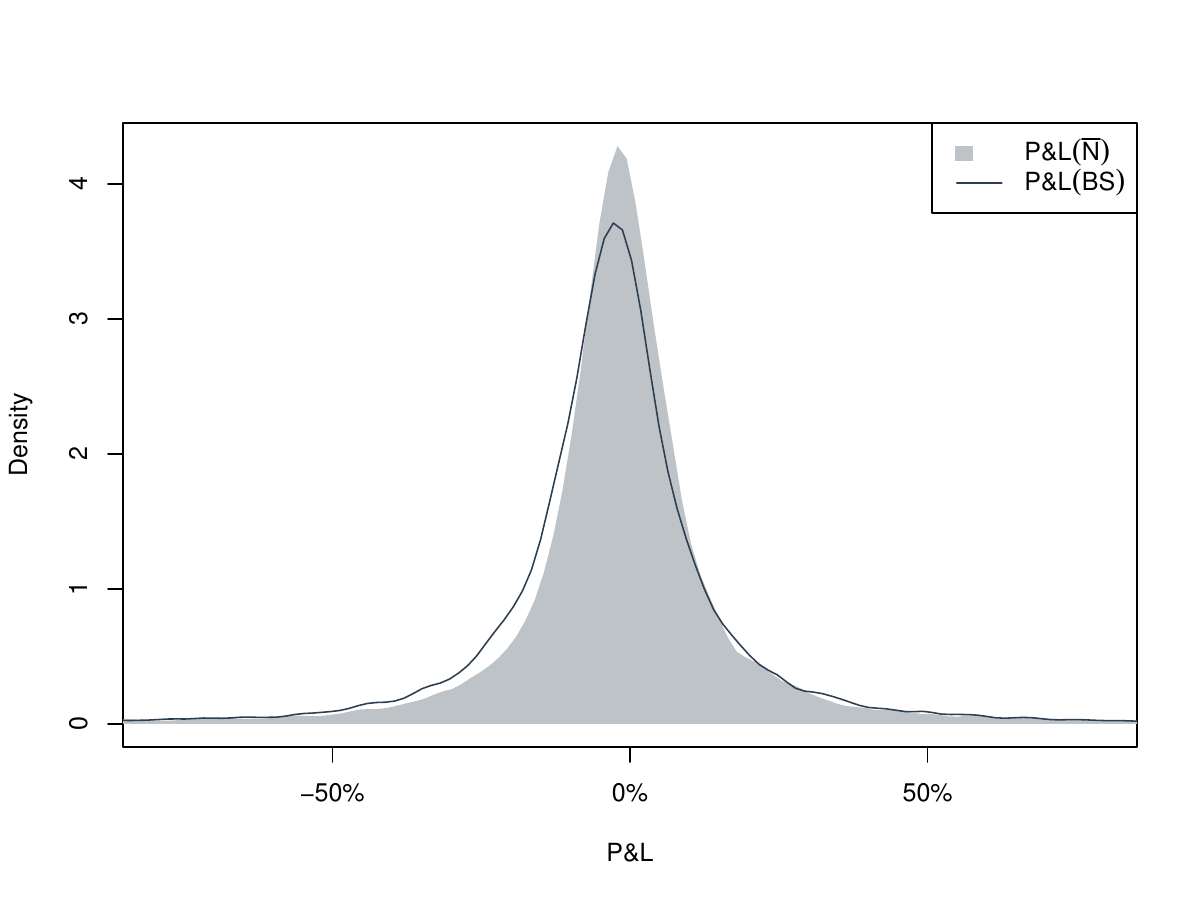}
    \caption{Constrained.}
    \label{fig:pl_bin_constrained}
\end{subfigure}
\caption{Empirical P\&L distributions for the neural network hedge and the Black–Scholes delta hedge for the digital call option with $T=2, K=1.2, P=1$ and $X_{0} = 1$.}
\label{fig:pl_bin}
\end{figure}

Figure~\ref{fig:price_bin} displays the learned pricing function, Figure~\ref{fig:pl_bin} the out-of-sample P\&L distributions, and Table~\ref{stats_bin} the corresponding summary statistics (P\&L expressed as percentages of the Black–Scholes digital call price).

\begin{table}[H]
\centering
\resizebox{0.9\textwidth}{!}{%
\begin{tabular}{|c|c|c|c|c|c|}
\hline
\text{Statistic} & \text{Black-Scholes} & \text{Unconstrained} & \text{Zero-target} & \text{Control-variate} & \text{Constrained} \\
\hline
\text{Mean} & -2.345 \% & -3.856 \% & -0.479 \% & -1.950 \% & -0.996 \%\\
\text{S.D.} & 30.06 \% & 28.41 \% & 28.59 \% & 28.11 \% & 29.46 \%\\
\hline
\text{Quantile $1\%$} & -99.54 \% & -122.30 \% & -85.39 \% & -95.96 \% & -94.63 \%\\
\text{Quantile $10\%$} & -22.73 \% & -22.55 \% & -15.89 \% & -18.57 \% & -19.08 \%\\
\text{Quantile $90\%$} & 17.76 \% & 15.60 \% & 18.17 \% & 15.13 \% & 17.31 \%\\
\text{Quantile $99\%$} & 98.29 \% & 76.56 \% & 99.94 \% & 95.45 \%  & 98.97 \%\\
\hline
\end{tabular}
}
\caption{Summary P\&L statistics for the digital option.}
\label{stats_bin}
\end{table}

The Unconstrained method gives the least accurate price (see Figure~\ref{fig:price_bin_unconstrained}). In terms of hedging, the standard deviations of the P\&L are very similar across all four approaches, with the Constrained method being marginally the highest. Overall, the three methods that embed the terminal condition do not show a decisive advantage in variance reduction for this strongly discontinuous payoff. Nevertheless, the Unconstrained approach still exhibits the worst downside risk, with the most negative 1\% quantile.

\subsubsection{Conclusion}

Our analysis of the square option, with its smooth payoff $(X_T - P)^2$, shows that the Unconstrained method, despite delivering a seemingly accurate initial price, performs the worst in terms of hedging effectiveness. Among the methods that incorporate the baseline function $f$ to embed the terminal condition, the Zero-Target approach achieves the best overall performance for this contract.

\smallbreak

For the digital option, which has a discontinuous payoff, the Unconstrained method struggles significantly with the terminal condition and produces the least accurate price. The other three methods also exhibit minor fitting imperfections for the payoff but outperform the Unconstrained case in pricing accuracy. In terms of P\&L standard deviation, all approaches yield comparable results; however, the Unconstrained method consistently displays the worst tail quantiles, indicating poorer protection in extreme scenarios.

\smallbreak

Although a call-spread approximation could have been used for the digital option, our purpose was precisely to handle a genuinely difficult payoff. We now proceed with the Equinox option, an exotic contract with a complex payoff structure, to further assess the robustness of the proposed neural network approaches on challenging terminal conditions.

\subsection{The Equinox option}

We introduce a more structured exotic contract that we call the Equinox option. Its payoff at the final horizon $T+R$ is defined, for parameters $R>0$, $B>0$, $P>0$, $G\geq 0$, as
\[
g(X_{T+R}, (B, P, G)) = \mathbf{1}_{\{X_T \leq B\}}(X_{T+R} - P)^+ + G\mathbf{1}_{\{X_T > B\}}.
\]
The contract therefore naturally decomposes into two components:
\[
g(X_{T+R}, (B, P, G)) = g_1(X_{T+R}, (R, B, P)) + G g_2(X_{T+R}, B),
\]
where:
\[
g_1(X_{T+R}, (R, B, P)) = \mathbf{1}_{\{X_T \leq B\}}(X_{T+R} - P)^+ \quad \text{and} \quad g_2(X_{T+R}, B) = \mathbf{1}_{\{X_T > B\}}.
\]

At time $T$ (time-to-maturity $R$) the barrier event is revealed, and:
\begin{itemize}
    \item The payoff of $g_1$ is either 0 or equivalent to a call option with time to maturity $R$ and strike price $P$.
    \item The payoff of $g_2$ at maturity is either 0 or 1, resembling a digital option, adjusted for interest rates over the period $[T, T+R]$, as estimated in Section \ref{digital}.
\end{itemize}
To price and hedge the Equinox option over the first period $[0,T]$ using our neural network framework requires addressing the terminal condition at time $T$:
\begin{equation}\label{eqcallpart}    
g_1'(X_T, (R, B, P)) = \mathbf{1}_{\{X_T \leq B\}} \text{Call}(R, X_T, C_T, K, P),
\end{equation}
where $\text{Call}(R,X_T,C_T,P)$ is the fair price at time $T$ of a call with remaining maturity $R$ and strike $P$. We propose the following practical implementation:
\begin{itemize}
    \item Train a first neural network $\text{Call}$ using $\overline{N}_{\theta^\circ}^{\circ}$ on vanilla calls using the methodology developed in Section~\ref{call}.
    \item Train a second neural network $\overline{N}_{\theta}$ with the payoff $\mathbf{1}_{\{X_T \leq B\}} \overline{N}_{\widehat{\theta}^\circ}^{\circ}(R, X_T, C_T, K, P)$.
\end{itemize}
A suitable function $f$ is:
\[
f(T-s, x, c, K, R, B, P) = h(T-s, x, B) \times \overline{N}_{\widehat{\theta}^\circ}^{\circ}(R + (T-s), x, c, K, P),
\]
where $h$ represents the price of a digital option in the Black-Scholes model. Note that this is not the standard Black-Scholes price, as $X_T$ and $X_{T+R}$ are not independent.

\medbreak

We then consider two modeling strategies:

\medbreak

\paragraph{Two separate networks.} Train independently:
\begin{itemize}
    \item $N_{\theta^1}^1(t, x, c, K, R, B, P)$ for the $g_1$ component (using the embedding $f$ above in the zero-target/control-variate/constrained setting),
    \item $N_{\theta^2}^2(t, x, c, K, R, B)$ for the pure digital $g_2$ component (Section~\ref{digital}).
\end{itemize}

\smallbreak

The Equinox price is then:
\begin{equation}\label{eq:equinox}
        N_{\theta^1}^1(R+T-t, x, c, K, R, B, P) + G e^{-r(R+T-t)} N_{\theta^2}^2(R+T-t, x, c, K, B).
\end{equation}

\paragraph{Single end-to-end network.} Train one global network $N_{\theta}(R+T-t, x, c, K, R, B, P, G)$ directly on the full payoff $g$ using the P\&L loss, with an appropriate embedding that combines the two baseline functions above (weighted by $G$).

\subsubsection{Equinox option with two neural networks}\label{deuxNN}

We train two independent neural networks trained with the P\&L loss.

\begin{itemize}
    \item $N_{\theta^1}^1$ is dedicated to the barrier-call component with payoff $g_1$,
    \item $N_{\theta^2}^2$ is the digital option with payoff $g_2$.
\end{itemize}
Together, these networks yield the Equinox option price and hedge, as specified in \eqref{eq:equinox}.

\medbreak

\begin{figure}[H]
\centering
\begin{subfigure}[b]{0.48\textwidth} 
    \hspace*{-0.5cm}\includegraphics[scale=0.36]{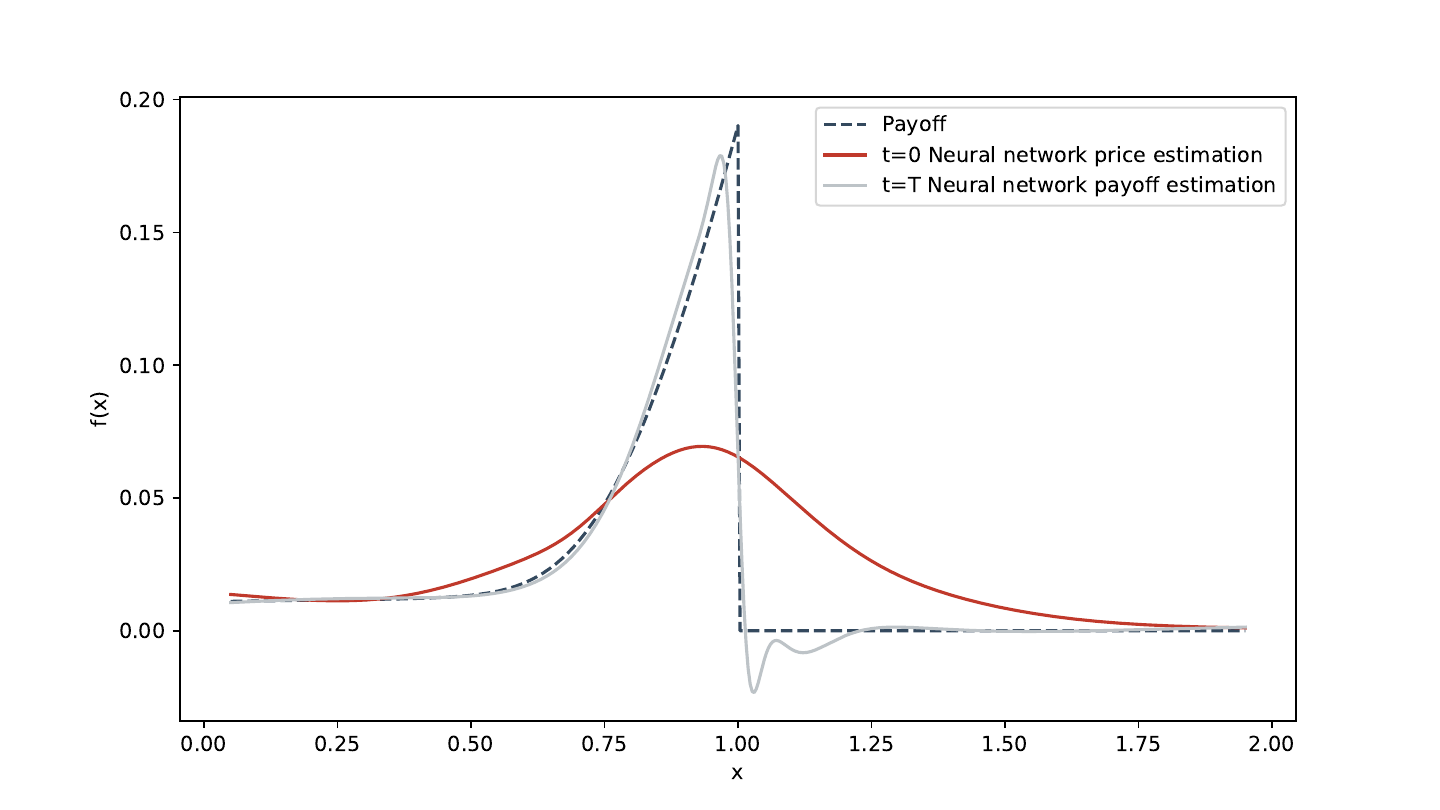}
    \caption{Unconstrained.}
\end{subfigure}
\hfill
\begin{subfigure}[b]{0.48\textwidth}
    \hspace*{-0.6cm}\includegraphics[scale=0.36]{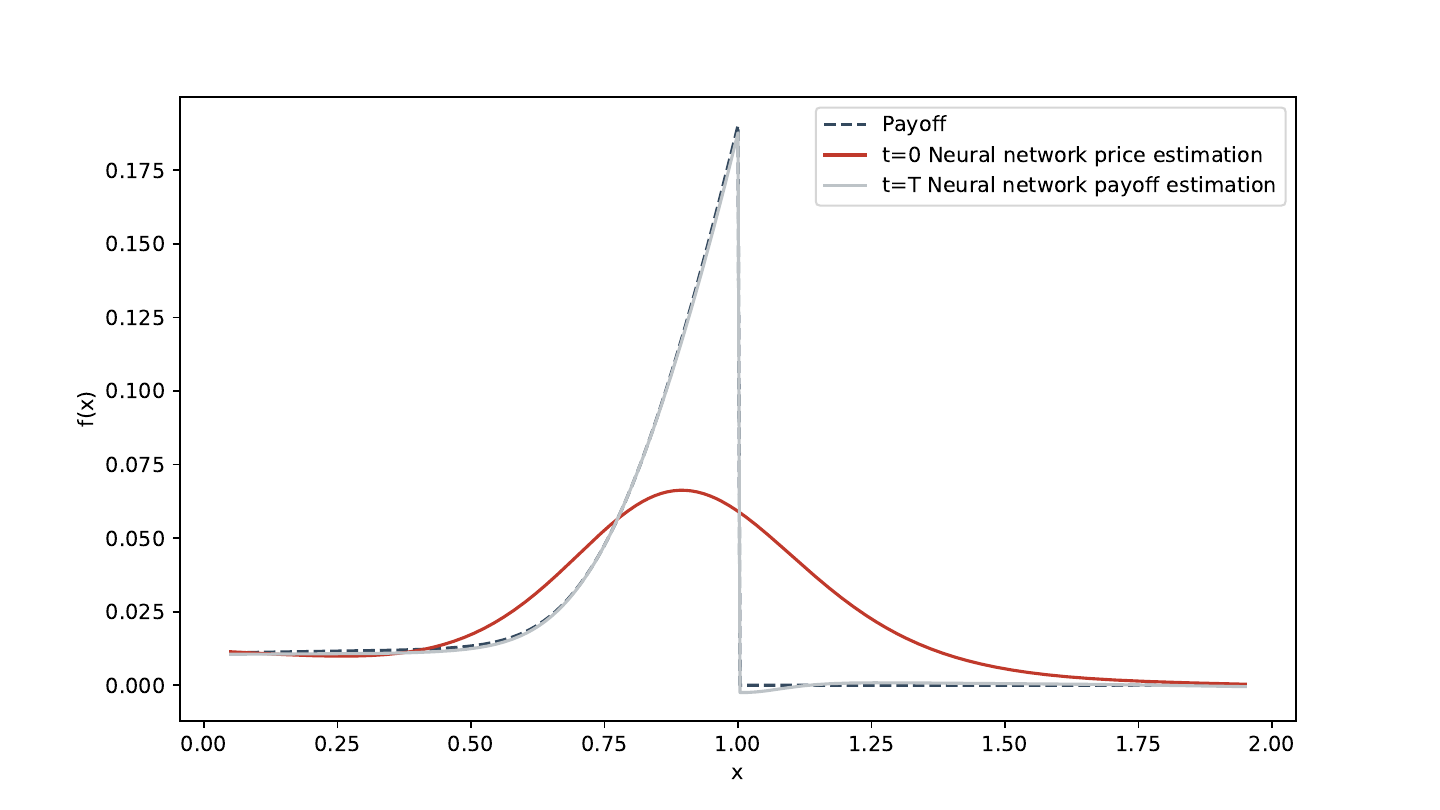}
    \caption{Zero-target.}
\end{subfigure}
\begin{subfigure}[b]{0.48\textwidth} 
    \hspace*{-0.5cm}\includegraphics[scale=0.36]{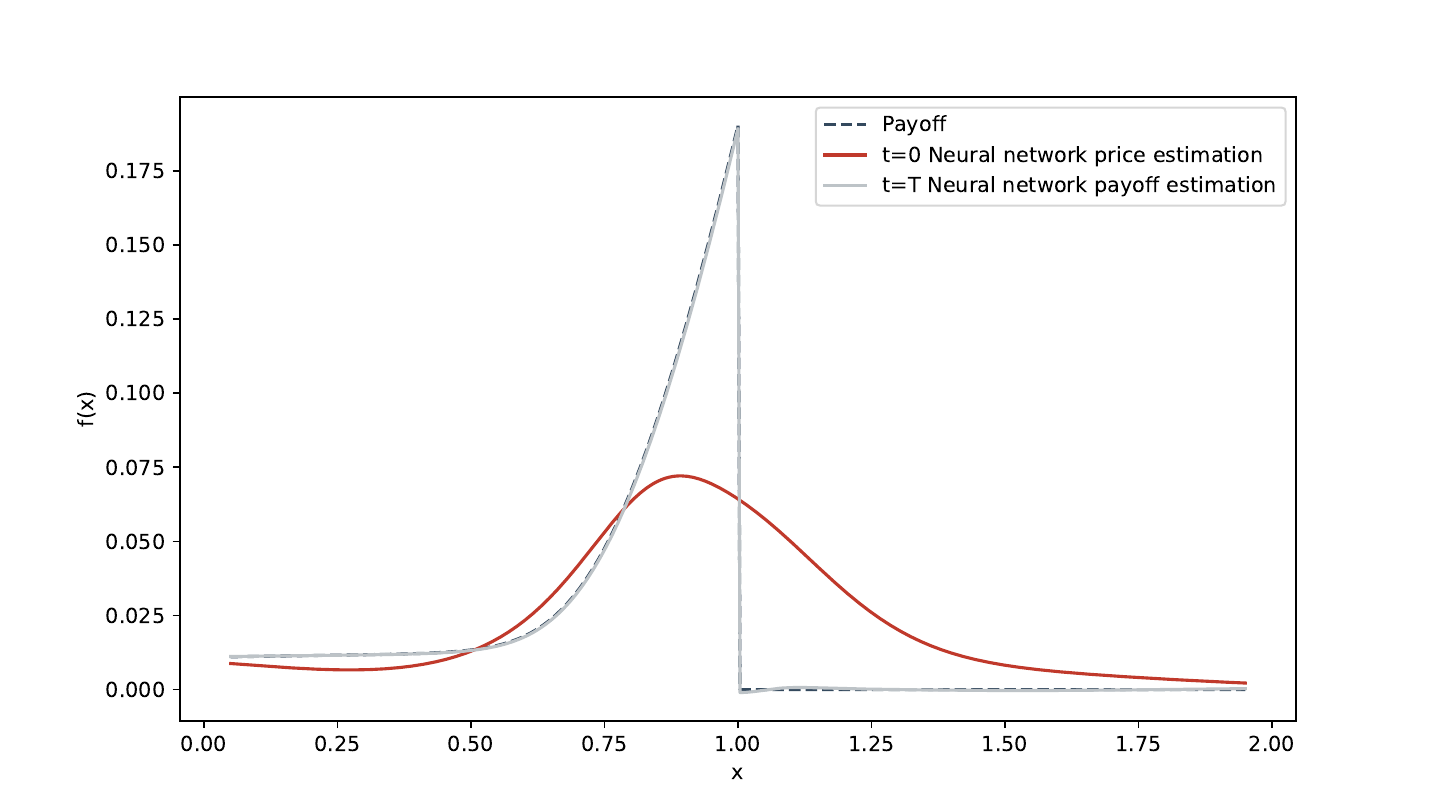}
    \caption{Control-variate.}
\end{subfigure}
\hfill
\begin{subfigure}[b]{0.48\textwidth}
    \hspace*{-0.6cm}\includegraphics[scale=0.36]{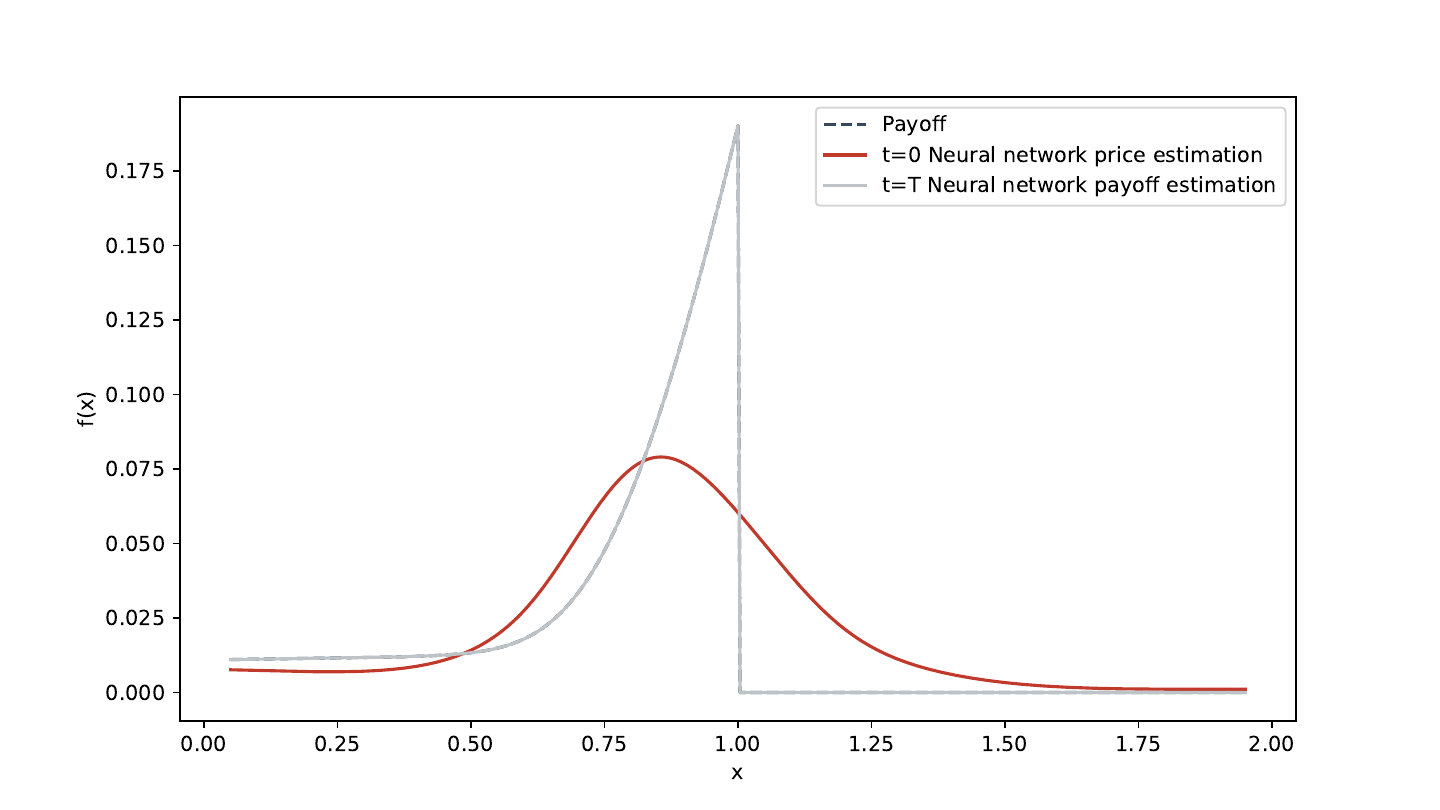}
    \caption{Constrained.}
\end{subfigure}
\caption{Pricing functions for the pure barrier-call component ($G=0$) with parameters $B=1$, $P=0.8$, $R=1$, $T=2, K=1$.}
\label{fig:price_eqcall}
\end{figure}

We present the results. Figure \ref{fig:price_eqcall} shows the estimated price for specific parameters of $N_{\theta^1}^1$, which addresses only the payoff $g_1$, excluding the digital option.

\begin{figure}[H]
\centering
\begin{subfigure}[b]{0.48\textwidth} 
    \hspace*{-0.5cm}\includegraphics[scale=0.36]{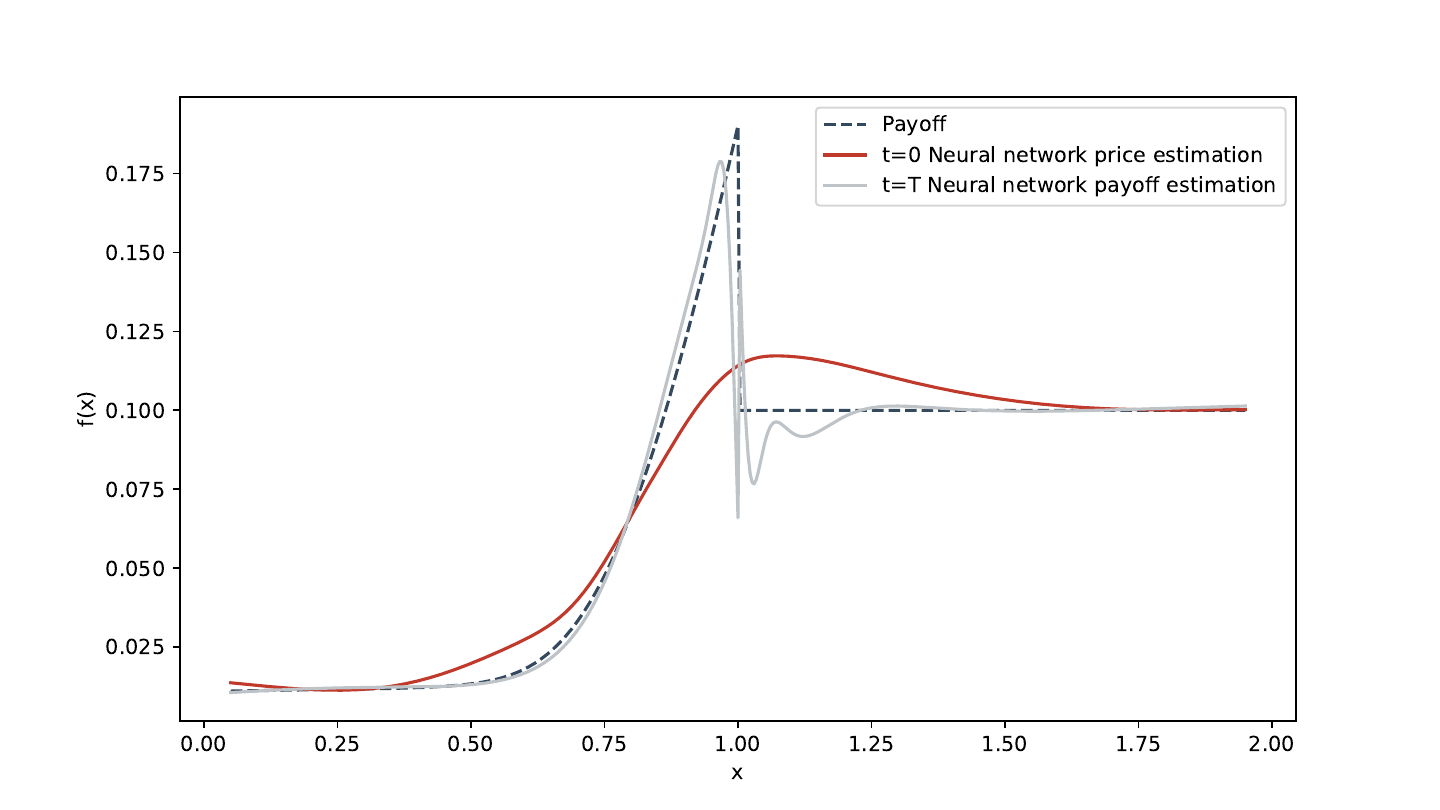}
    \caption{Unconstrained.}
\end{subfigure}
\hfill
\begin{subfigure}[b]{0.48\textwidth}
    \hspace*{-0.6cm}\includegraphics[scale=0.36]{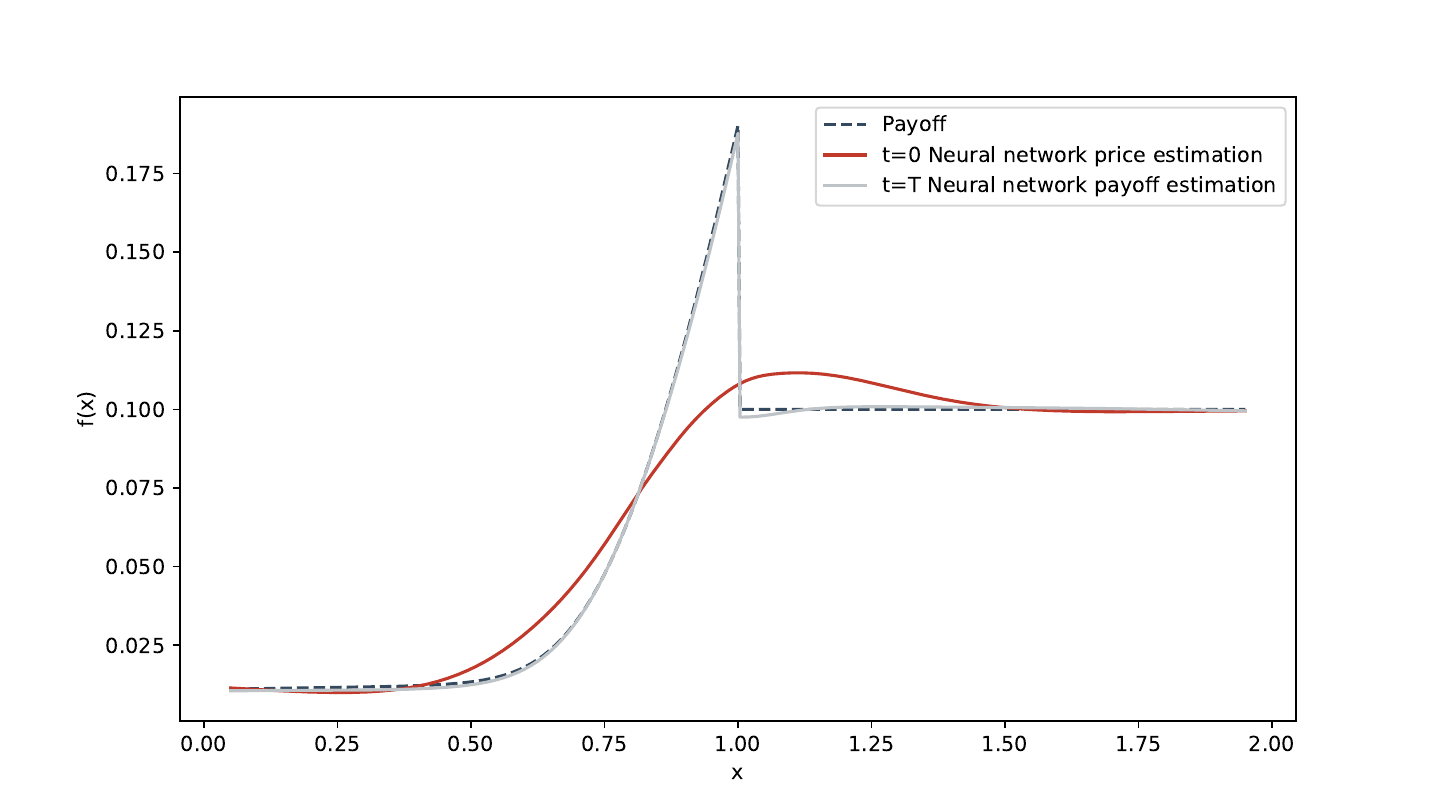}
    \caption{Zero-target.}
\end{subfigure}
\begin{subfigure}[b]{0.48\textwidth} 
    \hspace*{-0.5cm}\includegraphics[scale=0.36]{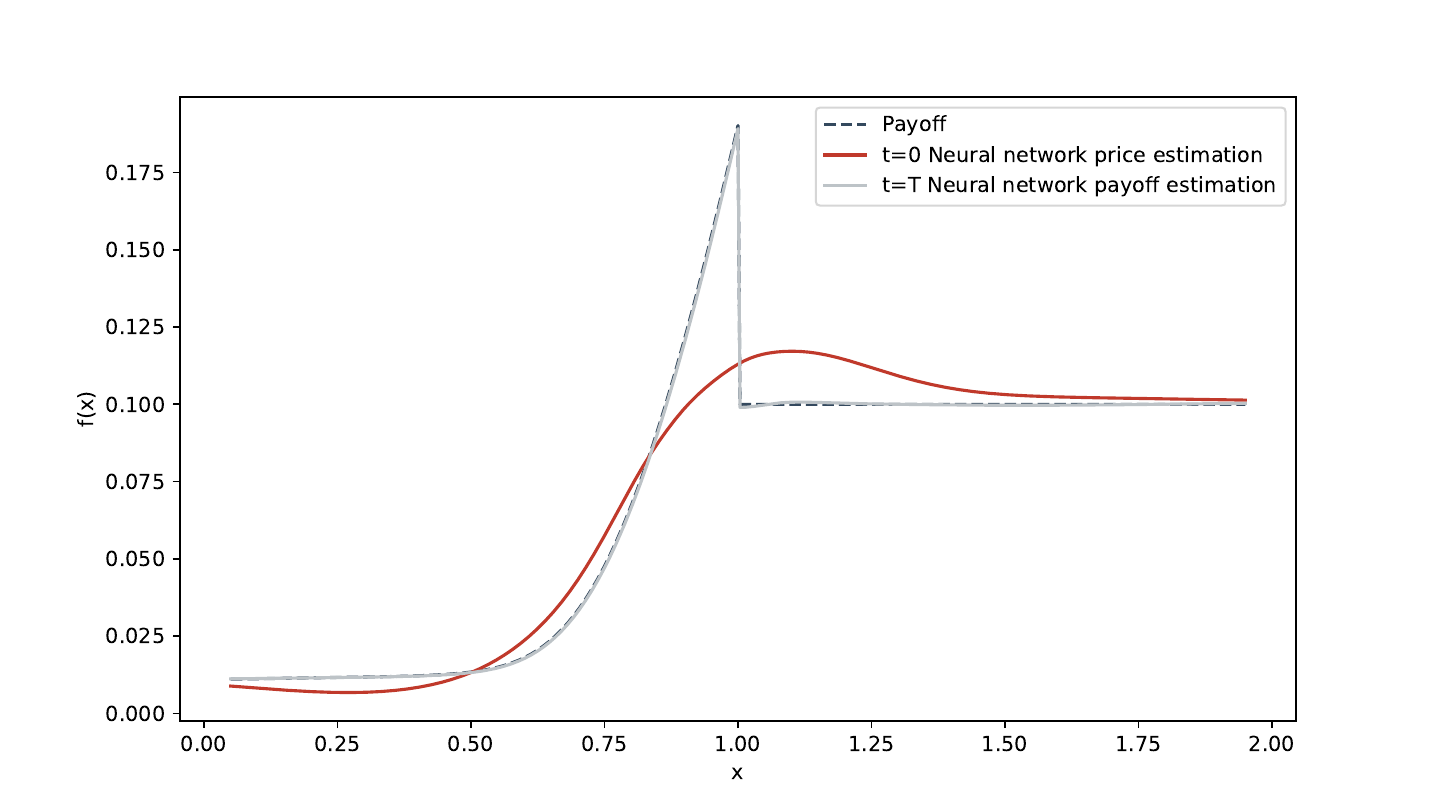}
    \caption{Control-variate.}
\end{subfigure}
\hfill
\begin{subfigure}[b]{0.48\textwidth}
    \hspace*{-0.6cm}\includegraphics[scale=0.36]{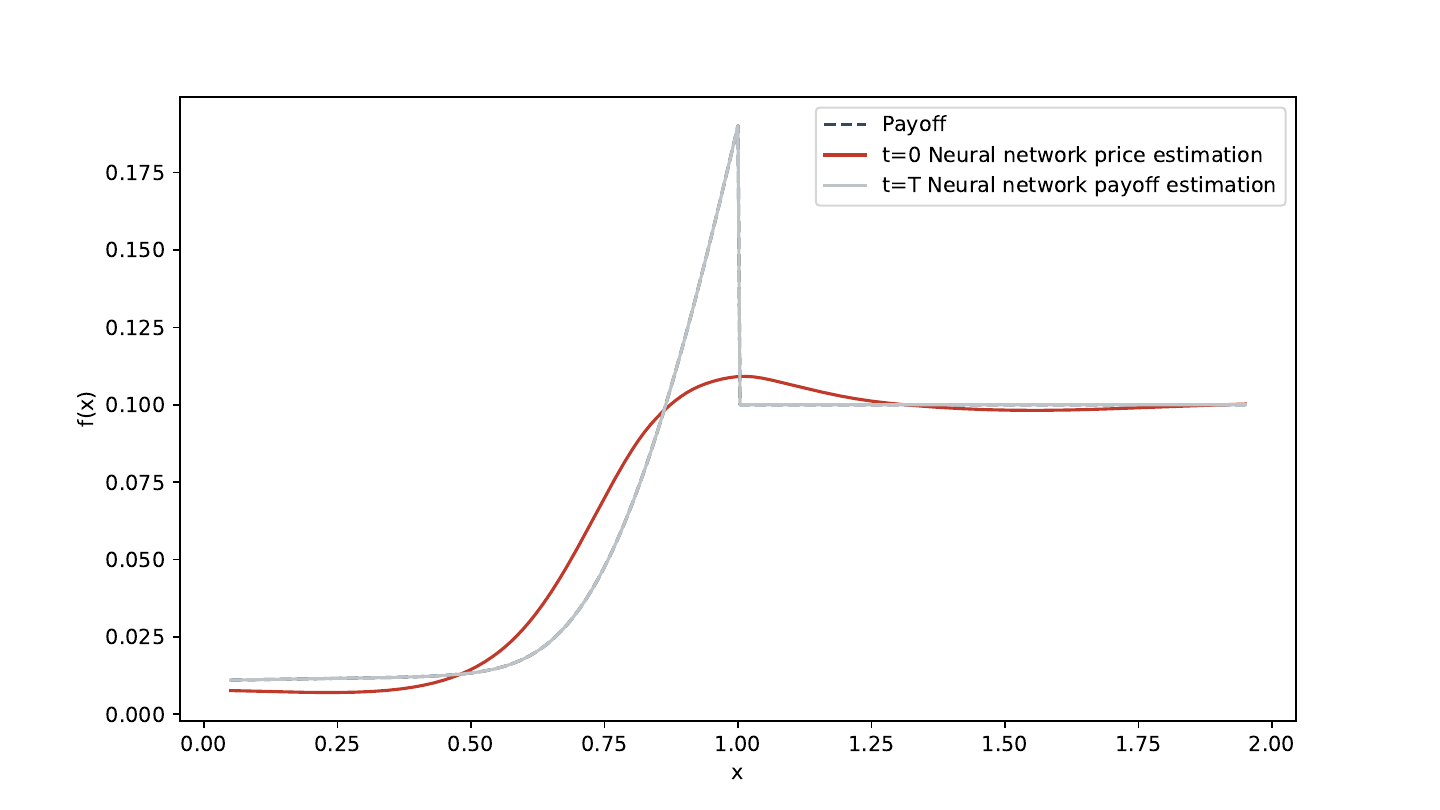}
    \caption{Constrained.}
\end{subfigure}
\caption{Pricing functions for the full Equinox option with the parameters $G=0.1$, $B=1$, $P=0.8$, $R=1$, $T=2, K=1$.}
\label{fig:price_eqcallport}
\end{figure}

The Unconstrained version again struggles severely with the complex terminal condition (especially visible when $G=0$), producing distorted pricing curves far from the true continuation value.

\begin{figure}[H]
\centering
\begin{subfigure}[b]{0.48\textwidth} 
    \includegraphics[scale=0.39]{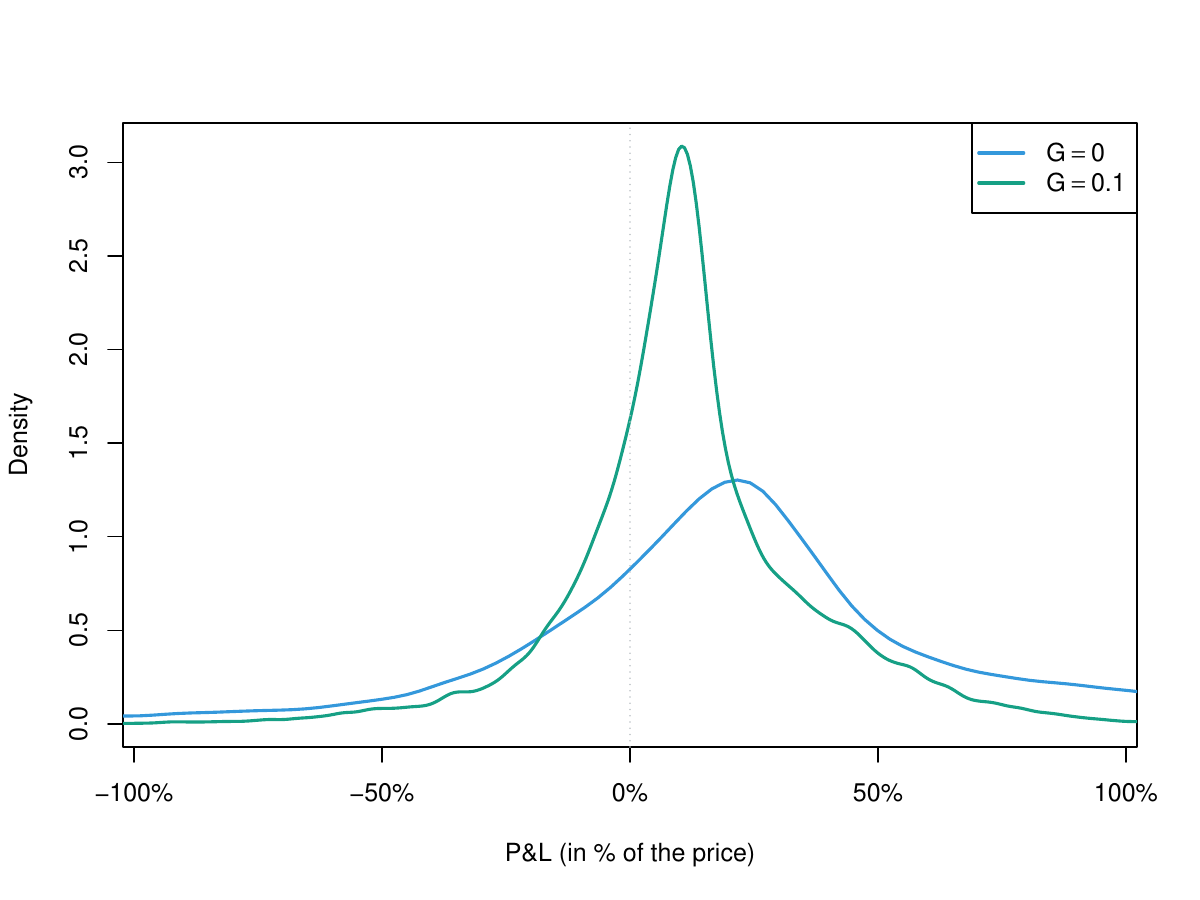}
    \caption{Unconstrained.}
\end{subfigure}
\hfill
\begin{subfigure}[b]{0.48\textwidth}
    \hspace*{-0.25cm}\includegraphics[scale=0.39]{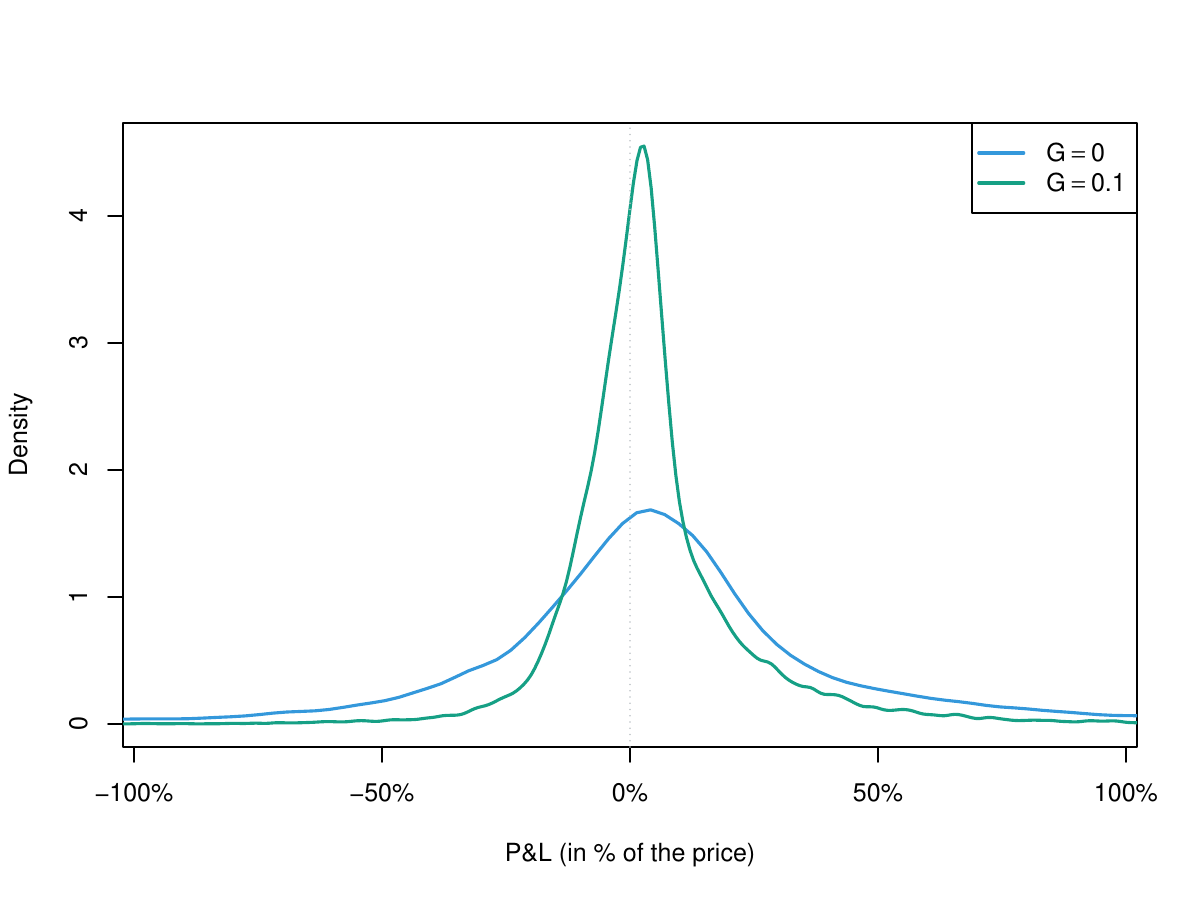}
    \caption{Zero-target.}
    \label{fig:pl_eqcall_zerotarget}
\end{subfigure}
\begin{subfigure}[b]{0.48\textwidth} 
    \includegraphics[scale=0.39]{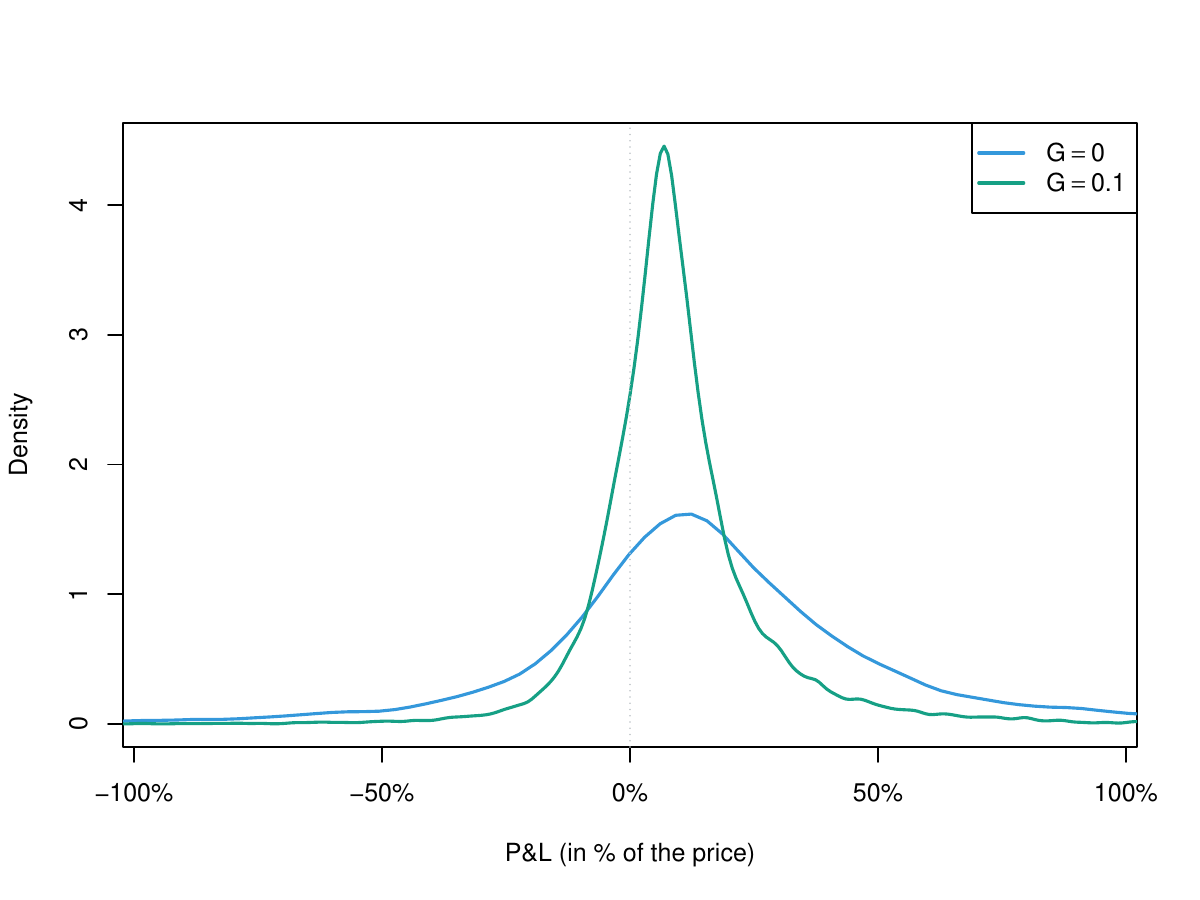}
    \caption{Control-variate.}
\end{subfigure}
\hfill
\begin{subfigure}[b]{0.48\textwidth}
    \hspace*{-0.25cm}\includegraphics[scale=0.39]{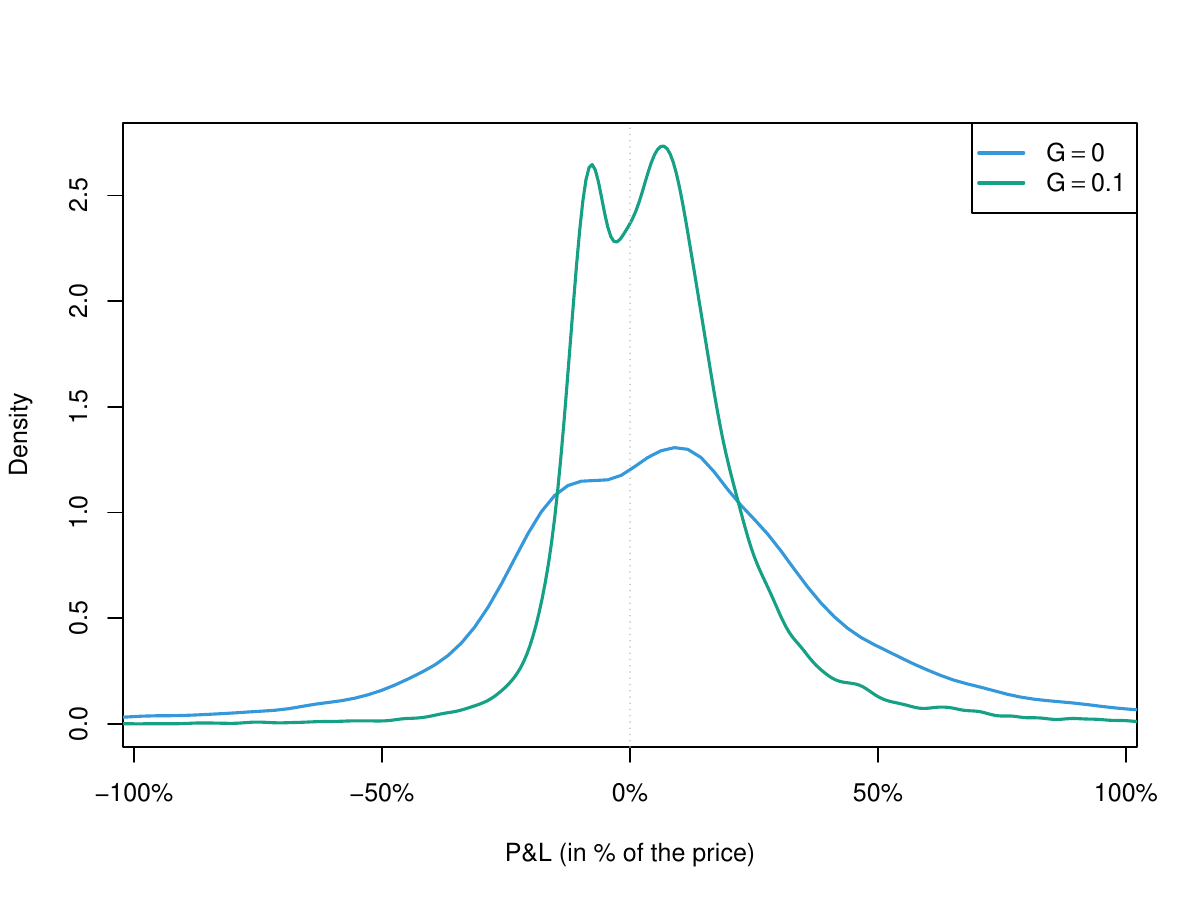}
    \caption{Constrained.}
\end{subfigure}
\caption{Empirical P\&L distributions for the neural network hedge for the Equinox option with parameters $B=1$, $P=0.8$, $R=1$, $T=2, K=1$, and $X_{0} = 1$.}
\label{fig:pl_eqcall}
\end{figure}

\begin{table}[H]
\centering
\resizebox{0.8\textwidth}{!}{%
\begin{tabular}{|c|c|c|c|c|c|}
\hline
\text{Statistic} & \text{Unconstrained} & \text{Zero-target} & \text{Control-variate} & \text{Constrained} \\
\hline
\text{Mean} & 23.273 \% & 8.903 \% & 18.306 \% & 11.396 \%\\
\text{S.D.} & 64.89 \% & 59.20 \% & 57.06 \% & 58.86 \%\\
\hline
\text{Quantile $1\%$} & -186.60 \% & -160.57 \% & -151.35 \% & -155.88 \%\\
\text{Quantile $10\%$} & -33.17 \% & -35.42 \% & -23.52 \% & -32.71 \%\\
\text{Quantile $90\%$} & 93.76 \% & 57.69 \% & 64.53 \% & 60.81 \%\\
\text{Quantile $99\%$} & 209.69 \% & 226.87 \% & 223.67 \% & 223.65 \%\\
\hline
\end{tabular}
}
\caption{P\&L statistics for the pure barrier-call component with $G=0$ (in \% of the Zero-target price).}
\label{stats_eqcall}
\end{table}

\begin{table}[H]
\centering
\resizebox{0.8\textwidth}{!}{%
\begin{tabular}{|c|c|c|c|c|c|}
\hline
\text{Statistic} & \text{Unconstrained} & \text{Zero-target} & \text{Control-variate} & \text{Constrained} \\
\hline
\text{Mean} & 12.269 \% & 4.738 \% & 9.689 \% & 6.192 \%\\
\text{S.D.} & 24.47 \% & 18.77 \% & 17.94 \% & 18.95 \%\\
\hline
\text{Quantile $1\%$} & -56.35   \% & -36.31 \% & -30.70 \% & -32.17 \%\\
\text{Quantile $10\%$} & -13.44 \% & -11.82 \% & -6.05 \% & -11.99 \%\\
\text{Quantile $90\%$} &  43.43 \% & 24.92 \% & 27.96 \% & 26.87 \%\\
\text{Quantile $99\%$} & 79.91 \% & 61.84 \% & 72.86 \% & 71.95 \%\\
\hline
\end{tabular}
}
\caption{P\&L statistics for the full Equinox option with $G=0.1$ (in \% of the Zero-target price).}
\label{stats_eqcallport}
\end{table}

The Unconstrained method produces the least accurate results. Methods that incorporate the payoff condition yield a more accurate price and a better hedge. In the unconstrained framework, the standard deviation for the call component of the Equinox option is 10\% to 14\% higher than other methods (Table \ref{stats_eqcall}), and for the full Equinox option, it is 29\% to 36\% higher (Table \ref{stats_eqcallport}). However, in all cases, the price appears to be overestimated.

\subsubsection{Equinox option with a single neural network}

We estimate the neural network $N_{\theta}$, that directly outputs the price and hedging strategy of the full Equinox option for any cash amount $G$ (in practice, $G \in [0,0.15]$). The network takes $G$ as an extra input dimension and is trained end-to-end with the P\&L loss. Although this approach does not exploit the exact linearity in $G$, it turns out to deliver the best overall performance.

\begin{figure}[H]
\centering
\begin{subfigure}[b]{0.48\textwidth} 
    \hspace*{-0.5cm}\includegraphics[scale=0.36]{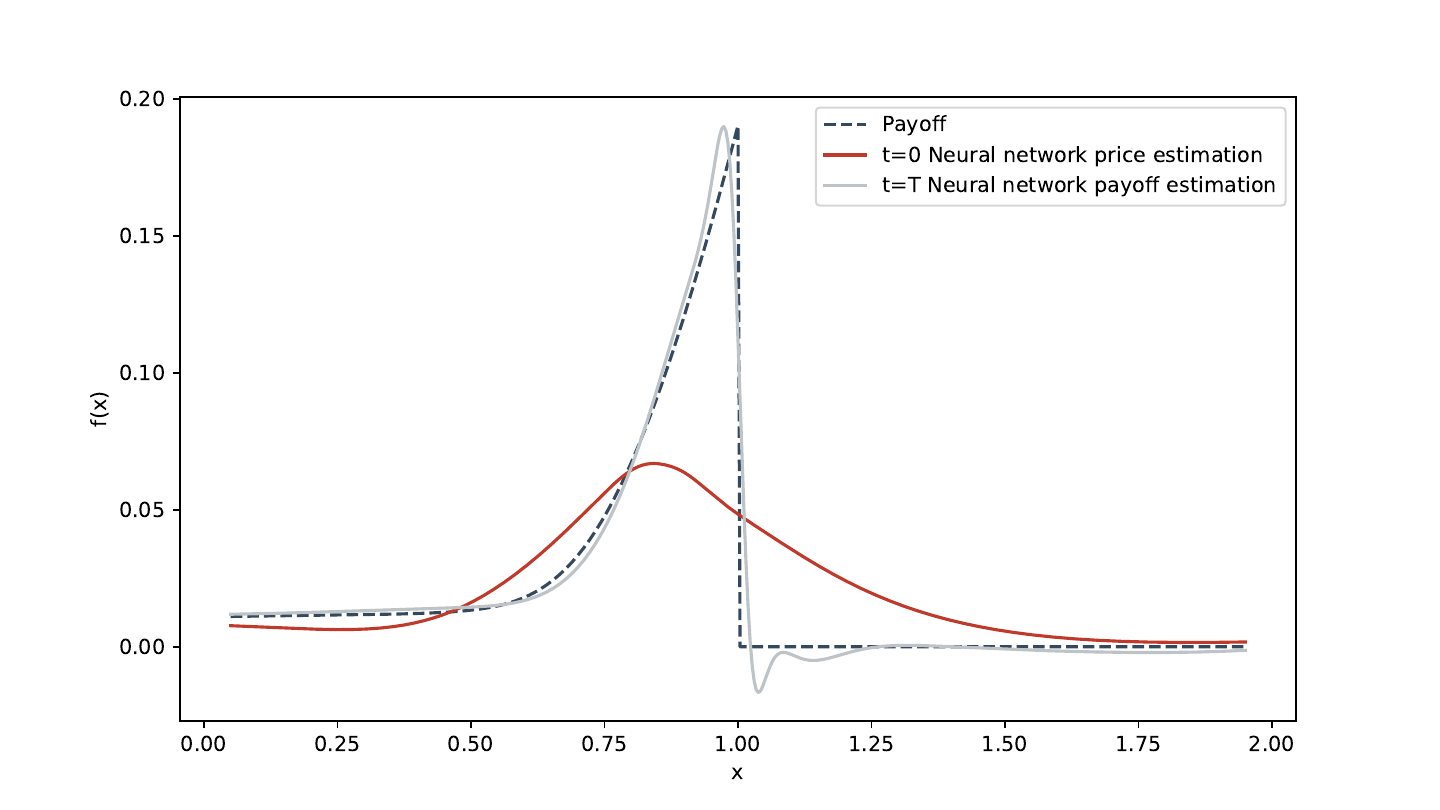}
    \caption{Unconstrained.}
\end{subfigure}
\hfill
\begin{subfigure}[b]{0.48\textwidth}
    \hspace*{-0.6cm}\includegraphics[scale=0.36]{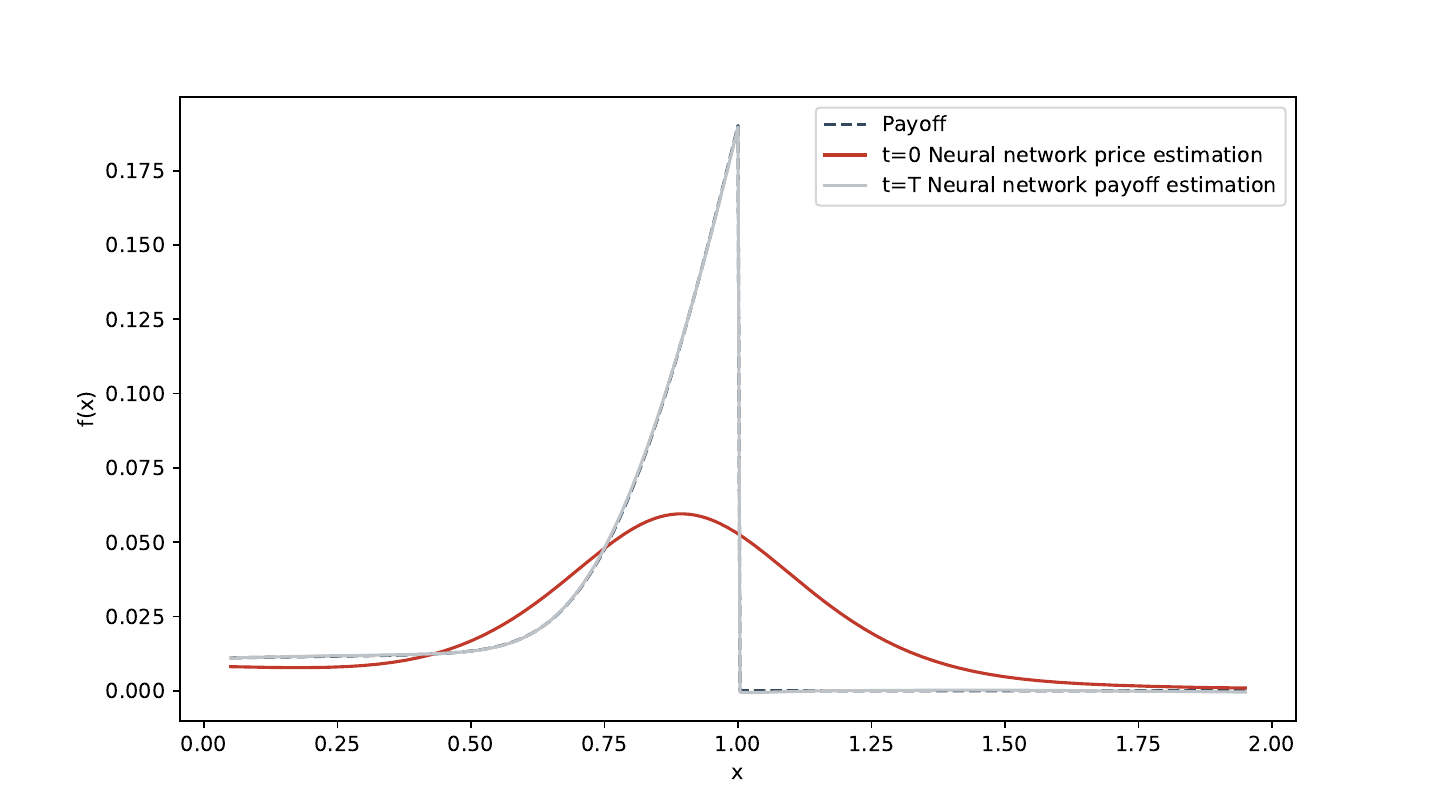}
    \caption{Zero-target.}
\end{subfigure}
\begin{subfigure}[b]{0.48\textwidth} 
    \hspace*{-0.5cm}\includegraphics[scale=0.36]{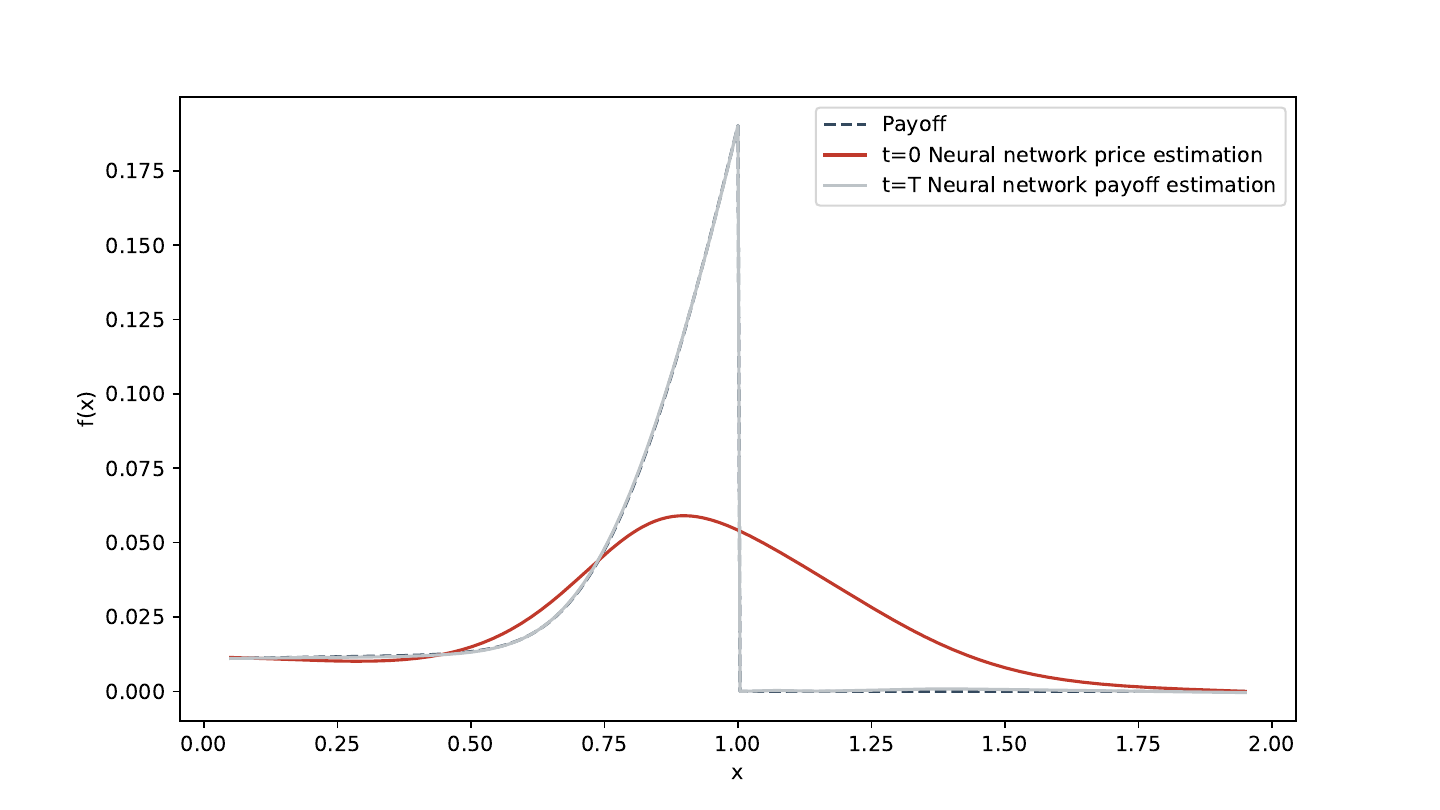}
    \caption{Control-variate.}
\end{subfigure}
\hfill
\begin{subfigure}[b]{0.48\textwidth}
    \hspace*{-0.6cm}\includegraphics[scale=0.36]{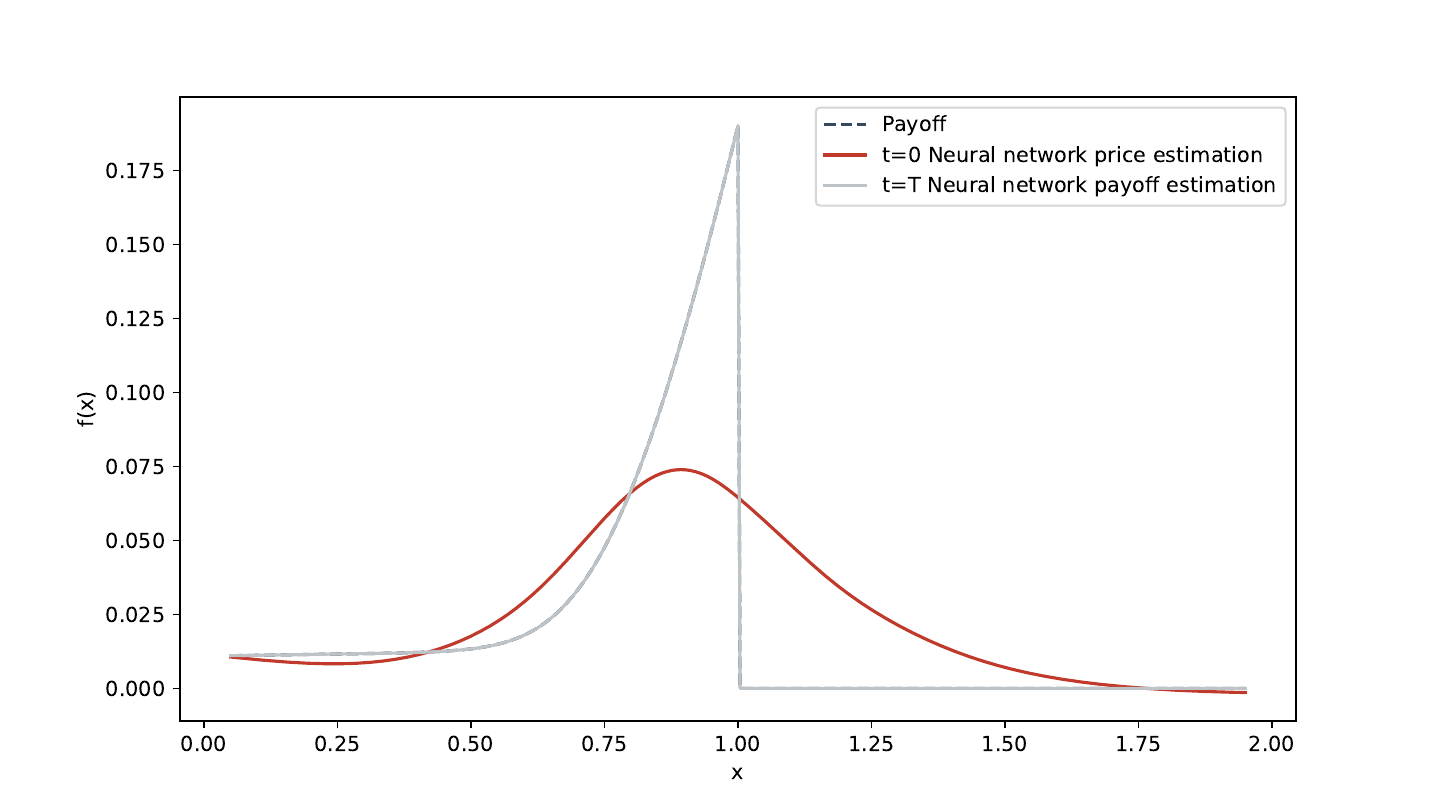}
    \caption{Constrained.}
\end{subfigure}
\caption{Pricing functions for the Equinox option with $G=0$ (pure barrier-call component) with parameters $B=1$, $P=0.8$, $R=1$, $T=2, K=1$.}
\label{fig:price_eqfullg00}
\end{figure}

\begin{figure}[H]
\centering
\begin{subfigure}[b]{0.48\textwidth} 
    \hspace*{-0.5cm}\includegraphics[scale=0.36]{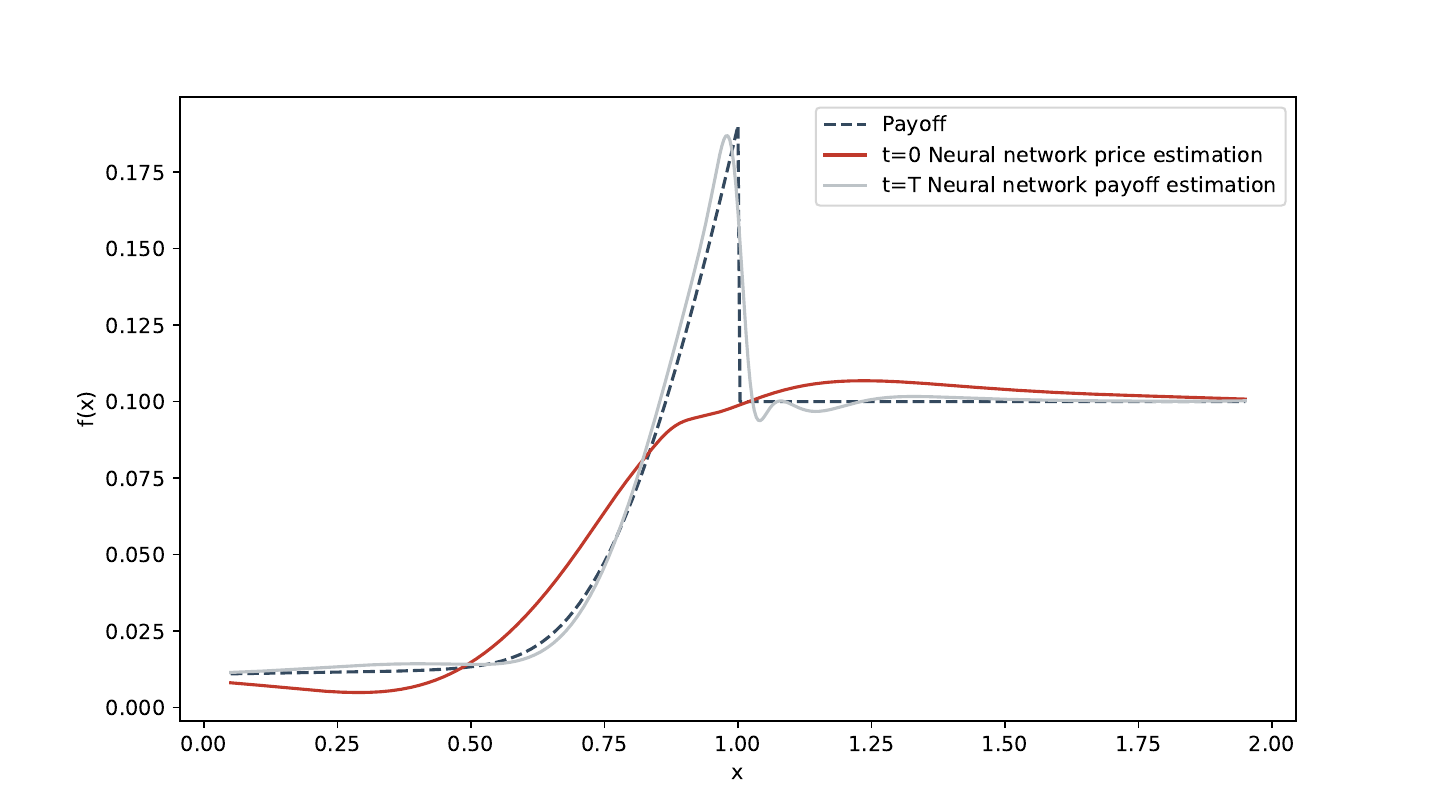}
    \caption{Unconstrained.}
\end{subfigure}
\hfill
\begin{subfigure}[b]{0.48\textwidth}
    \hspace*{-0.6cm}\includegraphics[scale=0.36]{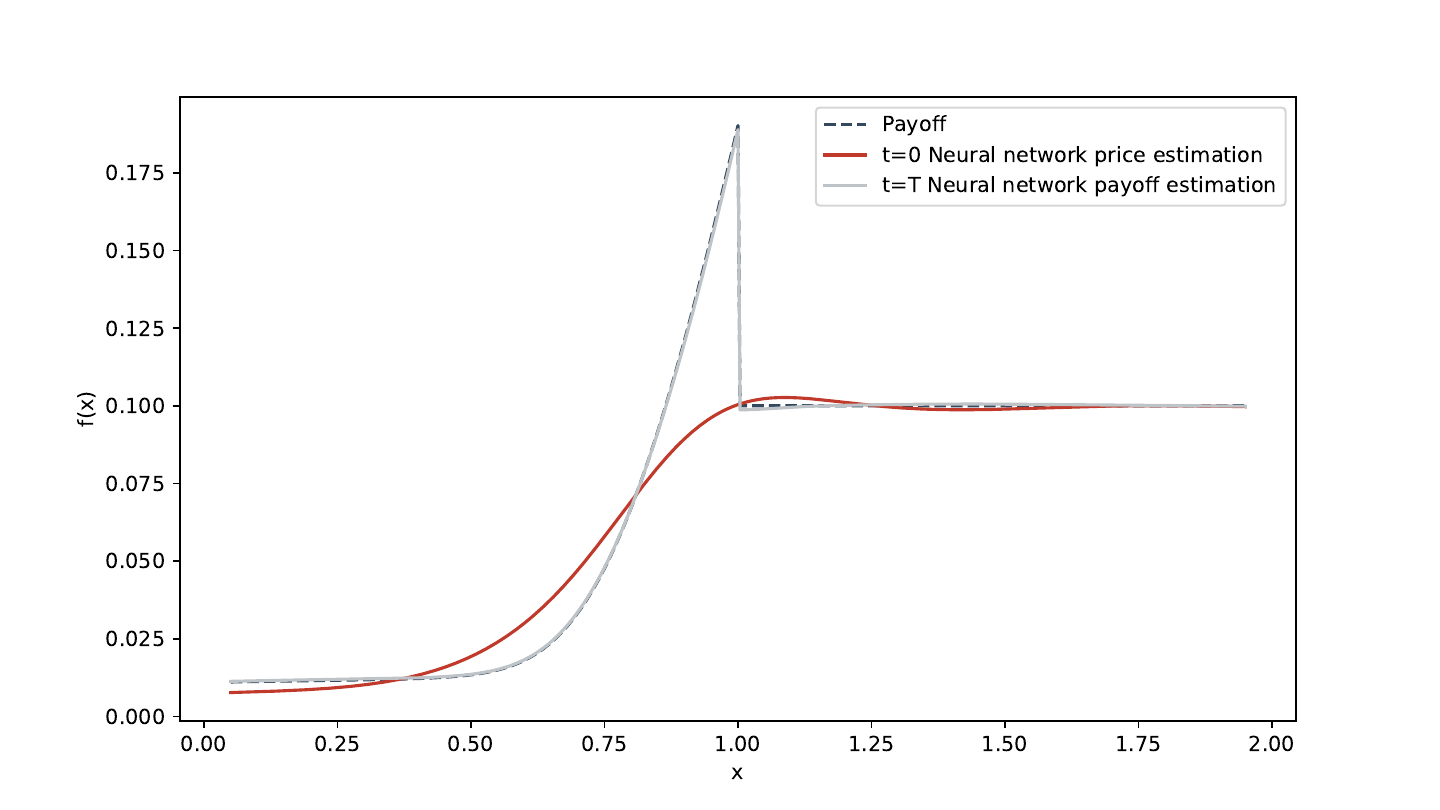}
    \caption{Zero-target.}
\end{subfigure}
\begin{subfigure}[b]{0.48\textwidth} 
    \hspace*{-0.5cm}\includegraphics[scale=0.36]{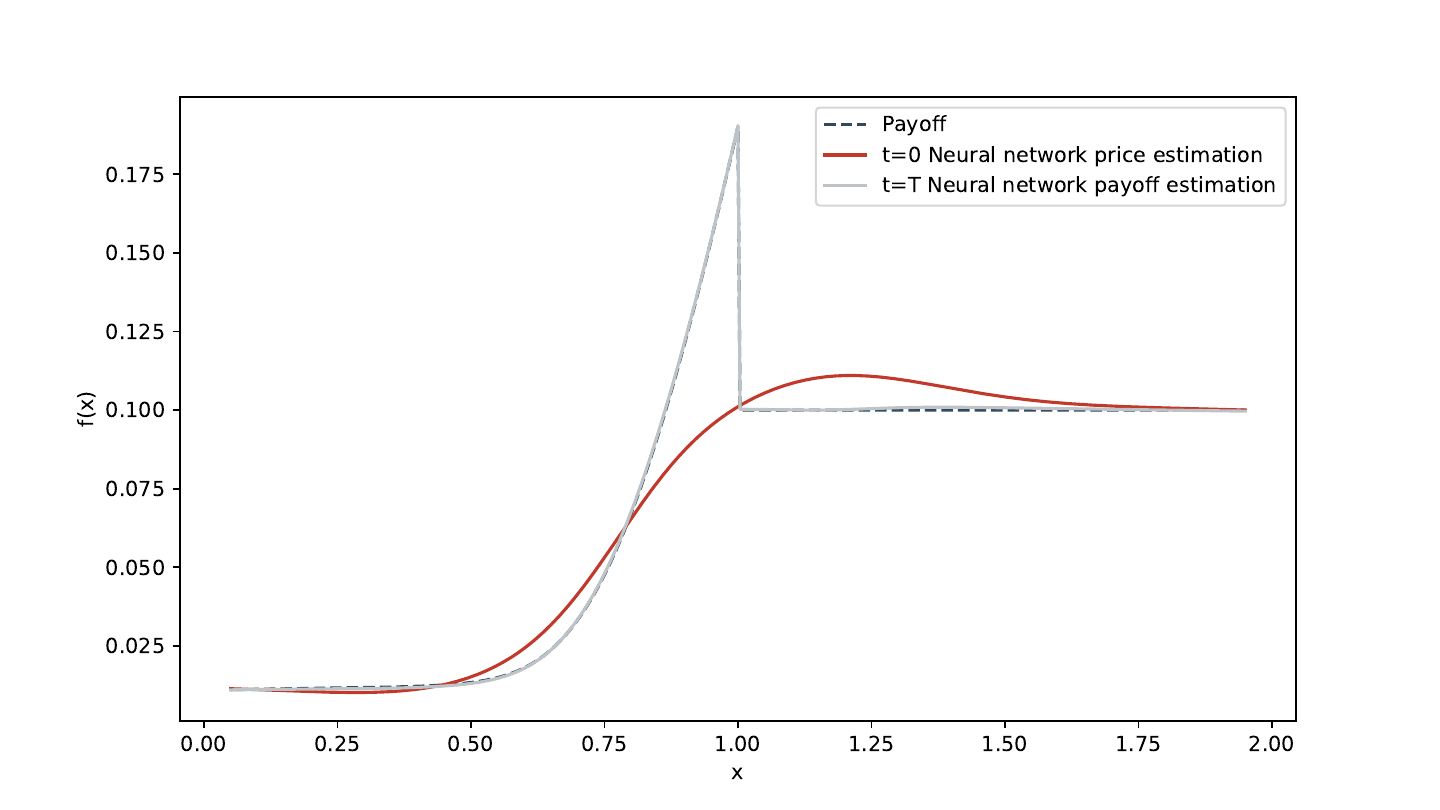}
    \caption{Control-variate.}
\end{subfigure}
\hfill
\begin{subfigure}[b]{0.48\textwidth}
    \hspace*{-0.6cm}\includegraphics[scale=0.36]{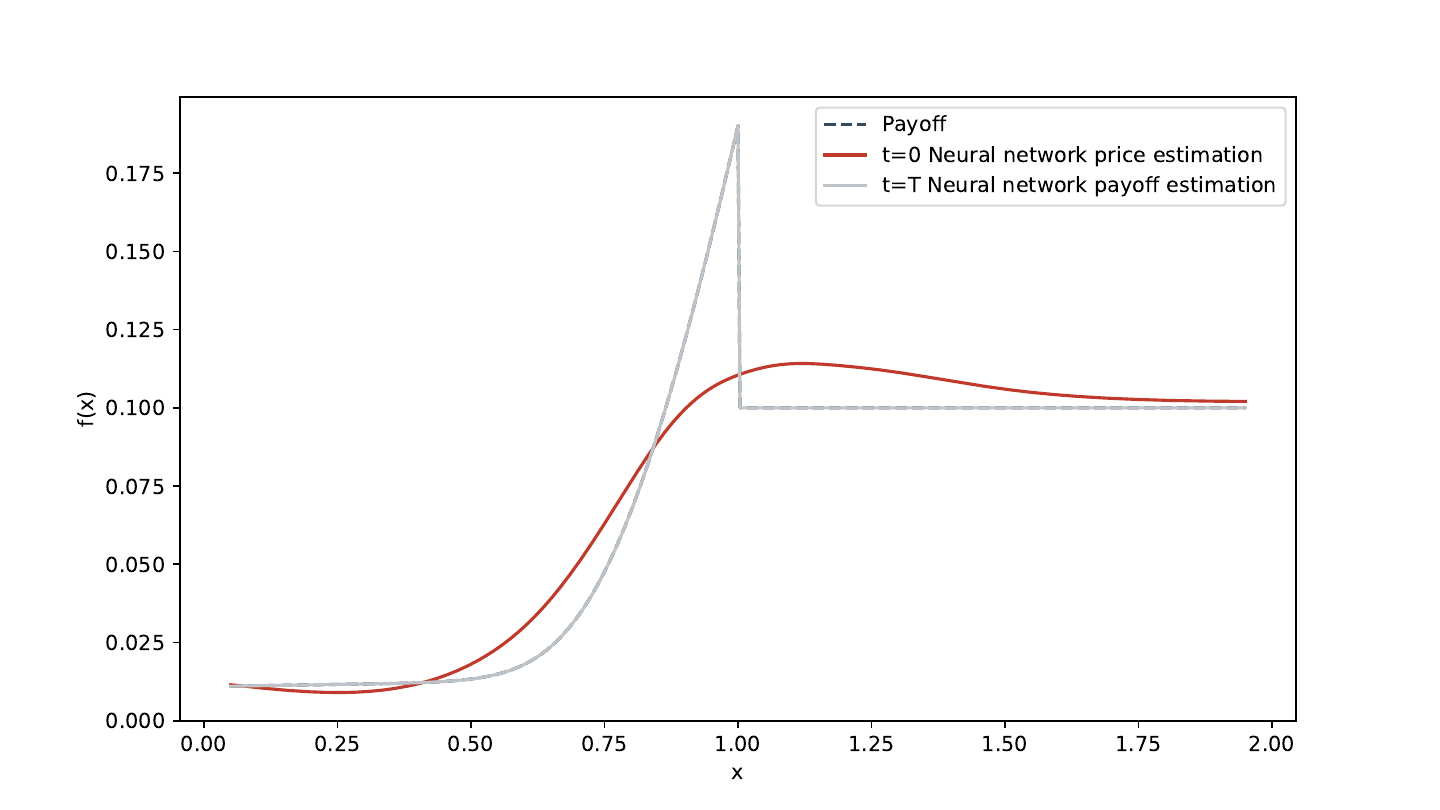}
    \caption{Constrained.}
\end{subfigure}
\caption{Pricing functions for the full Equinox option with $G=0.1$, and $B=1$, $P=0.8$, $R=1$, $T=2, K=1$.}
\label{fig:price_eqfullg01}
\end{figure}

Figures~\ref{fig:price_eqfullg00} and \ref{fig:price_eqfullg01} display the learned price $G=0$ and $G=0.1$, respectively.

Out-of-sample hedging P\&L are reported in Tables~\ref{stats_eqfullg00} ($G=0$) and \ref{stats_eqfullg01} ($G=0.1$). All P\&L figures are expressed as percentages of the Zero-target model price.

\begin{figure}[H]
\centering
\begin{subfigure}[b]{0.48\textwidth} 
    \includegraphics[scale=0.39]{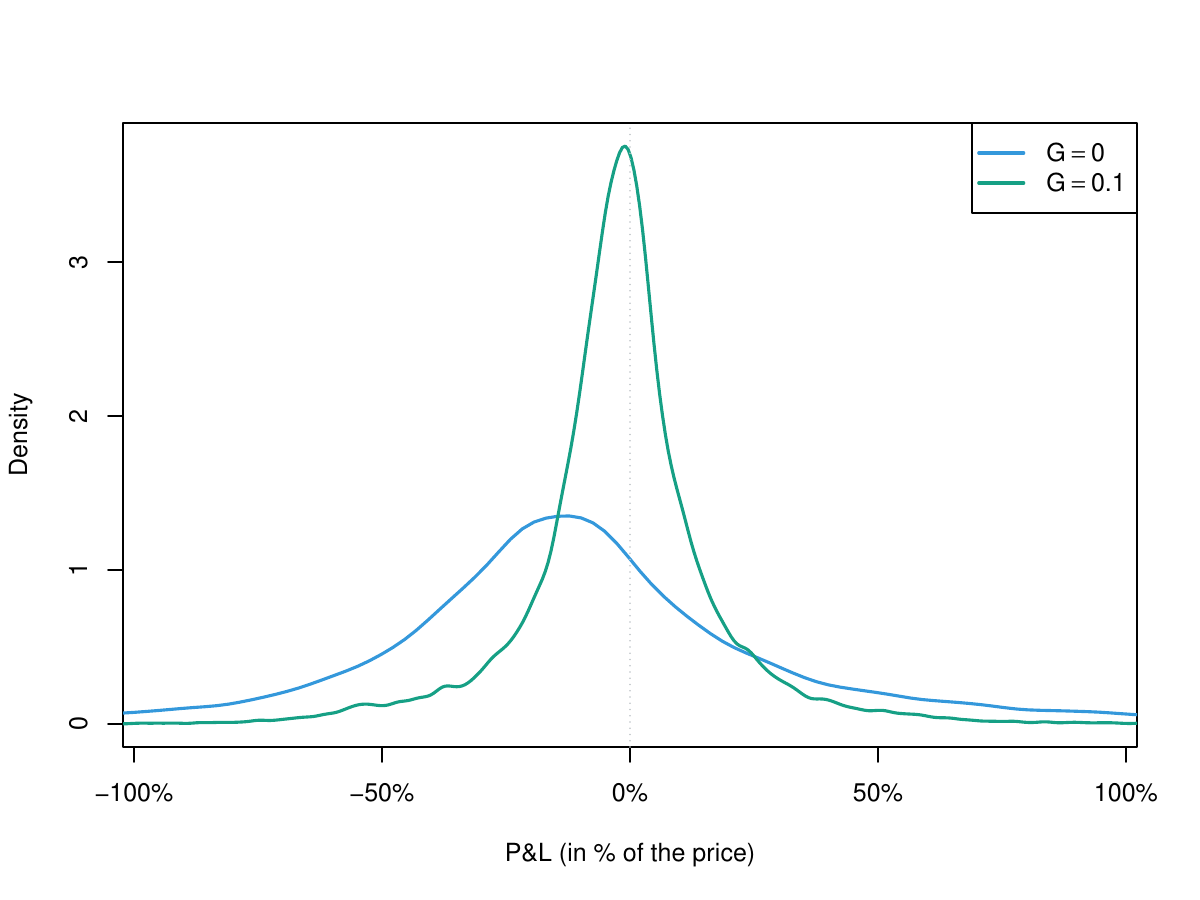}
    \caption{Unconstrained.}
\end{subfigure}
\hfill
\begin{subfigure}[b]{0.48\textwidth}
    \hspace*{-0.25cm}\includegraphics[scale=0.39]{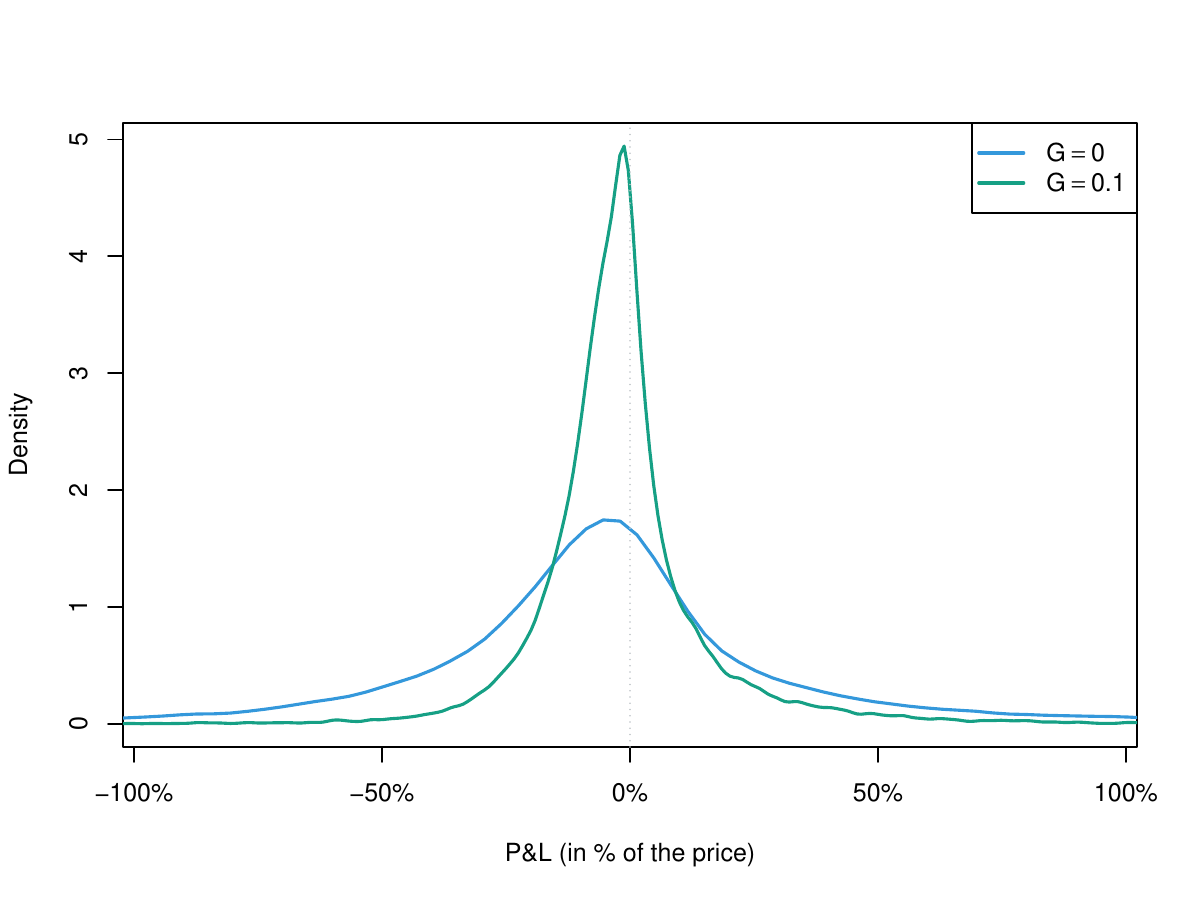}
    \caption{Zero-target.}
\end{subfigure}
\begin{subfigure}[b]{0.48\textwidth} 
    \includegraphics[scale=0.39]{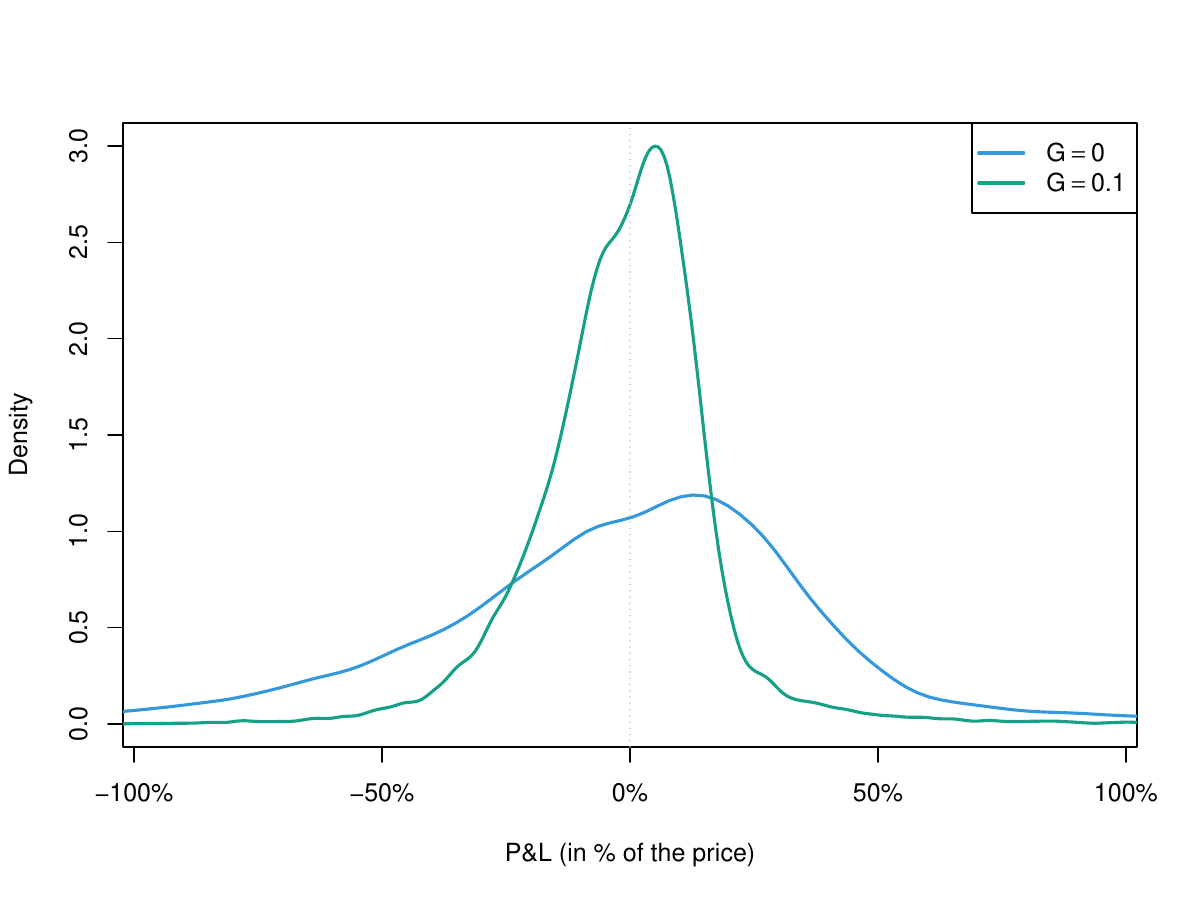}
    \caption{Control-variate.}
\end{subfigure}
\hfill
\begin{subfigure}[b]{0.48\textwidth}
    \hspace*{-0.25cm}\includegraphics[scale=0.39]{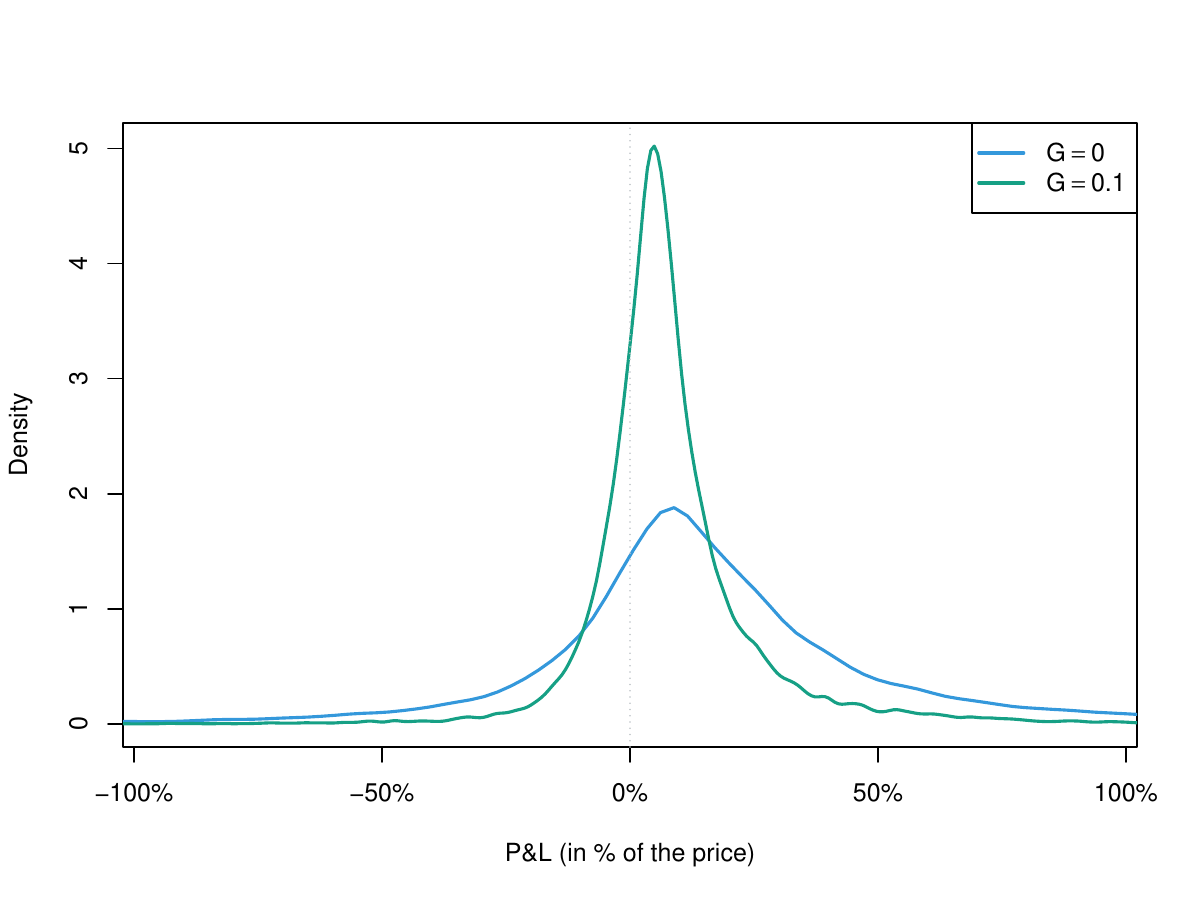}
    \caption{Constrained.}
\end{subfigure}
\caption{Empirical P\&L distributions for the neural network hedge for the Equinox option with parameters $B=1$, $P=0.8$, $R=1$, $T=2, K=1$, and $X_{0} = 1$.}
\label{fig:pl_eqfull}
\end{figure}

\begin{table}[H]
\centering
\resizebox{0.8\textwidth}{!}{%
\begin{tabular}{|c|c|c|c|c|c|}
\hline
\text{Statistics ($G=0$)} & \text{Unconstrained} & \text{Zero-target} & \text{Control-variate} & \text{Constrained} \\
\hline
\text{Mean} & -8.764 \% & -3.218 \% & -1.200 \% & 10.501 \%\\
\text{S.D.} & 61.34 \% & 58.92 \% & 58.32 \% & 57.37 \%\\
\hline
\text{Quantile $1\%$} & -189.42 \% & -170.88 \% & -183.38 \% & -147.91 \%\\
\text{Quantile $10\%$} & -60.36 \% & -49.30 \% & -56.85 \% & -27.48 \%\\
\text{Quantile $90\%$} & 48.99 \% & 43.70 \% & 43.63 \% & 60.20 \%\\
\text{Quantile $99\%$} & 198.69 \% & 208.58 \% & 186.13 \% & 231.32 \%\\
\hline
\end{tabular}
}
\caption{P\&L statistics for the pure barrier-call $G=0$.}
\label{stats_eqfullg00}
\end{table}

\begin{table}[H]
\centering
\resizebox{0.8\textwidth}{!}{%
\begin{tabular}{|c|c|c|c|c|c|}
\hline
\text{Statistics ($G=0.1$)} & \text{Unconstrained} & \text{Zero-target} & \text{Control-variate} & \text{Constrained} \\
\hline
\text{Mean} & -1.774 \% & -1.451 \% & -1.191 \% & 4.864 \%\\
\text{S.D.} & 19.05 \% & 17.64 \% & 17.94 \% & 17.19 \%\\
\hline
\text{Quantile $1\%$} & -57.45 \% & -43.04 \% & -49.76 \% & -30.20 \%\\
\text{Quantile $10\%$} & -21.64 \% & -17.48 \% & -21.90 \% & -8.78 \%\\
\text{Quantile $90\%$} &  18.17 \% & 15.44 \% & 15.55 \% & 23.15 \%\\
\text{Quantile $99\%$} & 55.47 \% & 61.84 \% &  51.53 \% & 71.23 \%\\
\hline
\end{tabular}
}
\caption{P\&L statistics for the pure barrier-call $G=0.1$.}
\label{stats_eqfullg01}
\end{table}

The single-network approach clearly dominates the two-network strategy of Section~\ref{deuxNN}. The Control Variate method achieves the highest price accuracy, while it also exhibits a slightly lower standard deviation, albeit with reduced price accuracy compared to other methods. The Unconstrained method again produces the largest standard deviation among the neural network hedges, exceeding the three methods that embed the terminal condition by 4\% to 7\% for $G=0$ and by 6\% to 11\% for $G=0.1$.
 
\subsection{Robustness with jumps}

To test the robustness of our deep-hedging framework to sudden regime shifts and model misspecification, we extend the stochastic-volatility model of Definition~\ref{model_croissant} by adding pure upward jumps to the volatility process:
\[
    \begin{aligned}
    \Sigma_{s}^{t, \sigma} = \sigma &+ \int_{t}^{s}-a(\Sigma_{u}^{t, \sigma} - \sigma_{\circ})du + \int_{t}^{s}\xi (\Sigma_u^{t, \sigma})^\gamma d(\rho_u^{t, p}W_u^1 + \sqrt{1-(\rho_u^{t, p})^2}W_u^2) \\
    &+ \int_{t}^{s}\kappa dN_u,
    \end{aligned}
\]
where $(N_t)_{0 \leq t \leq T}$ is an independent Poisson process with intensity $\lambda > 0$, and $\kappa > 0$ is the fixed jump size.

\smallbreak

We keep all other parameters from Table~\ref{model_param} and focus exclusively on the digital option (Section~\ref{digital}) priced and hedged with the Zero-target embedding. We train three separate networks with $\lambda \in \{0,\; 0.5,\; 2\}$ and test each of them out-of-sample under all three true intensities.

\begin{figure}[H]
\centering
\includegraphics[scale=0.7]{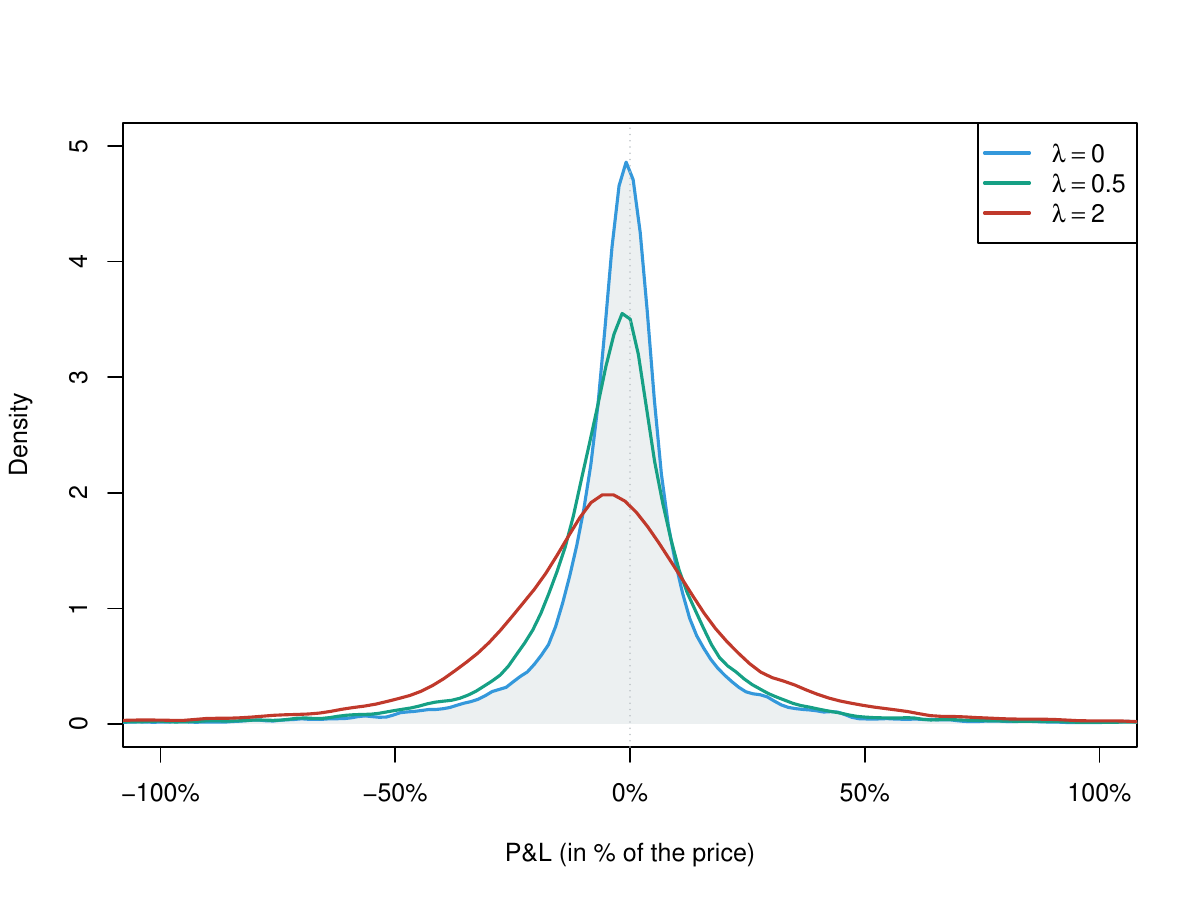}
\caption{Network trained with no volatility jumps $\lambda=0$. P\&L density under true intensity $\lambda \in \{0,\; 0.5,\; 2\}$.}
\label{fig:pl00}
\end{figure}

Figure~\ref{fig:pl00} shows the network trained in the original jump-free model. When a moderate number of jumps occurs in reality, $(\lambda=0.5)$, performance degrades gracefully: the distribution widens and the left tail becomes slightly heavier, but hedging remains reasonably effective. At high intensity ($\lambda=2$), however, the hedge collapses, with catastrophic losses on a non-negligible fraction of paths.

\begin{figure}[H]
\centering
\begin{subfigure}[b]{0.48\textwidth} 
    \hspace*{-0.25cm}\includegraphics[scale=0.36]{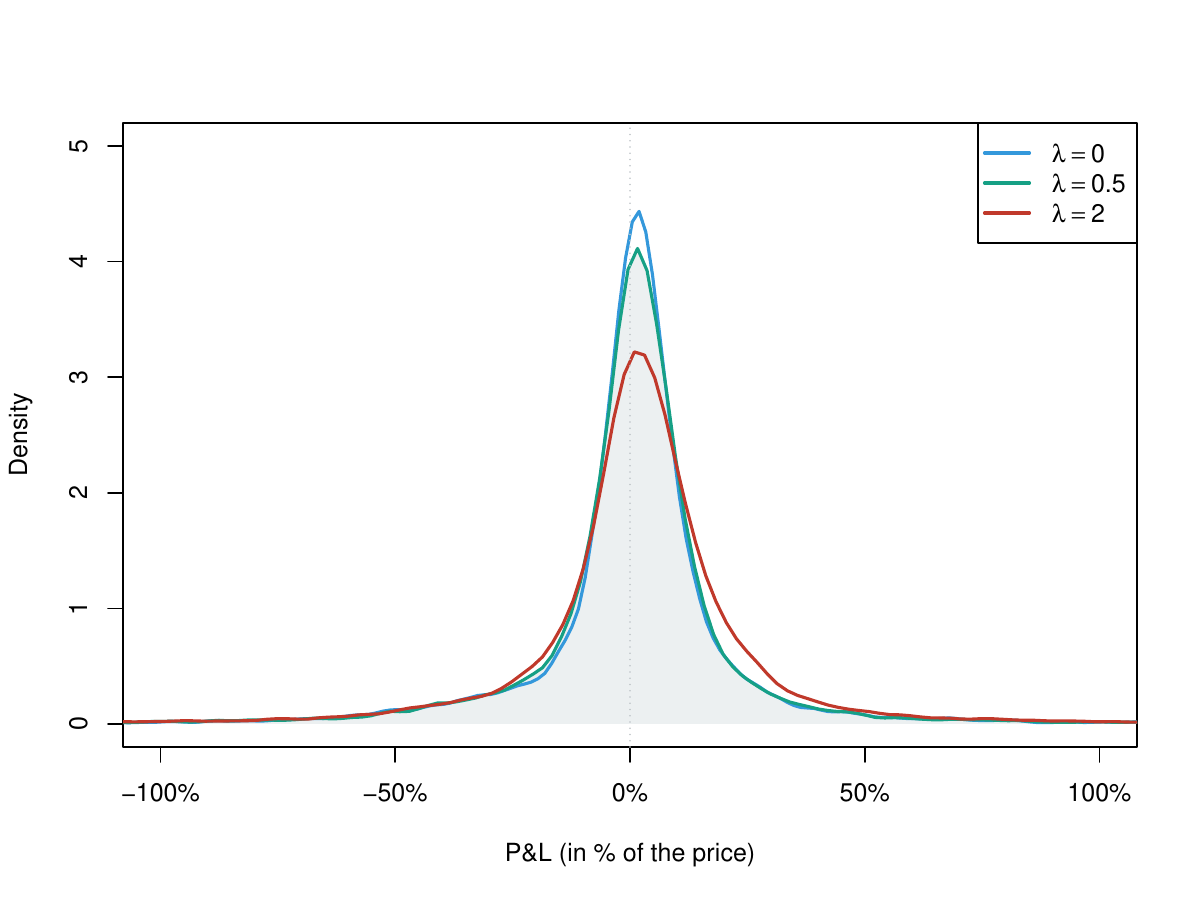}
    \caption{Trained with $\lambda=0.5$.}
    \label{fig:pl05}
\end{subfigure}
\hfill
\begin{subfigure}[b]{0.48\textwidth}
    \hspace*{-0.25cm}\includegraphics[scale=0.36]{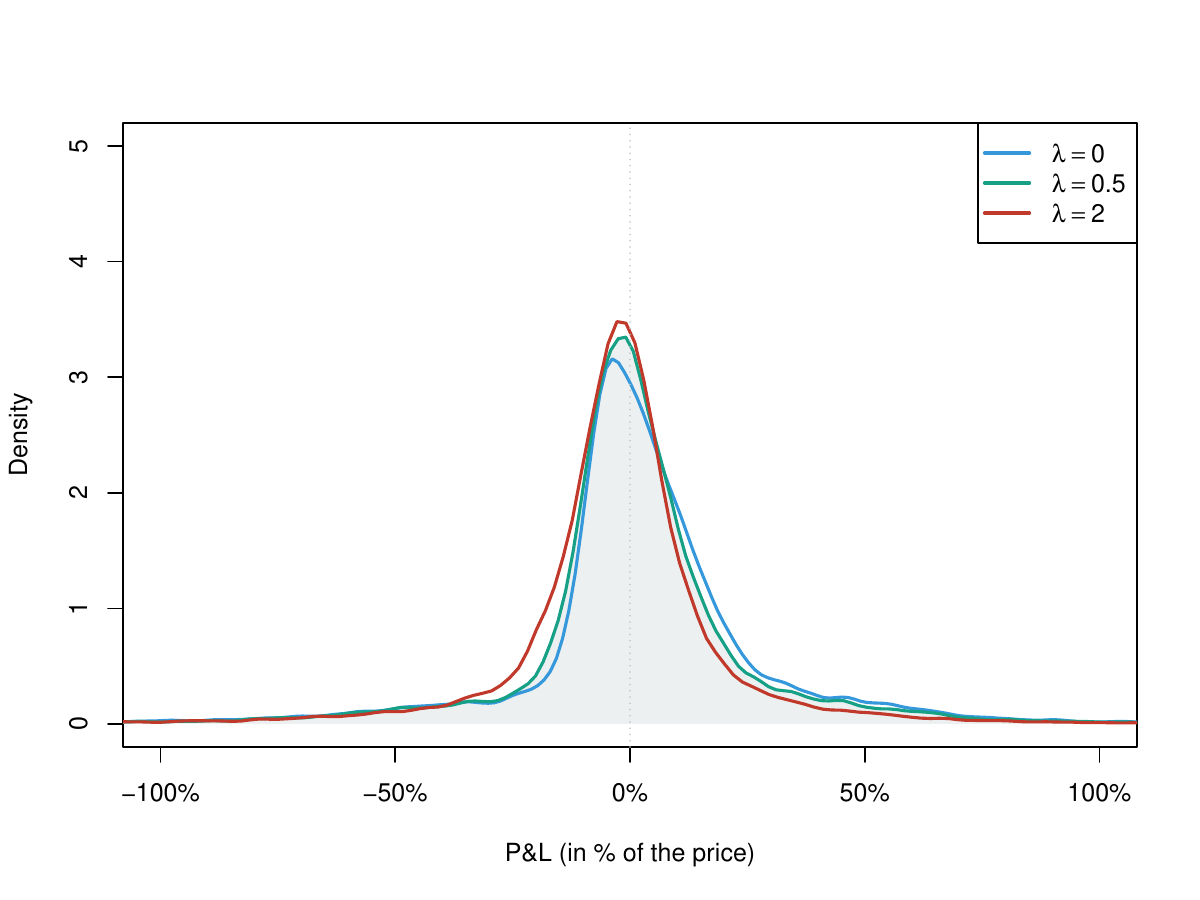}
    \caption{Trained with $\lambda=2$.}
    \label{fig:pl20}
\end{subfigure}
\caption{Networks trained with jumps. P\&L density under true intensity $\lambda \in \{0,\; 0.5,\; 2\}$.}
\label{fig:pl0520}
\end{figure}

In Figure \ref{fig:pl0520}, we examine a model with jumps to assess whether the neural network can still hedge the option effectively and to analyze the P\&L behavior under model misspecifications.

\smallbreak

The network trained with \(\lambda=0.5\) (Figure~\ref{fig:pl05}) delivers excellent hedging performance under the correctly specified model (green curve) and degrades only mildly when tested under \(\lambda=0\) (no jumps) or \(\lambda=2\) (four times more jumps). In particular, its behavior under severe upward misspecification (\(\lambda=2\) in reality) is dramatically better than the collapse observed in Figure~\ref{fig:pl00}.

\smallbreak

The network trained with high intensity \(\lambda=2\) (Figure~\ref{fig:pl20}) is naturally more conservative. Its in-sample performance is somewhat worse than the \(\lambda=0.5\) case (wider distribution), but it remains remarkably stable when tested on lower intensities, including the original jump-free world.

\section{Conclusion}

We modified the neural network to embed the terminal condition, particularly to handle non-smooth payoffs in incomplete markets, leveraging the self-financing property. For simple options, this approach significantly improved hedging performance, even for smooth payoffs, reducing the standard deviation of the P\&L by up to 20\%. For the exotic Equinox option, we explored two approaches: one separating the digital option component, using two simpler neural networks, and another using a single neural network. The single network approach proved more accurate, achieving a P\&L standard deviation up to 10\% lower when embedding the terminal condition. Robustness tests under a model with jumps further demonstrated the resilience of the Zero Target method, particularly when trained with moderate jump intensity, although performance declined under severe model misspecifications. Our approach advances the application of neural networks in quantitative finance by effectively addressing payoff non-smoothness and market incompleteness.

\section*{Acknowledgments}

Nicolas Baradel acknowledges the financial support provided by the \emph{Fondation Natixis} and is grateful to Olivier Croissant, Michel Crouhy, Noureddine Lehdili, Nadhem Meziou, and Denis Talay for numerous fruitful discussions and insightful comments that significantly improved the paper.

\bibliographystyle{plain}
\bibliography{bibliographie}

\end{document}